\documentclass[oldversion]{aa}
\usepackage{epsfig}
\usepackage{natbib}
\titlerunning{Jet-driven AGN winds in Radio Galaxies in the "Quasar Era"}
\authorrunning{Nesvadba et al.}
\begin{document}

\title{Evidence for powerful AGN winds at high redshift: Dynamics of galactic
  outflows 
in radio galaxies during the ``Quasar Era''
\thanks{Based on observations collected at the European Southern
Observatory, Very Large Telescope Array, Cerro Paranal, Chile (
076.A-0684(A))}} 

\author{N.~P.~H. Nesvadba\inst{1,2}, M.~D.~Lehnert\inst{1}, C.~De
  Breuck\inst{3},  A.~M.~Gilbert\inst{4}, W. van Breugel\inst{5}} 

\institute{GEPI, Observatoire de Paris, CNRS, Universite Denis Diderot; 5,
  Place Jules Janssen, 92190 Meudon, France 
\and 
Marie-Curie Fellow
\and
European Southern Observatory, Karl-Schwarzschild Strasse, D-85748 Garching
  bei M\"unchen, Germany
\and
The Aerospace Corporation, P.O. Box 92957, MS M2-266, El Segundo, CA 90245, USA
\and
University of California, Merced, PO Box 2039, Merced, CA 95344
}

\date{Recieved / Accepted}

\abstract{{}AGN feedback now appears as an attractive mechanism to resolve
some of the outstanding problems with the ``standard'' cosmological models, in
particular those related to massive galaxies. At low redshift, evidence is
growing that gas cooling and star formation may be efficiently suppressed by
mechanical energy input from radio sources. To 
directly constrain how this may influence the formation of massive
galaxies near the peak in the redshift distribution of powerful quasars,
z$\sim 2$, we present an analysis of the emission-line kinematics of 3
powerful radio galaxies at z$\sim 2-3$ (HzRGs) based on rest-frame optical
integral-field spectroscopy obtained with SINFONI on the VLT. The host
galaxies of powerful radio-loud AGN are among the most massive galaxies, and
thus AGN feedback may have a particularly clear signature in these galaxies. 

We find evidence for bipolar outflows in all HzRGs, with kinetic energies that
are equivalent to 0.2\% of
the rest-mass of the supermassive black hole. Observed total velocity offsets
in the outflows are $\sim 800-1000$ km s$^{-1}$ between the blueshifted and
redshifted line emission, and FWHMs $\sim 1000$ km s$^{-1}$ suggest strong
turbulence. Line ratios allow to measure electron temperatures, $\sim 10^4$ K
from [OIII]$\lambda\lambda\lambda$4363,4959,5007 at z$\sim 2$, electron
densities ($\sim 500$ cm$^{-3}$) and extinction (A$_V\sim 1-4$ mag).  Ionized
gas masses estimated from the H$\alpha$ luminosity are of order $10^{10}$
M$_{\odot}$, similar to the molecular gas content of HzRGs, underlining that
these outflows may indicate a significant phase in the evolution of the host
galaxy. The total energy release of $\sim 10^{60}$ erg during a dynamical time
of $\sim 10^7$ yrs corresponds to about the binding energy of a massive
galaxy, similar to the prescriptions adopted in galaxy evolution
models. Geometry, timescales and energy injection rates of order 10\% of
the kinetic energy flux of the jet suggest that the outflows are most likely
driven by the radio source. The global energy density release of $\sim
10^{57}$ erg s$^{-1}$ Mpc$^{-3}$ may also influence the subsequent evolution
of the HzRG by enhancing the entropy and pressure in the surrounding halo and
facilitating ram-pressure stripping of gas in satellite galaxies that may
contribute to the subsequent mass assembly of the HzRG through low-dissipation
``dry'' mergers.  }


\maketitle

\section{Introduction}
\label{introduction}
AGN feedback has now become a crucial process in most models of galaxy
formation and evolution. Current models postulate that massive galaxies
experienced an early epoch of vigorous star formation that was terminated by a
nearly instantaneous, powerful blow-out phase triggered by the AGN
\citep[e.g.][]{silk98, scannapieco04, springel05, dimatteo05, croton06,
bower06, hopkins06}. Such a mechanism would heat and remove most of the
ambient gas of the galaxy, and thus reconcile the discrepancy between the
hierarchical standard model of galaxy formation and observational constraints

Bolometric luminosities of
powerful AGN at high redshift correspond to an energy output of $10^{60-61}$
erg during an AGN lifetime of $10^{7-8}$ yrs.  Although this is equal, or even
exceeds the binding energy of a massive galaxy, this energy will only have an
impact on the evolution of the host galaxy, if it is efficiently transferred
to the ambient gas with a coupling efficiency of at least a few percent.

Observationally, evidence is growing that AGN feedback may be related mostly
to radio-loud AGN. Perhaps the most spectacular examples at low redshift are
the giant cavities in the X-ray halos of massive galaxy clusters, which appear
to inject few $\times 10^{60}$ erg into the surrounding gas within a few
$\times 10^7$ yrs, sufficient to suppress gas cooling
\citep{boehringer93,birzan04,rafferty06,mcnamara07}. \citet{best05,best06}
find more 
subtle, yet general evidence that radio sources may balance gas cooling in
$\sim 2000$ radio-loud early-type galaxies taken from the Sloan Digital Sky
Survey 
and the NVSS and FIRST radio surveys. Interestingly, the fraction of radio
loud galaxies seems to be a strong function of mass, suggesting that such
radio-driven feedback will act predominantly in the most massive galaxies, in
broad agreement with the predictions of galaxy evolution models. 
This is also
supported by the observation that powerful radio galaxies are among the most
massive galaxies at all redshifts \citep{debreuck01,willott03,rocca04}. Since
AGN feedback is expected to act predominantly in massive galaxies, we may
expect that its influence will be most clearly expressed in particularly
massive galaxies.

However, since massive early-type galaxies appear to have completed most of
their growth at high redshift \citep[e.g.,][and references
therein]{rudnick03}, directly investigating the impact of jet-driven AGN
feedback during the formation of massive galaxies requires direct observations
of radio-loud galaxies at z$\ge 1$.  Interestingly, longslit spectroscopy
of powerful radio galaxies at z$\ge 1$ (and in a few cases at lower redshifts)
revealed strongly distorted emission line kinematics, with velocities often
exceeding $\sim 1000$ km s$^{-1}$ along the axis of the radio jet
\citep[e.g.,][]{tadhunter91,mccarthy96,evans98,baum00,villar99,
inskip02,clark98,best97}.  Similarly, [OIII]$\lambda$5007 narrow-band imaging
of 4C19.71 at z$\sim 3.6$ revealed a giant emission line region extending over
$\sim 74$ kpc \citep{armus98}. This emission line region, corresponding to an
ionized gas mass of $10^{8-9}$ M$_{\odot}$ aligns roughly with the axis of the
radio jet, which led \citeauthor{armus98} to suspect that the jet and the
nebulosity may be physically related. \citet{heckman91a,heckman91b} also find
this for radio-loud quasars. Moreover, Ly$\alpha$ spectroscopy and line imaging
revealed luminous emission line halos with sizes that appear related to the
size of the radio jet, sometimes with kinematically more quiescent gas beyond
\citep[e.g.][]{villar03,villar06,villar07}.

However, the complexity of the
velocity fields and relatively large spatial extent of the emission line
regions made it difficult to robustly infer the global characteristics of
these nebulae.
Integral-field spectrographs now open a new avenue to study the underlying
dynamical mechanisms with unprecedented robustness and across the full
two-dimensional surface of the emission line nebulae. Over recent years, our
understanding of galaxy evolution in the early universe has significantly
advanced due to the discovery of large numbers of galaxies
at similar redshifts \citep[e.g.,][]{steidel96,smail02}, and during this
time, modeling of galaxy and structure formation has improved. This
allows us to reassess the emission line kinematics of powerful high-redshift
radio galaxies (HzRGs). Near-infrared studies are particularly well suited to
study the inner regions of HzRG extended emission line regions, allowing for
relatively high spatial resolution of $\sim 0.5$\arcsec\ even with
seeing-limited observations, and opening a window to the rest-frame optical
line emission where extinction is less severe than in the rest-frame UV. 

\citet{nesvadba06} find strong evidence for energetic outflows of ionized
gas from rest-frame optical integral-field spectroscopy of the powerful radio
galaxy MRC1138-262 at z$\sim 2.2$. With a spatial 
extent of 30 kpc, relative velocities and line widths of FWHM$\sim 1000$ km
s$^{-1}$ 
across the source, corresponding to kinetic energies of few$\times
10^{60}$ erg, this gas may be experiencing a feedback
episode as dramatic as necessary to unbind a significant fraction of the
ambient gas from the halo of a massive galaxy. The kinetic energy of the
outflow corresponds to a few percent of the jet kinetic energy, and the
dynamical timescale of a few $\times 10^7$ yrs appears roughly similar to the
age of the radio source. Moreover, the turbulent, luminous gas extends to the
size of the radio jets. In contrast, very similar observations do not show
evidence for such outflows in compact radio galaxies at similar redshifts
\citep{nesvadba07b}, which are likely the younger analogs to extended
HzRGs. This agrees with expectations. If the entrainment rates are
similar, then only a few percent of the ISM will already be
affected by the radio source for ages less than $\sim 10^6$ yrs.

Currently, the host galaxies of powerful, radio-loud AGN (HzRGs and quasars)
represent the only galaxy population at z$\ge 2$ with such extreme
kinematics. To address the question whether MRC1138-262 is a ``one-off'', or
whether AGN feedback is a common phenomenon in HzRGs, we continue the analysis
of \citet{nesvadba06,nesvadba07b}, adding another three HzRGs with
near-infrared integral-field data at z$\approx 2-3$. 
Comparing with the
samples of \citet[e.g.,][]{iwamuro03,debreuck00,humphrey08,baum00}, we find
that their rest-frame UV-optical properties are within the typical range of
powerful HzRGs.
These galaxies have
extended radio morphologies, and relatively similar radio power to
MRC1138-262, ${\cal P}_{1.4 GHz} \sim 10^{26}$ W Hz$^{-1}$.  
Thus, they are among the most powerful radio galaxies observed at all
redshifts. Given the observational difficulties in studying high-redshift
galaxies, this seems the most promising way of identifying the fingerprints of
powerful AGN-driven feedback and underlying physical processes directly from
the gas kinematics of powerful AGN host galaxies at z$\sim 2$. 

However, these radio
powers are not exceptional in the early universe. \citet{willott01} studied
the redshift evolution of the radio luminosity function and find a population
of galaxies with radio powers ${\cal P}_{151 MHz} \sim 10^{26.5-29.5}$ W
Hz$^{-1}$ at 151 MHz, which shows a strong peak near z$\sim 2$ and rapidly
declines towards lower redshifts. Finally, the radio emissivity is likely a
strong function of environment and evolutionary stage of the synchrotron
plasma \citep[e.g.,][]{kaiser97a,kaiser97b}, causing changes in the radio
luminosity of more than an order of magnitude. Hence, by selecting
particularly powerful sources, we may bias our sample more in terms of the
evolutionary stage or environment than the kinetic luminosity of the radio
source. 

Specifically, we observed MRC0316-257 at z$=3.13$, MRC0406-244 at z$=2.42$,
and TXS0828+192 at z$=2.57$. Entrained ionized gas masses and kinetic energies
seem very similar to MRC1138-262 in all three cases. This leads us to propose
that AGN feedback may be a common phenomenon in powerful radio galaxies during
the ``Quasar Era''. 

\section{Observations and Data Reduction}
Observations in the H and K bands were carried out with the integral-field
spectrograph SINFONI \citep{bonnet04} on the VLT in December 2005 under good
conditions.

SINFONI is a medium-resolution, image-slicing integral-field spectrograph,
with 8\arcsec$\times$ 8\arcsec\ field of view at a 0.25\arcsec$\times$
0.25\arcsec\ pixel scale, and spectral resolution of $R\sim
3000$, and $R \sim 4000$ in the H, and K bands, respectively. Individual
exposure times were 450 s in H and 600 s in K, respectively. Total
exposure times are given for each band and each source in Table
\ref{tab:exposure}.

We use the IRAF \citep{tody93} standard tools for the reduction of
longslit-spectra, modified to meet the special requirements of integral-field
spectroscopy, and complemented by a dedicated set of IDL routines. Data are
dark-frame subtracted and flat-fielded. The position of each slitlet is
measured from a set of standard SINFONI calibration data, measuring the
position of an artificial point source. Rectification along the spectral axis
and wavelength calibration are done before night sky subtraction to account
for some spectral flexure between the frames. Curvature is measured and
removed using an arc lamp, before shifting the spectra to an absolute (vacuum)
wavelength scale with reference to the OH lines in the data. To account for
variations in the night sky emission, we normalize the sky frame to the
average of the object frame separately for each wavelength before sky
subtraction, masking bright foreground objects, and correcting for residuals
of the background subtraction and uncertainties in the flux calibration by
subsequently subtracting the (empty sky) background separately from each
wavelength plane. In the H band, where the suppression of night sky lines is
particularly difficult, we additionally clipped all strongly deviant pixels at
wavelengths near prominent night sky lines.

The three-dimensional data are then reconstructed and spatially aligned using
the telescope offsets as recorded in the header within the same observing
block (a sequence of typically 5 to 10 individual exposures with a total
length of about an hour), and by cross-correlating the line images from the
combined data in each observing block, to eliminate relative offsets between
different observing block.  Telluric correction is applied to each individual
cube before the cube combination. Flux scales are obtained from standard star
observations. From the light profile of the standard star, we measure the FWHM
spatial resolution in the combined cube, which is typically $\sim 0.6$\arcsec.

\subsection{Complementary Data Sets and Alignment}
\label{ssec:alignment}

We complement our SINFONI data through deep, A-array 1.4 GHz radio maps of
\citet{carilli97} with a spatial resolution of $\sim 0.4$\arcsec. These data
were kindly provided to us by 
C. Carilli. 
Studying the interplay between radio jets and the ISM of high-redshift
galaxies requires a relative alignment that is accurate to $\le$ 0.5\arcsec,
which is better than the accuracy of the absolute astrometry reached with the
VLT. We therefore assume that the radio 
core falls onto the center of the continuum emission for all sources. Note
that this coincides with the position of the largest velocity gradient in the
velocity maps of TXS0828+193 and MRC0406-244 to within less than a few tenths of
a second of arc. Remaining small offsets are likely due to projection
effects. Since we did not detect MRC0316-257 in the continuum, we align the
radio core with the steepest gradient in the velocity map, implying that the
basic physical mechanism of the outflow is similar in the 3 sources -- which
is justified given their overall similarity.  We also note that, unlike in
MRC1138-262 \citep{nesvadba06}, we do not find a nuclear broad-line region
related to the AGN in any of the three sources.

\section{Continuum and Line Emission Morphologies and Integrated Spectra}   
\label{sec:morphologies}
We constructed continuum-free line images by summing over the spectral regions
of the emission lines in the integrated spectra, and removing the underlying
continuum emission. Rather than adopting a fixed aperture, we selected all
pixels where the [OIII]$\lambda$5007 emission line was detected at a
signal-to-noise ratio greater than 3. This helps to maximize the
signal-to-noise ratio, which may otherwise be affected by night-sky residuals,
whose strength often varies across the detector. To extract line-free
continuum images, we fitted the spectrum of each spatial pixel where it is
unaffected by bright emission lines or strong night sky line residuals. We
interpolated over the contaminated parts before collapsing the spectra.

\subsection{MRC0316-257}

\begin{figure}
\centering
\epsfig{figure=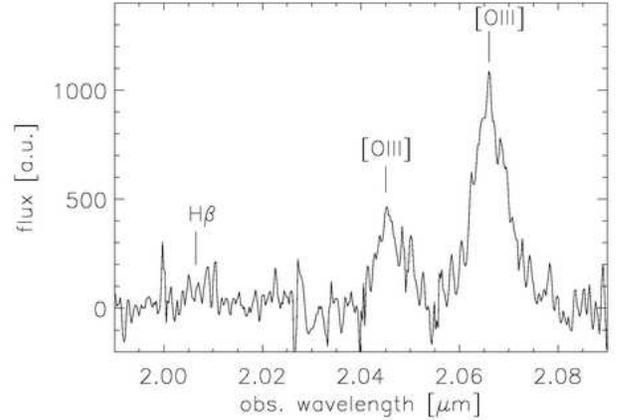,width=0.48\textwidth}
\caption{Integrated K band spectrum of MRC0316-257 at z$=$3.13, extracted from
all pixels in which the [OIII]$\lambda$5007 line emission exceeded 3 $\sigma$,
and covering the [OIII]$\lambda$$\lambda$4959,5007 doublet and H$\beta$.}
\label{fig:0316intspec}
\end{figure}

The integrated spectrum of MRC0316-257 is shown in
Fig.~\ref{fig:0316intspec}. The [OIII]$\lambda\lambda$4959,5007 doublet and
H$\beta$ are detected at redshifts of z$=$3.1273\footnote{Adopting the flat
$\Omega_{\Lambda}=0.7$ cosmology with H$_{0}=70$ km s$^{-1}$ Mpc$^{-1}$ we use
a luminosity distance D$_{L}=26.7$ Gpc and angular size distance D$_{A}=$ 1.5
Gpc. The size scale is 7.6 kpc/\arcsec. The age of the universe at z$\sim
3.13$ for this cosmological model is 2.0 Gyr.} and z$=$3.1292 for
[OIII]$\lambda$5007, respectively, and have line widths of FWHM$=$1465 km
s$^{-1}$ and FWHM$=$900 km s$^{-1}$ for [OIII]$\lambda$5007 and H$\beta$,
respectively\footnote{Unless stated explicitly, all kinematic parameters
    are in the rest-frame.}. See Table~\ref{tab:emlines0316} for a summary of the emission
line properties and related uncertainties.

We show the [OIII]$\lambda$5007 emission line morphology in
Fig.~\ref{fig:maps0316}. Line emission is detected over an area of
$\approx 2.8\arcsec \times 1.2\arcsec$ (21 kpc $\times$9 kpc), and is
elongated along the axis of the radio jet (the 1.4 GHz VLA map is shown as
contours in Fig.~\ref{fig:maps0316}). H$\beta$ and continuum emission are
very faint, so that their morphology could not be measured.

\subsection{MRC0406-244}
\label{ssec:0406}
MRC0406-244 at z=2.44\footnote{The luminosity distance is D$_{L}=19.8$ Gpc and
angular size distance D$_{A}=$ 1.7 Gpc. The size scale is 8.1 kpc/\arcsec. The
age of the universe for this redshift and cosmological model is 2.6 Gyr.} has
a splendid hour-glass shaped, rest-frame UV and optical morphology which has
previously been studied with NICMOS and WFPC imaging on-board the Hubble Space
Telescope \citep{pentericci01}, as well as with ground-based observations 
\citep{mccarthy91,mccarthy92, rush97} in the optical and near-infrared. As
pointed out by \citet{rush97}, the morphology is a strong function of
waveband. In particular, the J-band morphology is much more compact than,
e.g., the morphology in the Ks or g-band. \citet{rush97} suspected that this is
most likely due to contamination with strong emission lines (H$\alpha$ in the
Ks band, Ly$\alpha$ in the g band).

\begin{figure}
\centering
\epsfig{figure=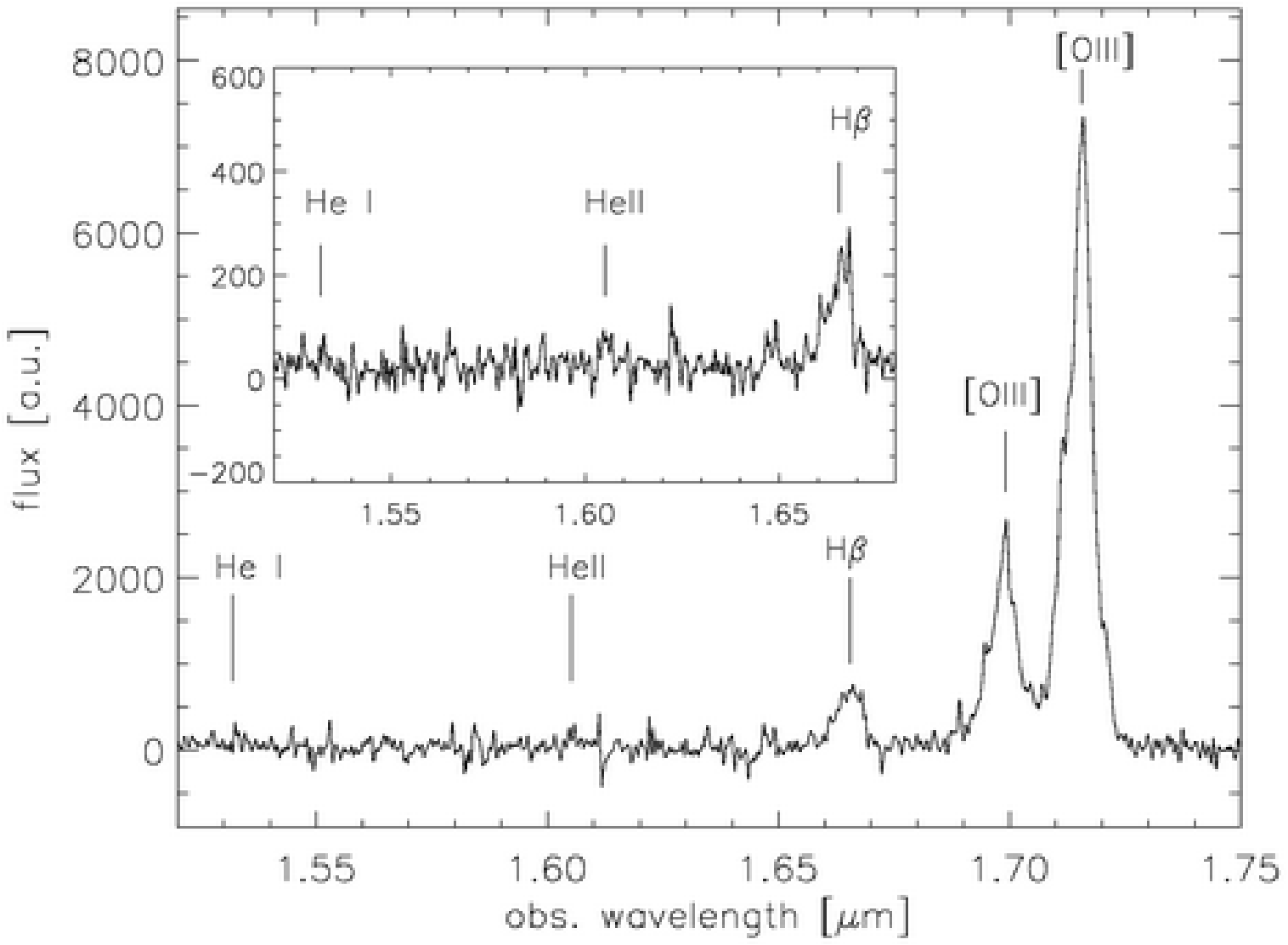,width=0.48\textwidth}
\epsfig{figure=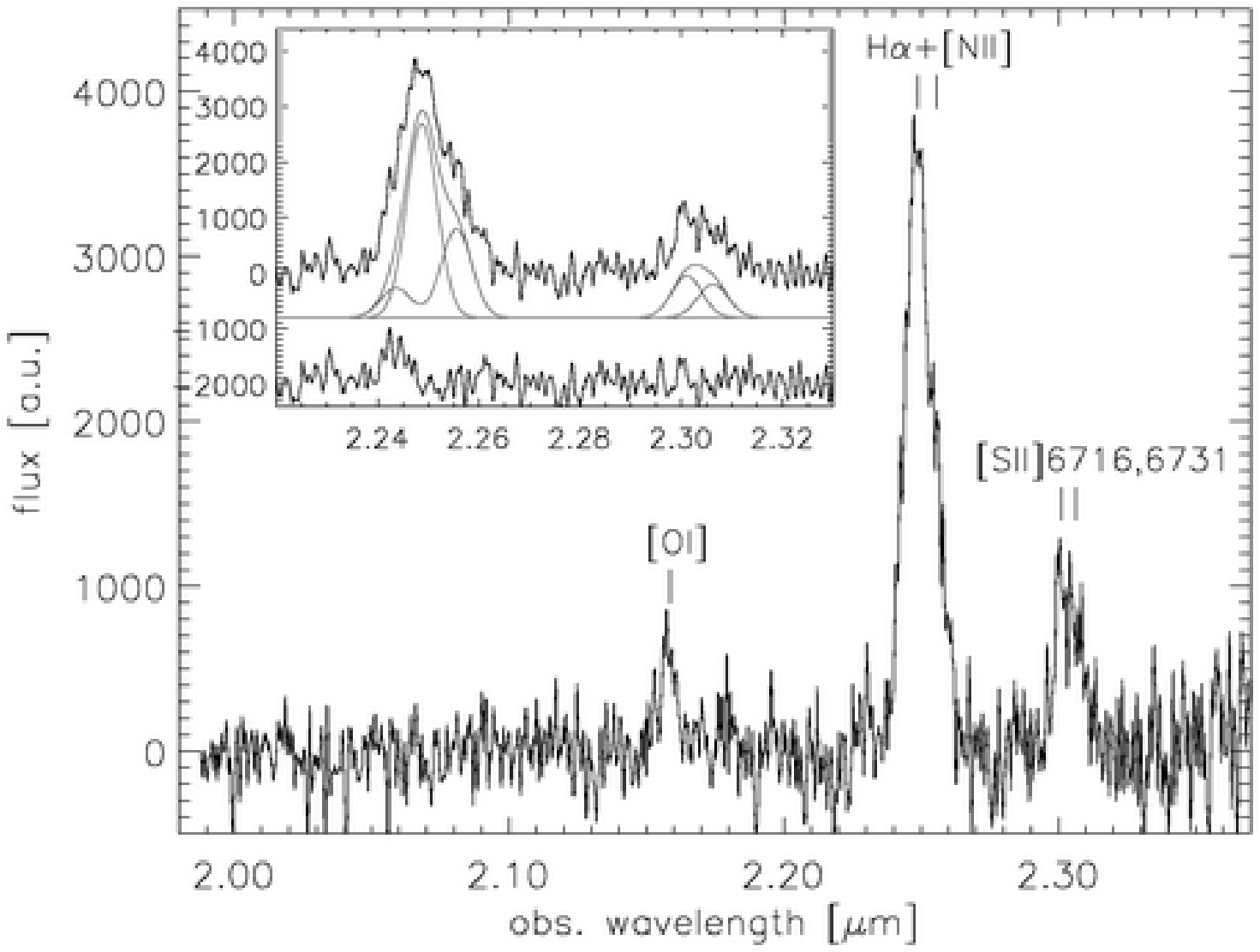,width=0.48\textwidth}
\caption{Integrated H (top) and K band spectrum (bottom) of MRC0406-244 at
z$=$2.42, extracted from all pixels in which the [OIII]$\lambda$5007 line
emission exceeds 3 $\sigma$. 
The inset in the upper panel shows the detected,
relatively faint H$\beta$, and [HeII]$\lambda$4685 lines, and the expected
wavelength of [HeI]$\lambda$4471, which is not detected.
The inset in the
right panel shows (from top to bottom): (1) the blended H$\alpha$ and
[NII]$\lambda\lambda$6548,6583 lines, and the [SII]$\lambda\lambda$6716,6731
doublet (2) the best Gaussian fit to these lines (see text) (3) fit
residuals. The fitted lines and residuals are shifted along the ordinate by
arbitrary amouts.}
\label{fig:0406intspec}
\end{figure}

Our SINFONI data illustrate that this is indeed the case. In
Fig.~\ref{fig:maps0406} 
we show the [OIII]$\lambda$5007 emission line morphology. Contours indicate
the line-free K-band continuum. Line emission is clearly extended over
3.7\arcsec $\times$ 1.1\arcsec\ (30.1 kpc$\times$ 9.3 kpc) and is reminiscent
of two edge-brightened bubbles with several brighter knots superimposed. HST
NICMOS imaging of \citet{pentericci01} in the F160W filter (which covers
[OIII]$\lambda$5007 and is dominated by line emission) reveals a more complex
morphology, with several bright knots and fainter filaments inbetween.

The line-free continuum morphology extracted from the SINFONI data cube shows
that continuum emission is faint and relatively compact, but spatially
resolved (contours in Fig.~\ref{fig:maps0406}). We detect the continuum over
an area of approximately 1.3\arcsec$\times$0.6\arcsec\ (10.6 kpc$\times$5
kpc), corresponding to an integrated magnitude of K$=17.7$ mag in a
3\arcsec\ aperture centered on the continuum peak (compared to K$=17.5$ mag if
not correcting for the line contamination). In the H band, we 
measure H$=18.9$ mag and H$=19.3$ mag for the uncorrected and
corrected magnitude, respectively. We also estimate an azimuthally-averaged
half-light radius, which corresponds to the area of brightest continuum
emission, that contains half the total continuum flux above a 3$\sigma$
threshold. We find a half-light radius of r$_{1/2} \sim$ 0.6\arcsec\ ($\sim 5$
kpc).

The integrated spectra of MRC0406-244 in the H and K bands are shown in
Fig.~\ref{fig:0406intspec}. Rest-frame optical emission lines are very bright,
in particular the [OIII]$\lambda\lambda$4959,5007 doublet and H$\beta$ in the
H band, and the blended H$\alpha$ and [NII]$\lambda$6583 emission lines in the
K band. In the K band, we also detect the [SII]$\lambda$6716,6731 doublet and
[OI]$\lambda$6300. Moreover, we marginally detect the faint
HeII$\lambda$4686 in the H band spectrum. The inset in
the left panel of Fig.~\ref{fig:0406intspec} shows the fainter lines in the
H-band. The profiles of some lines are affected by residuals of night-sky
lines (e.g., on the red wing of H$\beta$). See Table~\ref{tab:emlines0406} for
the observed wavelengths, redshifts, line widths and fluxes of the detected
emission lines in the integrated spectrum.

\subsection{TXS0828+193} 
\label{sssec:0828}

TXS0828+193 at z=2.57\footnote{The luminosity distance is D$_{L}=21.1$ Gpc and
angular size distance D$_{A}=$ 1.7 Gpc. The size scale is 8.0 kpc/\arcsec. The
age of the universe for this redshift and cosmological model is 2.5 Gyr.} is
the galaxy with the most extended radio lobes in our sample (103 kpc for our
cosmology). F675W WFPC2 imaging indicates a clumpy,  
irregular and asymmetric morphology at rest-frame UV wavelengths
\citep{pentericci01}.  

\begin{figure}
\centering
\epsfig{figure=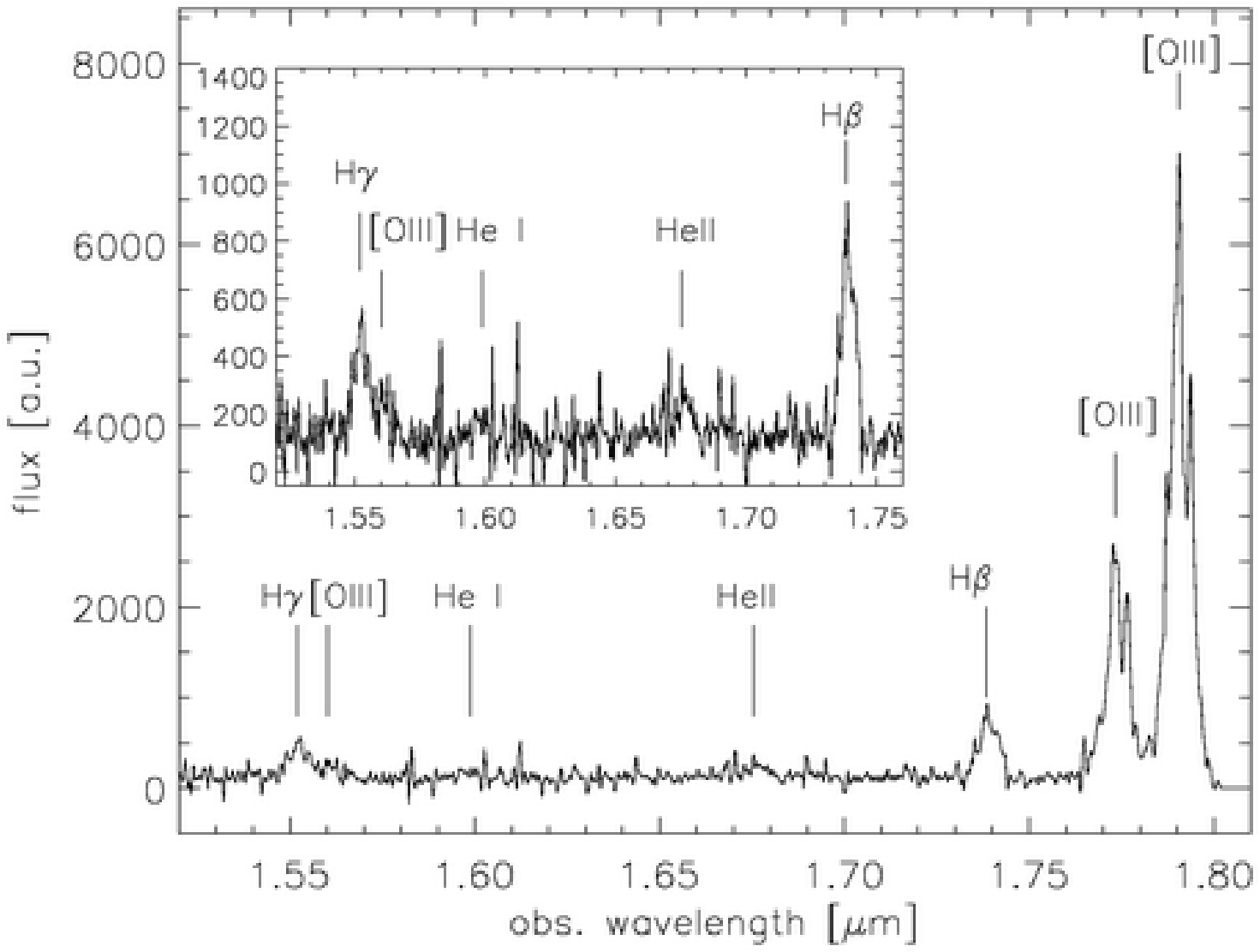,width=0.48\textwidth}
\epsfig{figure=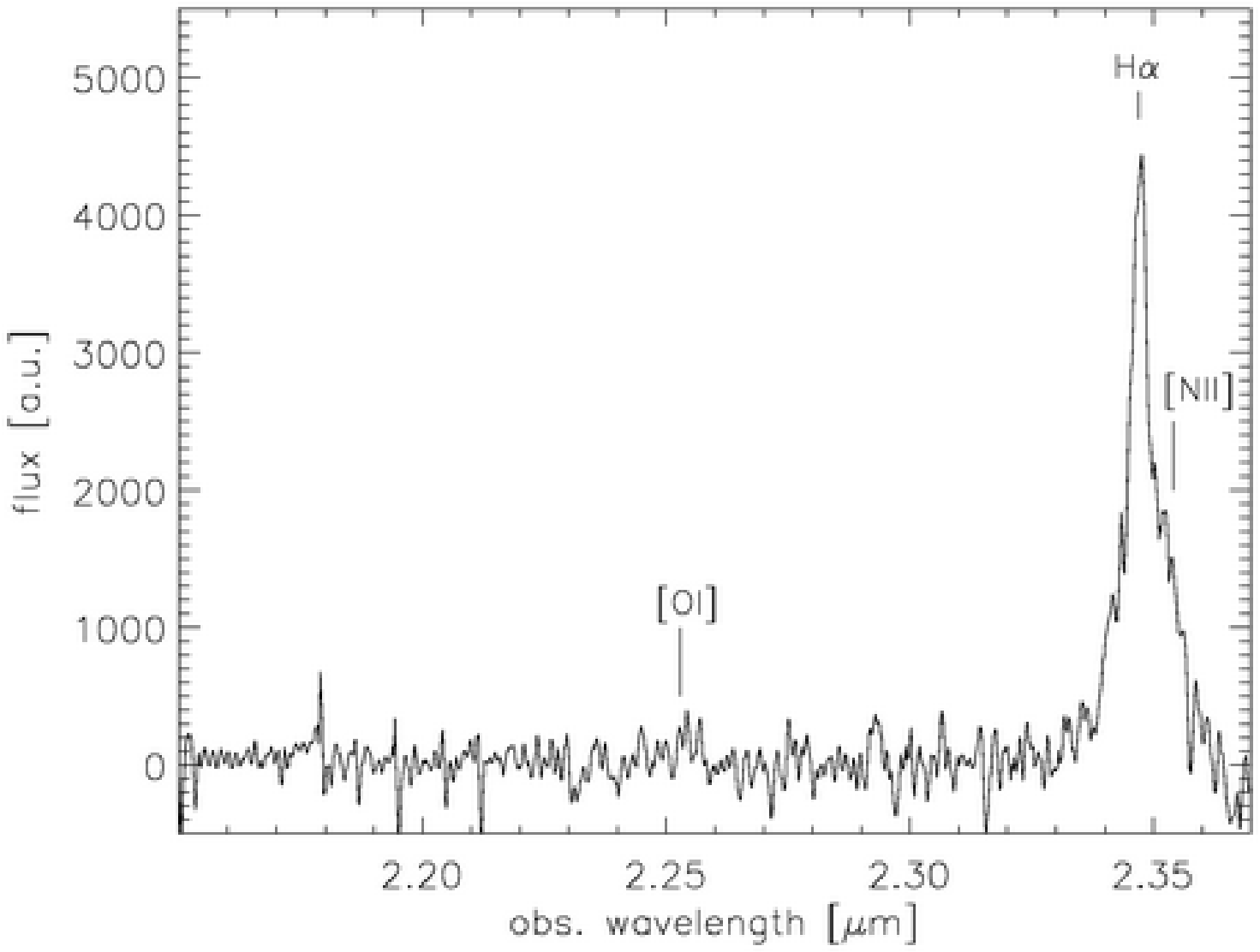,width=0.48\textwidth}
\caption{Integrated H (top) and K band spectrum (bottom) of TXS0828+193 at
z$=$2.57, extracted from all pixels in which the [OIII]$\lambda$5007 line
emission exceeds 3 $\sigma$. The inset in the upper panel shows the detected,
relatively faint H$\beta$, H$\gamma$, and [HeII]$\lambda$4685 lines, and the
expected wavelength of [HeI]$\lambda$4471, which is not
detected. [OIII]$\lambda\lambda$4959,5007 have complex profiles, where the
different components arise from the blue and redshifted bubble, respectively.
[OIII]$\lambda$6300 is faint, but detected with a significance of
5$\sigma$. In the K band, we detect H$\alpha$ and
[NII]$\lambda\lambda$6548,6583, and the fainter [OI]$\lambda$6300 line. The
[SII]$\lambda\lambda$6716,6731 doublet is at observed wavelengths
$>2.4$$\mu$m, which are dominated by telluric absorption.}
\label{fig:0828intspec}
\end{figure}

TXS0828+193 is a particularly luminous line emitter. In
Fig.~\ref{fig:0828intspec} we show the integrated spectrum of TXS0828+193 in
the H and K bands. [OIII]$\lambda\lambda$4959,5007, H$\beta$,
H$\gamma$ are detected in the H band. [OIII]$\lambda$4363
is faint, but detected with a significance of 3.6 $\sigma$. In the K band we
observe [OI]$\lambda$6300, H$\alpha$ and [NII]$\lambda$6583. The
[SII]$\lambda\lambda$6716,6731 doublet falls at observed wavelengths
$\lambda$$>$2.4, 
which are dominated by telluric absorption. The emission line
properties of the integrated spectrum of TXS0828+193 are summarized in
Table~\ref{tab:emlines0828}.

We show the [OIII]$\lambda$5007 emission line morphology of TXS0828+193 in
Fig.~\ref{fig:maps0828}. The bright extended emission line regions are
significantly more extended than the continuum (shown as contours in
the left panel of Fig.~\ref{fig:maps0828}), covering an area of 3.5\arcsec
$\times$ 1.2\arcsec\ 
( 28 kpc$\times$9.6 kpc).  The line image is dominated by two bright regions,
an unresolved peak to the south, and an extended, more diffuse emission line
region to the north-east. Comparison with the WFPC2 F675W image of
\citet{pentericci01} (their Fig.~7) indicates that the [OIII]$\lambda$5007
emission line morphology and the rest-frame UV morphology are very similar.

The continuum is faint in the H and K band (approximately corresponding to the
rest-frame V and R bands) and significantly more compact than the line
emission. We show the line-free, combined H and K band continuum morphology as
contours in Fig.~\ref{fig:maps0828}. Continuum emission is detected over an
area of $\sim 1.4$\arcsec$\times$0.8\arcsec, corresponding to about 11.3
kpc$\times$6.4 kpc. Estimating an azimuthally-averaged half-light radius
similar to \S\ref{ssec:0406}, we find 0.47\arcsec\ (3.8 kpc).
region. We estimate integrated magnitudes in the H and K band using 3\arcsec\
apertures, K$=$19.4 mag, H$=$19.7  mag (compared to K$=$18.5 mag, and H$=$19.1
mag, without correcting for line contamination).

\section{Characteristics of the extended emission line regions}
\subsection{Kinematics} 
\label{ssec:kinematics}

We extracted spectra from 3$\times$3 pixel square apertures across the
sources (0.375\arcsec$\times$0.375\arcsec) and fitted Gaussian line profiles
to each spectrum. Relative velocities were estimated from the measured
wavelengths of the line cores at each position (core of the strongest
component in case of complex line profiles).

The large line widths observed in the extended emission line regions of HzRGs
complicate the analysis of the spectra due to the blending of emission lines,
in particular of H$\alpha$ and [NII]$\lambda$6548,6583. In order to obtain a
consistent fit of all lines within each spectrum, we therefore estimated the
redshift and line width from the [OIII]$\lambda\lambda$4959,5007 doublet
(whose components are luminous and can be isolated relatively easily) and
imposed the same redshift and line width on all other emission lines. As a
result, the line fluxes are the only free parameter in the fits to all other
lines except [OIII].

However, such an approach implies that all emission lines trace the same gas
kinematics. We tested this explicitly by comparing velocities and line widths
obtained from [OIII]$\lambda\lambda$4959,5007 in TXS0828+193 with those
measured from H$\beta$, in areas where H$\beta$ is detected at a significant
level ($> 3\sigma$), and find good agreement between the measured velocities
and line widths.

For much of the extended emission, the residuals suggest that fitting a
single Gaussian component is adequate. However, in the central
regions in particular, we do observe more complicated profiles and line
splitting (see, 
e.g., Fig.~\ref{fig:linesplit}). In these cases, we fitted 2$-$3 components
with redshifts and line widths determined from the
[OIII]$\lambda\lambda$4959,5007 doublet, and leaving the flux of each line and
each component as a free parameter.

We present the maps of rest-frame relative velocities and line widths
(full width at half 
maximum) of the 3 galaxies in Fig.~\ref{fig:maps0316} to~\ref{fig:maps0828}
. Overall, the velocity 
fields are remarkably similar in all targets. In each galaxy we identify two
regions with relatively homogeneous internal kinematics, which extend from
near the radio core, and projected velocities relative to each other of $\sim
700-1000$ km s$^{-1}$. Velocities change abruptly in the central region of the
velocity field, within about a spatial resolution element of our data sets,
corresponding to $\sim 0.6$\arcsec\ or $\le$ 5 kpc. The velocity gradients for
the individual galaxies are given in Table~\ref{tab:outflow}. 
 
The velocity maps show substructure within each region, although the velocity
gradients are significantly smaller than the large velocity shift near the
center of the galaxies. Velocity gradients amount to $\sim$ 100$-$200 km
s$^{-1}$. We emphasize that the three targets have similar velocity fields,
in spite of different morphologies. 

Figures~\ref{fig:maps0316} to ~\ref{fig:maps0828} also show maps of
the FWHM line widths. Overall, 
line widths are FWHM$\sim 500-1200$ km s$^{-1}$ in the three galaxies. Note
that much of this range of FWHMs is due to variations in the line widths
within individual areas, and not so much due to large variation between
galaxies in our sample.

Apparent peaks in the central regions with FWHM$\ge 1400$ km s$^{-1}$ are
likely artefacts due to the large velocity gradients and partial overlap of
line emission from the two sides within one seeing disk. We also observe line
splitting (see Fig.~\ref{fig:linesplit}) in this area, and the observed line
widths are consistent with what 
would be expected from blended lines with the observed line widths and
velocity gradient in and between the two regions. In other regions of the
galaxies, the emission line widths likely reflect a large intrinsic turbulent
motion, since the widths exceed the velocity gradients within each bubble (at
least on kpc scales).

\subsection{Extinction}
\label{ssec:extinction}
MRC0406-244 and TXS0828+193 are bright H$\alpha$ and H$\beta$ emitters and
allow to construct maps of the H$\beta$/H$\alpha$ line ratios and thus 
extinction. For MRC0316-257 this is not possible, due to the larger redshift,
z$=$3.13, at which H$\alpha$ cannot be observed from the ground, and because
H$\beta$ is relatively faint.  

We measured the H$\beta$/H$\alpha$ line ratios from 0.375\arcsec$\times$
0.375\arcsec\ box apertures (3$\times$3 pixels).
To derive extinctions, we use a galactic extinction law and an intrinsic
(unreddened) Balmer decrement H$\alpha$/H$\beta$ = 2.9. (See
\S\ref{ssec:kinematics} for a description of how we correct for
blending between H$\alpha$ and the [NII]$\lambda\lambda$6548,6583 doublet.)

\begin{figure}
\centering
\epsfig{figure=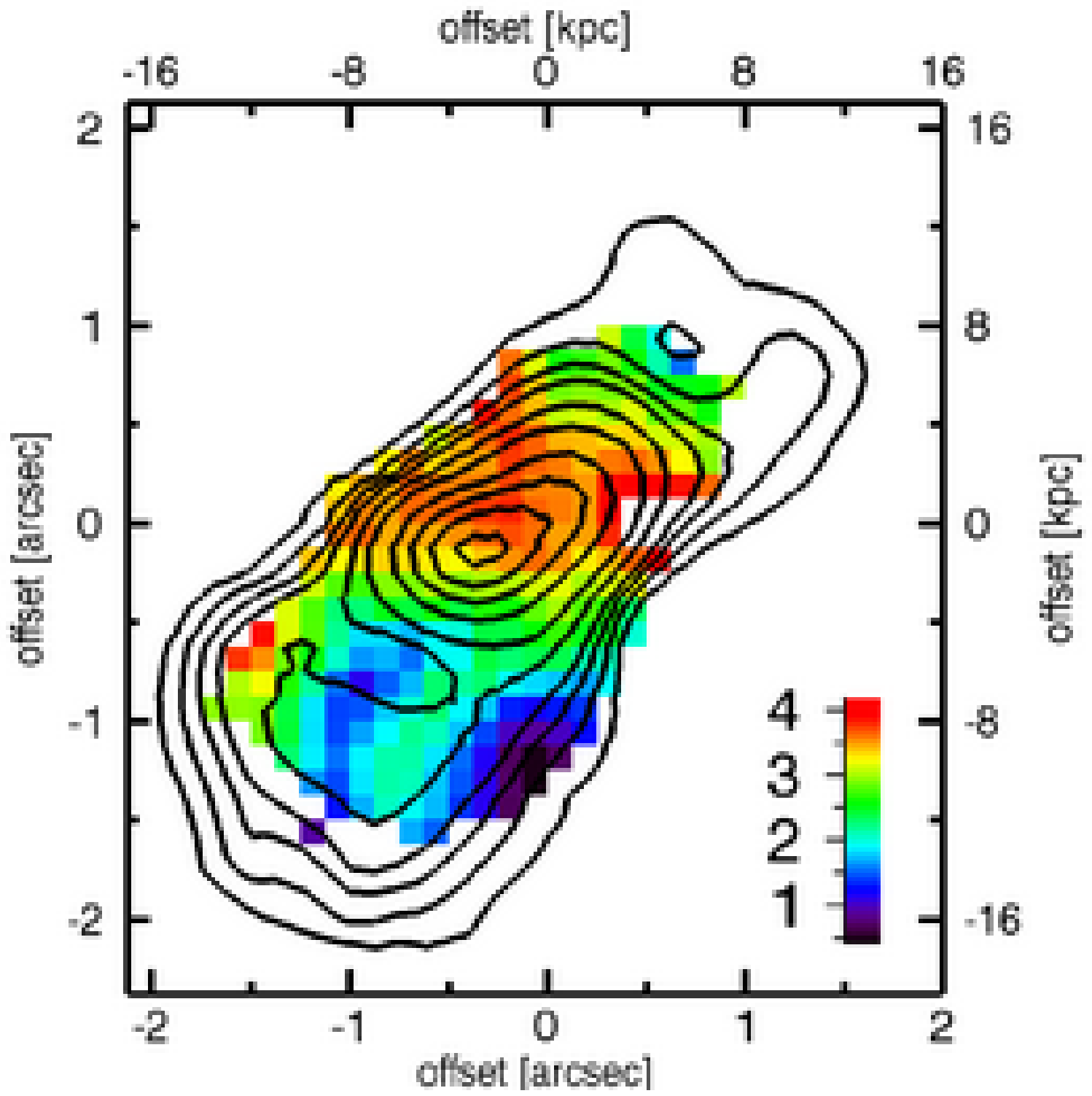,width=0.48\textwidth}
\epsfig{figure=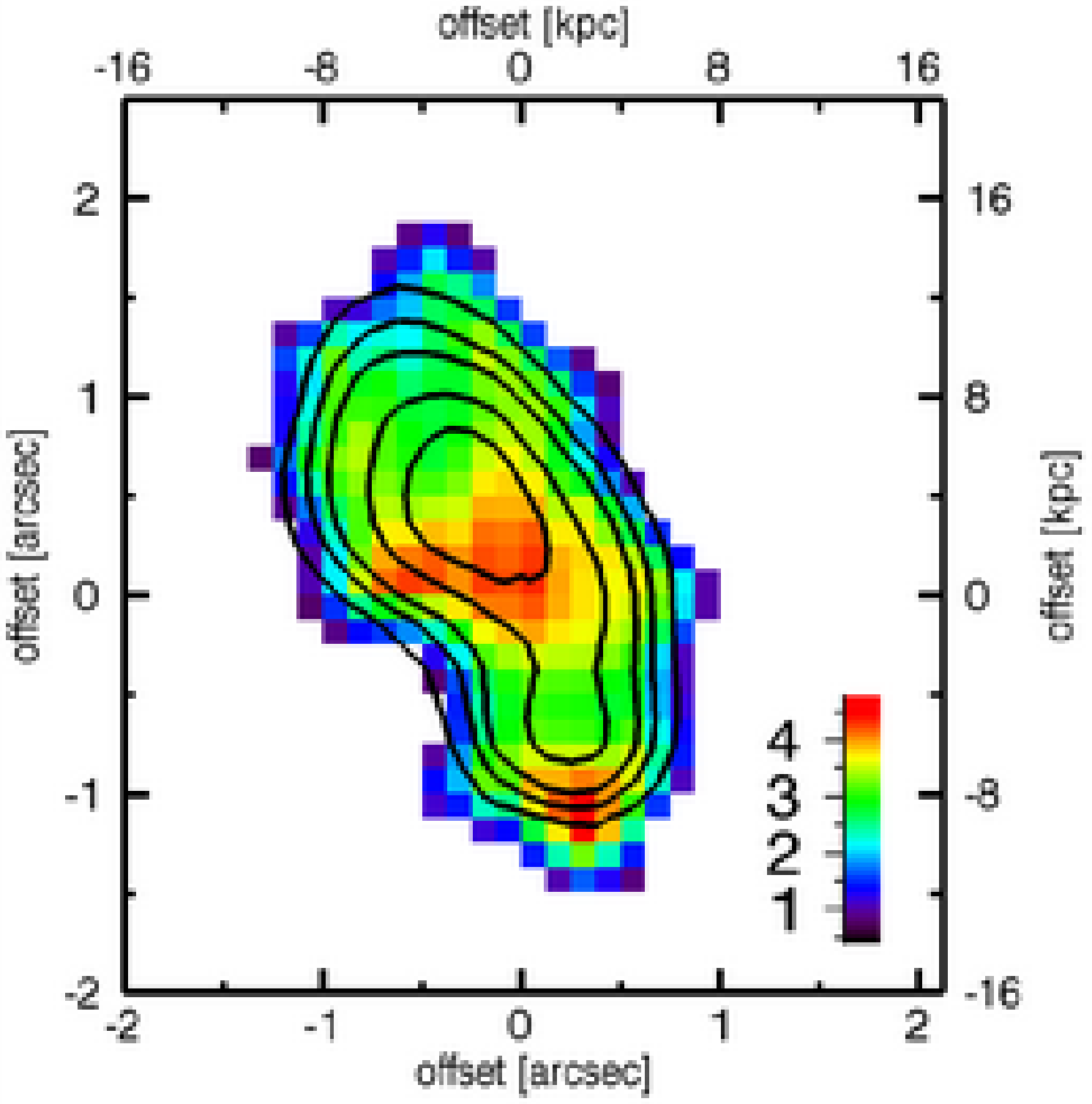,width=0.48\textwidth}
\caption{A(H$\beta$) extinction maps for MRC0406-244 (top) and TXS0828+193
(bottom). Color bars show the extinctions (in magnitudes) estimated from the
observed H$\beta$/H$\alpha$ emission line ratios, assuming a galactic
extinction law. Contours show the emission line morphology to ease
orientation.}
\label{fig:extinction}
\end{figure}

We show the extinction maps of MRC0406-244 and TXS0828+193 in the top and
bottom panel of Fig.~\ref{fig:extinction}, respectively. Contours indicate the
[OIII]$\lambda$5007 emission line morphology to ease orientation. Extinctions
at the wavelength of H$\beta$, $A(H\beta)$ vary between $A(H\beta)\sim 1-4.5$
mag in both galaxies. (We give A(H$\beta$)$=$A(V)$+$0.14 mag, because we
measured the extinction from the emission lines.)  The most heavily extincted
regions are near the center of each source, where A(H$\beta$)$\sim 4-4.5$ ;
the spatial extent of these areas corresponds approximately to the size of the
continuum emission. Extinction in the extended emission line regions is less
severe, with $A(H\beta)^{0406}_{ext}\sim 1-2$ mag in MRC0406-244, and
$A(H\beta)^{0828}_{ext}\sim 2-3$ mag in TXS0828+193.

Actively star-forming galaxies may have strong stellar Balmer absorption
lines, in particular for the higher-order Balmer lines. This may result in
underestimating the H$\beta$ emission line flux relative to H$\alpha$, and
thus in overestimating the extinction. Since the continuum in our galaxies is
faint we cannot measure the effect of underlying stellar absorption directly.
However, the large H$\alpha$ and H$\beta$ equivalent widths, the faintness of
the continuum, and the significantly larger spatial extent of the line
emission indicate that stellar absorption will have a negligible effect.

\subsection{Electron Densities}
\label{ssec:densities}
The [SII]$\lambda\lambda$6716,6731 doublet in TXS0828+193 is relatively faint
and falls into a wavelength range that is dominated by telluric absorption
bands; however, in MRC0406-244, the doublet is clearly detected in the K band
at signal-to-noise ratios $\approx 10$. (see Fig.~\ref{fig:0406intspec} for the
integrated K band spectrum of MRC0406-244). Due to the large line widths, the
doublet is partially blended. To minimize the blending, we extracted the
integrated spectrum of both bubbles individually, avoiding the central region
where lines are particularly broad due to the overlap of the two
bubbles. This is justified by the overall very uniform line ratios of the
[SII]$\lambda\lambda$6716,6731 doublet extracted from smaller apertures.
\begin{figure}
\centering
\epsfig{figure=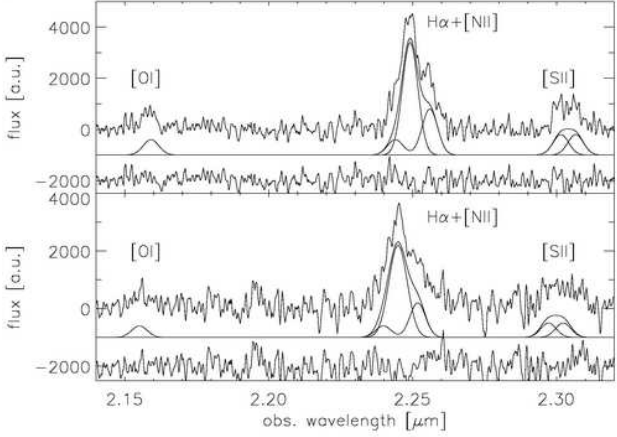,width=0.45\textwidth}
\caption{Integrated spectra of the two bubbles in MRC0406-244, the redshifted
  southern bubble is shown on top, the blueshifted northern bubble is
  shown below. Upper panels show the spectra, Gaussian fits to the emission
  lines and fit residuals are shown in the mid and lower panels,
  respectively. Spectra are smoothed by 5 pixels spectrally, corresponding to
  $\sim 1/5$ of the total line width. The measured
  [SII]$\lambda\lambda$6716,6731 line ratios are consistent with electron
  densities of a few 100 cm$^{-1}$, with a best-fit 
value of 500 cm$^{-1}$ (see text). Redshifts and line widths were measured
from the [OIII]$\lambda\lambda$4959,5007 line doublet, so that the only free
parameters to all other lines, including [SII], are the line fluxes.}
\label{fig:s2spec}
\end{figure}

We show the K band spectra of the two bubbles
in Fig.~\ref{fig:s2spec}. The upper panels show the measured spectra, fits and
fit residuals are shown below. The profiles of all lines in the K-band are
relatively narrow and regular. We used again [OIII]$\lambda$4959,5007 in the H
band to constrain the redshift and line width, so that the line fluxes are the
only free parameters.

[SII] line ratios, R$_{[SII]} =$ F(6716)$/$F(6731), in the two bubbles are
similar within the uncertainties; we measure R$_{[SII]}^{0406b} =1.0 \pm 0.15$
in the blueshifted bubble, and R$_{[SII]}^{0406r} =1.0 \pm 0.10$ in the
redshifted bubble. For a temperature of T$=10^4$ K
(\S\ref{ssec:temperatures}), this corresponds to a range in electron density
of $n_e^{0406b} =300-1045$ cm$^{-3}$ and $n_e^{0406r} =370-830$ cm$^{-3}$ in
the blueshifted and redshifted bubble, respectively, with a best value of
about 500 cm$^{-3}$ in either case, which is the value we will adopt in the
following.

Only one z$\sim 2$ HzRG has an electron density estimate given in the
literature, 
which is MRC1138-262 \citep{nesvadba06} with densities in the range of
$240-570$ cm$^{-3}$. The values in both galaxies are very similar, suggesting
that densities of a few $\times 100$ cm$^{-3}$ are likely typical for the
luminous optical emission line regions in HzRGs. 

\subsection{Electron Temperatures}
\label{ssec:temperatures}
\begin{figure}
\centering
\epsfig{figure=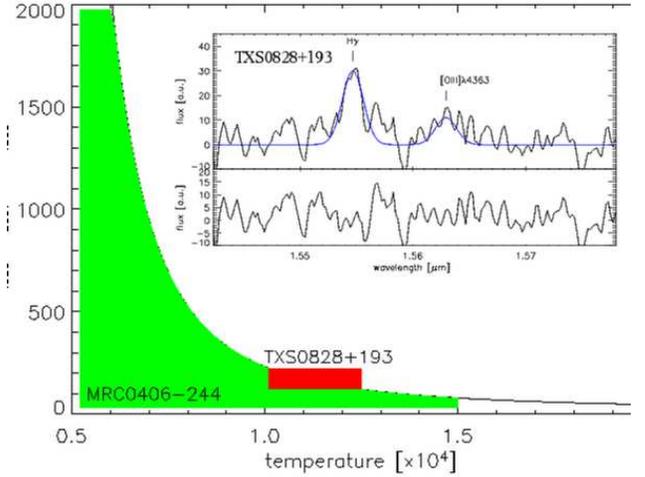,width=0.45\textwidth}
\caption{Relationship between electron temperature and [OIII] line ratios for
  an electron density of 500 cm$^{-3}$. The green shaded area shows the
  temperature range for MRC0406-244 derived from a 3$\sigma$ limit on
  [OIII]$\lambda$4363, the red area shows the range of temperatures from the
  ratio based on the measured [OIII]$\lambda$4363 flux in TXS0828+193,
  allowing for the 30\% uncertainty. The inset shows a zoom on H$\gamma$ and
  [OIII]$\lambda$4363 in the H-band spectrum of TXS0828+193. Blue lines
  indicate the expected emission lines with the redshift and line width
  measured from [OIII]$\lambda$5007. The lower panel shows the fit rediduals.}
\label{fig:spec4363}
\end{figure}

We detect the faint [OIII]$\lambda$4363 emission line in the H-band spectrum
of TXS0828+193, which is the first time that this line has been measured in
the spectrum of a z$\ge 2$ HzRG at $>3 \sigma$ significance. We detected the
line in a $5\times5$ pixel aperture in the H band centered on the bright
southern knot. The spectrum around the wavelength of [OIII]$\lambda$4363 is
shown in the inset of Fig.~\ref{fig:spec4363}.  We use the ratio of emission
line fluxes $F$ of the 3 [OIII] lines at 4363\AA, 4959,\AA, and 5007\AA,
respectively, F$_{4363}$, F$_{4959}$, and F$_{5007}$, R$_{5007+4959,4363}
= (F_{4959}+F_{5007})/F_{4363}$ to estimate the electron temperature of
TXS0828+193. We find R$_{5007+4959,4363}\sim 173\pm52$, correcting for
extinction. For an electron density n$_e\sim 500$ cm$^{-3}$, and allowing for
$\sim 30$\% uncertainty of the [OIII]$\lambda$4363 flux measurement, this
corresponds to an electron temperature of $\sim 10100-12500$ K (see
Figure~\ref{fig:spec4363}). 

We do not detect [OIII]$\lambda$4363 or H$\gamma$ in MRC0406-244. This may be
due to the somewhat lower redshift of this source, which 
causes both lines to fall at a wavelength where the transparency of the night
sky and SINFONI's efficiency are not very favourable. Placing a 3$\sigma$
limit 
on [OIII]$\lambda$4363 in MRC0406-244, we find an extinction-corrected
R$_{5007+4959,4363}=55.6$, which implies an electron temperature T$^{0406}_e  
\le 1.5\times 10^{4}$ K. 

Measuring electron temperatures from R$_{[OIII]}$ is the most direct method,
but obviously requires high-quality spectroscopy of high-redshift galaxies,
and exposure times that are often prohibitively large. \citet{humphrey08}
explored alternative, temperature-dependent line ratios using longslit spectra
of rest-frame UV and optical emission lines. For MRC0406-244 and TXS0828+193,
which are both included in their sample, they find T$_{e}^{0406}
=11700^{+1800}_{-1500}$ K and for several line ratios in TXS828+193, they find
temperatures in the range T$_{e}^{0828} =14000-22000$ K. While
their results differ from ours if taken at face value, the differences are
insignificant, given the large uncertainties of either measurement.

\subsection{Ionized gas masses}
\label{ssec:ionizedgas}
We use the extinction corrected H$\alpha$ fluxes (\S\ref{ssec:extinction}) to
estimate ionized gas masses. Similar to \citet{nesvadba06}, we assume simple
case B recombination, and use the n$_e\sim 500$ cm$^{-3}$ electron density
measured in MRC0406-244 \citep[see also \S~4.2 of][in particular
equation~1]{nesvadba06}. Total H$\alpha$ fluxes in MRC0406-244 and TXS0828+193
correspond to $\log{[L/L_{\odot}]}^{0406}_{corr} = 45.2$ erg s$^{-1}$ and 
$\log{[L/L_{\odot}]}^{0828}_{corr} = 45.3$ erg s$^{-1}$, respectively
(or $\log{[L/L_{\odot}]}^{0406} = 44.5$ erg s$^{-1}$ and
$\log{[L/L_{\odot}]}^{0828}$ = 44.4 erg s$^{-1}$ without extinction
correction). Estimating the gas masses in each spatial resolution
element, and summing over the full emission of the two galaxies, we find $\log
{[M_{HII}/M_{\odot}]}^{0406}\sim 10.6$ and
$\log{[M_{HII}/M_{\odot}]}^{0828}\sim 10.7$ for MRC0406-244 and TXS0828+193,
respectively. All values are also given in Table~\ref{tab:gasmasses}. 

We caution that ionized gas mass estimates at z$\ge 2$ can be precise to
an order-of-magnitude level only. In particular, extinction and electron
densities involve measuring relatively faint lines, H$\beta$ and
[SII]$\lambda\lambda$6716,6731, and the physical parameters are strongly
non-linear function of the line ratios. In spite of signal-to-noise ratios
$\sim 10$ in the faint lines, which is excellent compared to
IFU studies of galaxies at high redshift \citep[e.g.][]{nmfs06}
this will add uncertainties of factors $\ge 2$ in each estimate and for
typical values measured in our sample. Correcting for the blend of the
H$\alpha$+[NII] lines, we minimized the contribution of H$\alpha$ in the fit,
which will result in  underestimating the H$\alpha$ flux. Overall,
we find that these uncertainties will amount in total to less than $\pm 0.5$
dex, safely placing our estimates into the range of a few$\times 10^{10}$
M$_{\odot}$.

A few studies have previously estimated ionized gas masses in HzRGs based on
slit-spectra and narrow line images, although with significantly larger
uncertainties. \citet{armus98} report on a giant nebulosity in 4C+19.71 at
z$\sim 3.6$, which they found through [OIII]$\lambda$5007 narrow band imaging,
estimating gas masses of $\sim 10^{8-9}$ M$_{\odot}$. From [OII]$\lambda$3727
longslit spectroscopy of a sample of 52 HzRGs taken from the 3CR,
\citet{baum00} find that the ionized gas masses in HzRGs increase linearly
with redshift, which may be related to larger quantities of cold gas in and
around massive galaxies at high redshift. They also find that the 10 HzRGs at
$z\ge 1$ have ionized gas masses of several $10^9 M_{\odot}$, which is likely
a lower limit, due to partial coverage of the source with the slit, and
because [OII]$\lambda$3727 is 
susceptible to extinction. (In \S\ref{ssec:extinction} we found A$_V =1-4$ mag
for the extended emission line regions of our targets.) Comparing their result
for MRC0406-244 and ours, we can loosely estimate that correction factors of a
few are required to 
compensate for systematic offsets. This suggests that the intrinsic masses of
their sample will plausibly be of order $10^{10} M_{\odot}$.

\citet{humphrey08} measured H$\alpha$ and H$\beta$ fluxes for 9
powerful radio galaxies during the ``Quasar Era'' with the ISAAC longslit
spectrograph on the VLT. Unfortunately, H$\beta$ is not or only marginally
detected in most of their sample, making a direct comparison with the
extinction-corrected gas masses of our sample difficult. However, for the
uncorrected H$\alpha$ fluxes, we find similar values within factors of a few
as in our sample, which suggests ionized gas masses of order $10^{10}$
M$_{\odot}$ may be common in HzRGs.

\section{The nature of the emission line region}

Rest-frame optical integral-field spectroscopy has been published for a few
dozen galaxies at z$\sim 2$ of different types and masses
\citep[e.g.][]{nmfs06, nesvadba07a, nesvadba06b, swinbank06, law07,
nesvadba08, genzel06, wright07,bournaud08}, including galaxies undergoing
mergers, candidates for relatively quiescent disk rotation and galaxies that
are likely disks in formation, and galaxies showing the signs of
starburst-driven winds. The sample includes galaxies with and without
AGN. Compared to these samples, our radio galaxies have several unique
properties which allow us to constrain the mechanism which is driving the gas
dynamics. We will briefly discuss the different possible scenarios.

\subsection{Merging} 

If the two bubbles represented the interstellar medium of two massive galaxies
in course of a merger, we should find a continuum peak in each bubble,
representing the stellar component of each galaxy of the merging
pair. However, we find only one continuum source in MRC0406-244 and
TXS0828+193, respectively, which is in the center of each source, where the
velocity of the ionized gas changes abruptly between the blueshifted and the
redshifted bubble. (For MRC0316-257, we did not detect the continuum.) This
does not rule out that these galaxies are related to a major merger, however,
the two components must be closer than the resolution of our data, about 5
kpc, and the two bubbles cannot be gravitationally bound to one of the
galaxies each. (Note that our continuum images are not deep enough to
rule out minor mergrs.)

Broad-band morphologies of HzRGs are in agreement with this
interpretation. Broad-band morphologies of HzRGs are known to be often
irregular \citep[e.g.,][]{vanbreugel98,pentericci01}, however, in many cases
and depending on redshift, this may be due to emission line contamination, as
has been suspected before \citep[e.g.][]{rush97}, and is now evident from the
line and continuum morphologies extracted from our data cubes. A number of
HzRGs, including galaxies with HST-NICMOS imaging, have very compact
morphologies and elliptical light profiles consistent with being merger
remnants \citep[][]{vanbreugel98,pentericci01}.

\begin{figure}
\centering
\epsfig{figure=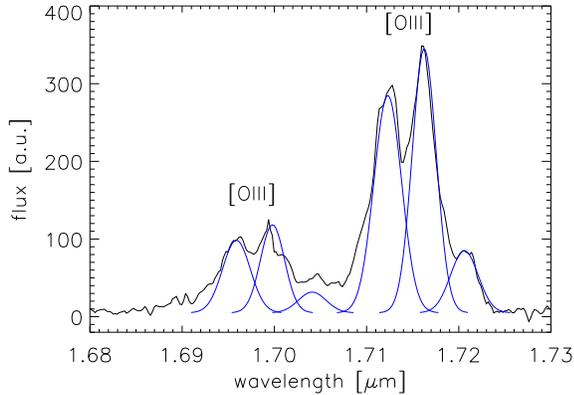,width=0.45\textwidth}
\caption{[OIII] emission lines extracted from a small aperture
(0.375\arcsec$\times$0.375\arcsec) near the center of MRC0406-244. Line
splitting is evident in both lines, with three narrow components and an
underlying, broad component. The mid component in [OIII]$\lambda$4959 is
affected by a night sky line residual. To guide the eye, we also show
Gaussians for each component of the [OIII]$\lambda$4959 and [OII]$\lambda$5007
line respectively. Note that these are not strictly fits, but have the same
line width and redshift for both [OIII] lines. The red component of this line
appears fainter than the corresponding component of the [OIII]$\lambda$5007
line, because of an underlying broad component of the [OIII]$\lambda$5007 line
superimposed, which we ignored.}
\label{fig:linesplit}
\end{figure}

\subsection{Disk rotation} 

Disk rotation and merging are difficult to distinguish at z$\sim 2$ due to the
relatively coarse spatial resolution of the data relative to the relevant size
scales, typically several seconds of arc \citep[see][for a more detailed
discussion]{nesvadba08}. However, all candidates for disk rotation at high
redshift have more gradual velocity offsets. Unlike other high-redshift galaxy
populations, the emission line regions of HzRGs are well resolved spatially,
and should show a characteristic velocity pattern 
known as ``spider diagram'' if their kinematics were dominated by disk
rotation \citep[see, e.g. Fig.~1 of][]{kronberger07}. However, the observed
abrupt velocity offset within a spatially unresolved region characterized by
strong line splitting (Fig.~\ref{fig:linesplit}) is at odds with galaxies
showing disk rotation at low and at high redshift. In addition, the emission
line morphologies are often irregular or filamentary, which would not be
typical for rotating gaseous disks (see Fig.~\ref{fig:maps0316} to
~\ref{fig:maps0828} ).  

Moreover, the large velocity offsets and line widths suggest unplausibly large
enclosed masses of often $10^{13-14}$ M$_{\odot}$ within few 10s of kpc
\citep[compared to stellar masses of few $\times 10^{11}$ M$_{\odot}$;
][]{seymour07}, if the gas was to be bound gravitationally \citep[see, e.g.,
Table 1 of][]{baum00}. For HzRGs that are surrounded by galaxy overdensities,
\citet{venemans07} estimate virial masses of few $\times 10^{14}$ M$_{\odot}$
for the entire dark matter structure out to radii of $\sim 1-2$ Mpc, based on
observations of the
velocity dispersion of the companion galaxies. If the observed emission line
kinematics of HzRGs were due to rotation, then a significant fraction of this
mass would have to be within the inner 20-30 kpc of the structure, at odds
with what we know about dark matter halo mass profiles. \citet{nesvadba06}
argued that the ionized gas of MRC1138-262 has a kinetic energy which is about
similar to the binding energy of a galaxy with a mass of few $\times 10^{11}$
M$_{\odot}$, in which case much of the gas would
not be gravitationally bound. We conclude that the observed velocity fields
cannot be related to gravitationally driven motion, they are
super-gravitational.

\subsection{Outflows or infall?}
Directly measuring the direction of a gas flow in extragalactic objects is
inherently difficult, in particular at high redshift. Observed velocity
gradients are near the escape velocity of the dark halo surrounding a massive
galaxy \citep[see][for a more detailed discussion]{nesvadba06}, which suggests
that we may be witnessing the ionized gas being unbound from the host
galaxy. However, in the next section we will argue that the gas dynamics are
most likely related to the expanding cocoon of hot, overpressurized gas
surrounding the radio jet in agreement with previous studies based on
longslit spectroscopy \citep[e.g.,][]{clark98,villar99,inskip02,best97}.

 Such an interaction will result in complex
kinematics, in agreement with the observed large line widths. The cocoon
likely forms from backflowing gas which has previously been heated by shocks
at the working surface of the jet. Although \citet{alexander96} pointed out
that this gas will likely not move far from the hotspot and may mostly
contribute to inflate the cocoon, this makes a more detailed assessment of the
flow direction of the ionized gas worthwhile.

One way to constrain this direction is through a careful study of the radio
spectral properties of the two lobes in each source \citep[see
also][]{humphrey08}.  \citep[e.g.,][]{alexander87}.  Radio emission from a
source embedded in a magnetized plasma will be depolarized by a degree that
depends on the column of the transversed medium. Thus, the near side of the
radio source will appear more strongly polarized than the far side\citep[the
``Laing-Garrington effect'';][]{laing88,garrington88}. Moreover, the observed
brightness of the radio hotspots on the near side of the radio source may be
enhanced due to Doppler-boosting \citep[e.g.,][]{alexander87}. The hotspots
have younger ages and accordingly flatter spectral indices than the
surrounding lobes. Thus, at high frequencies, the relative contribution of the
hotspot will contribute more strongly  to the overall emission from the lobe,
and the spectral index on the near side of the radio source will appear
flatter than that on the far side.

Comparing with the
observations of \citet{carilli97}, we find in each of our galaxies that the
northern radio lobe is more strongly polarized at 4.7 GHz and at 8.2 GHz than
the southern lobe. Moreover, the northern lobe has a somewhat flatter spectral
index in all galaxies. In all three cases, this is also the side of the
blueshifted bubble of ionized gas. This strongly advocates that the blueshift
represents a net outflow relative to the central galaxy. 

Outflows of gaseous material are also observed in powerful radio galaxies at
lower redshifts, z$<0.1$. \citet{morganti05} find broad, blueshifted HI
absorption lines at relative velocities of $\ge 1000$ km s$^{-1}$ from the
systemic redshift in 7 out of 8 sources. This is consistent with AGN-driven
outflows of neutral gas, corresponding to mass loss rates of up to $\sim 50
M_{\odot}$ yr$^{-1}$.

With the simple premise that the extended emission line regions in HzRGs
represent energetic outflows, we will now constrain the basic dynamics of the
gas, before arguing in the next section, that the outflows are most likely
related to the radio jet. 

\subsection{Dynamical Time Scales} 
\label{ssec:dynamicaltimes}
Emission line regions extend over fairly similar radii in all three galaxies,
10 kpc $\times$9 kpc, 15 kpc $\times$9 kpc, and 14 kpc $\times$9 kpc for
MRC0316-257, MRC0406-244, and TXS0828+193, respectively (see
Table~\ref{tab:outflow}). We can use the size of the outflows along the axis
of the radio jet and the relative velocity between the blueshifted and the
redshifted bubble in each galaxy to estimate the crossing time. Using the
velocities given in Table~\ref{tab:outflow}, we find dynamical timescales of
$\tau_{ion}\sim$ few $\times 10^{7}$ yrs, which is similar to ages inferred for
radio jets \citep[e.g.,][]{kaiser97}. See Table~\ref{tab:outflow} for the ages
derived for individual galaxies. These estimates can only be accurate
within factors of a few, given the unknown radial density profile of the gas 
prior to the radio-loud phase, the effects of projection and relatively coarse
spatial resolution of our data.

Alternatively, we can assume that the velocity gradients observed in
individual bubbles are dominated by the lateral expansion of the bubbles, as
observed in low-redshift narrow-line regions being inflated by interactions
with the overpressurized, expanding cocoon of the radio jet
\citep[e.g.,][]{capetti99}. HzRGs have very irregular environments with
gaseous halos 
\citep[e.g.,][]{villar03}, satellite galaxy overdensities
\citep[e.g.][]{venemans07}, and perhaps infalling gas clouds, which will
influence the distribution and density of the sourrounding medium. This
will make it difficult to reliably measure the relatively small velocity
gradients within an individual bubble. However, if we only use clouds with
relatively smooth internal velocity gradients, in particular the blue
(northern) bubbles of MRC0316-247 or of MRC0406-244, we find internal velocity
gradients of $\sim 120-300$ km s$^{-1}$, corresponding to dynamical time
scales of a few $\times 10^7$ yrs, which is consistent with the dynamical time
scales estimated from the velocity gradients along the radio jet axis.

\subsection{Kinetic Energies}
\label{ssec:kinenergy}
In \S 5.2 and \S 5.3 of \citet{nesvadba06}, we applied two methods to estimate
the injection of kinetic energy into the outflows:

\noindent
(1) A strict, but
presumably not very precise lower limit, based on the observed ionized gas
mass, the bulk motion, $v$, and the turbulent motion, $\sigma$, i.e., $E = 1/2
m (v^2 + \sigma^2)$. 

\begin{figure}
\centering
\epsfig{figure=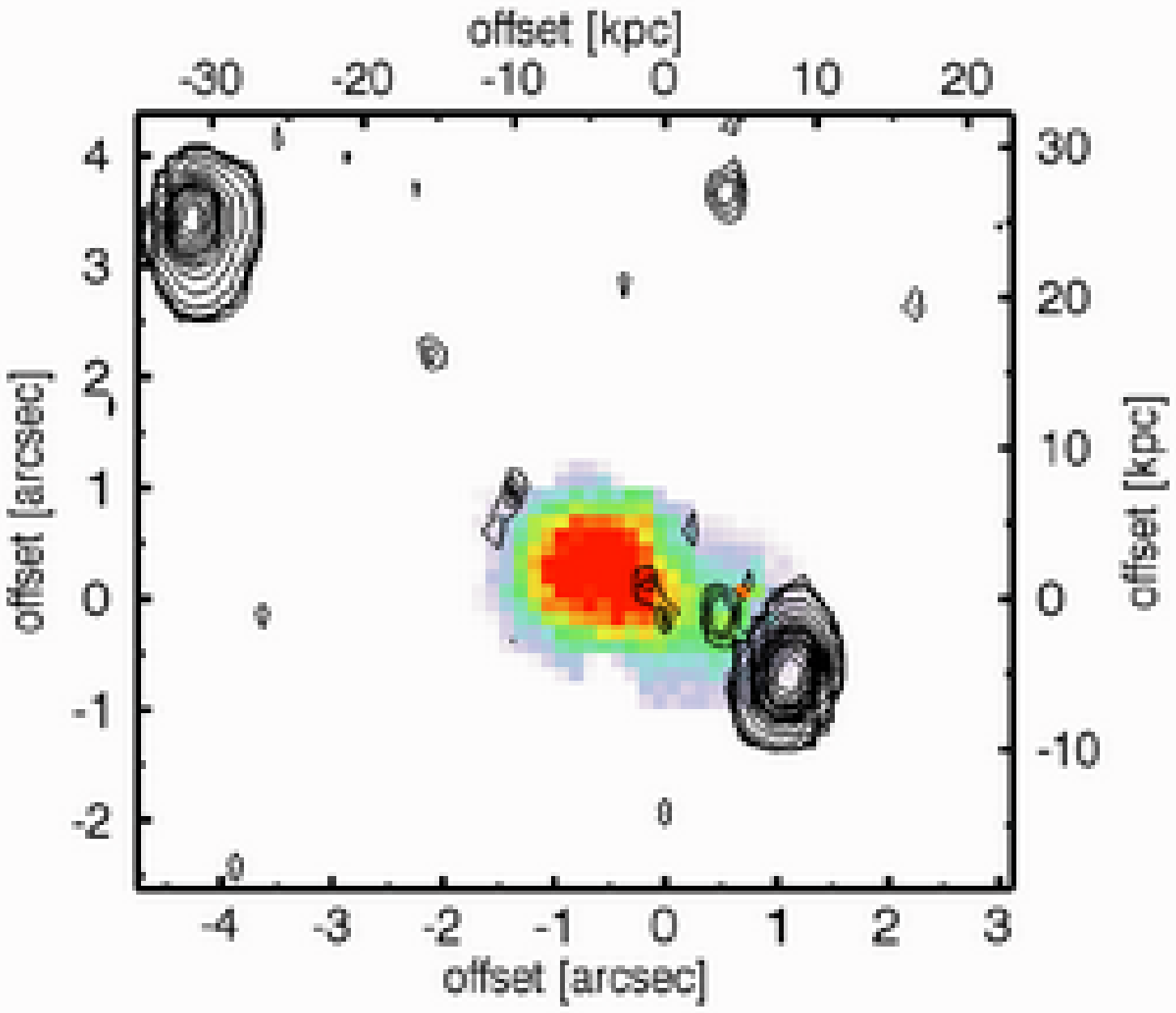,width=0.48\textwidth}\\
\epsfig{figure=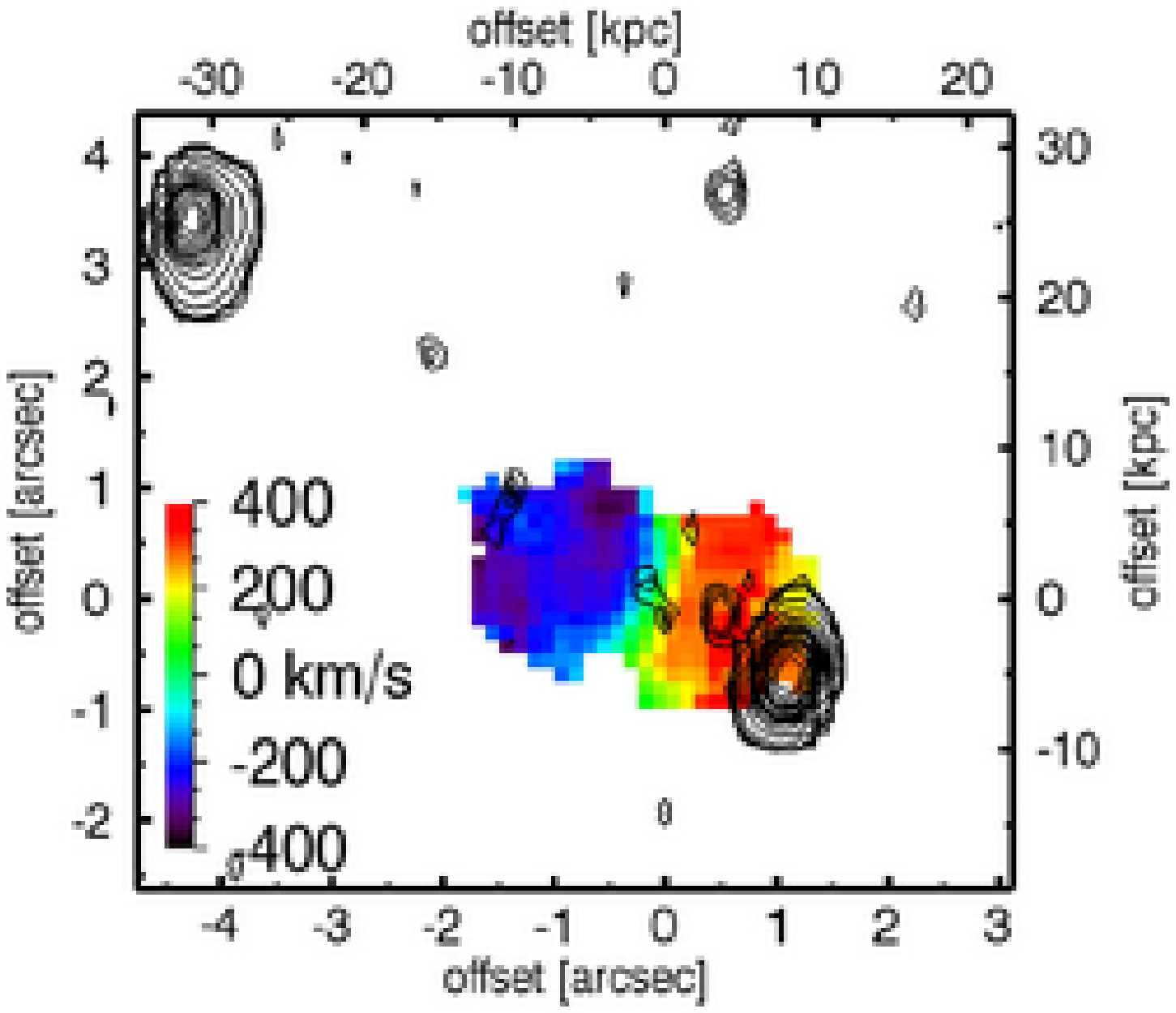,width=0.48\textwidth}
\epsfig{figure=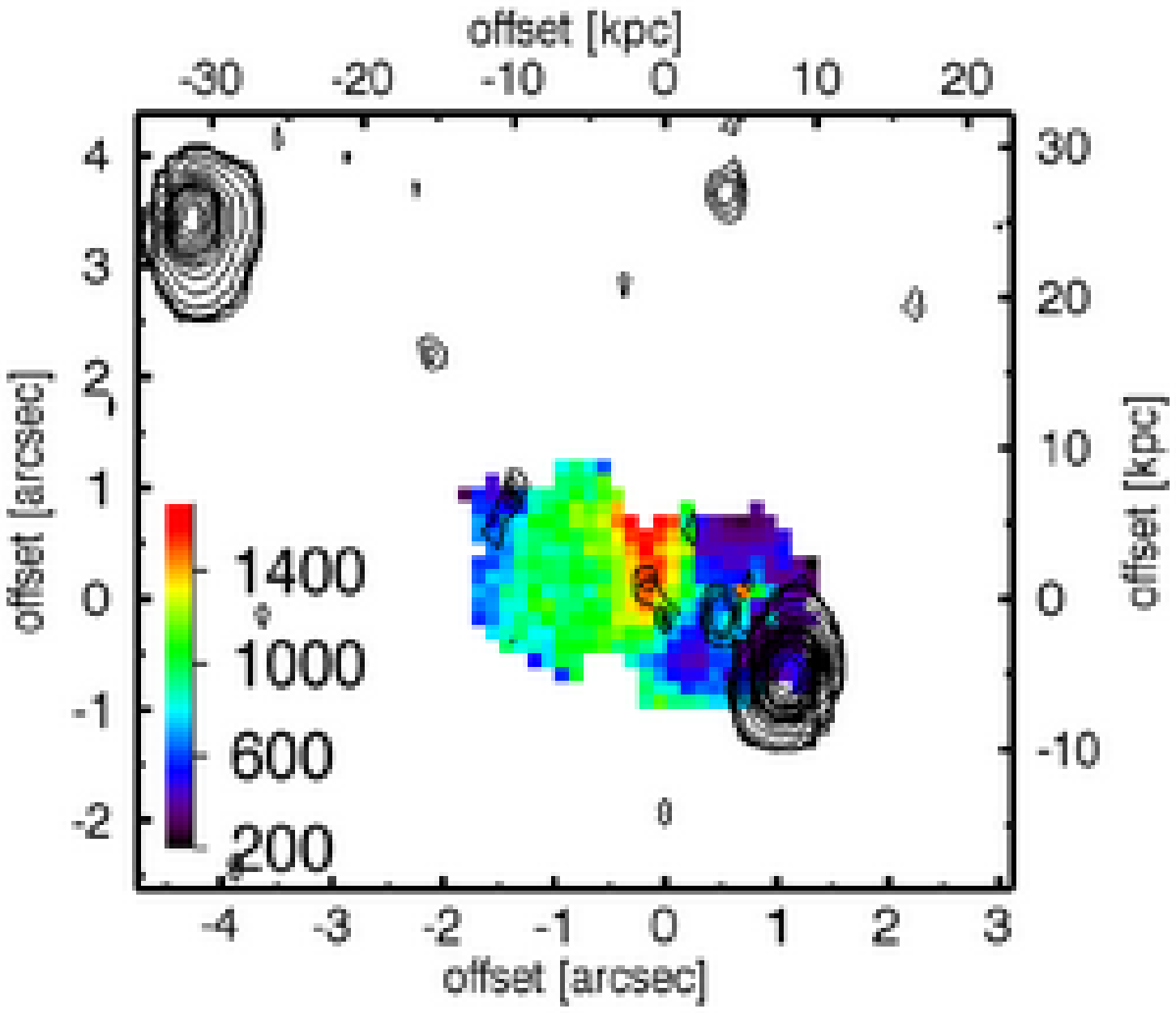,width=0.48\textwidth}
\caption{{\it top:} [OIII]$\lambda$5007 emission line morphology of
  MRC0316-257 at z$=3.13$ with contours showing the 1.4~GHz morphology. 
  {\it bottom left:} Velocity map. Color bar shows the relative velocities in 
  km s$^{-1}$, contours indicate the 1.4~GHz morphology. {\it bottom right:}
  Maps of the line widths, color bars show the FWHM in km 
  s$^{-1}$, contours indicate the 1.4~GHz morphologies. North is up, east to
  the left in all images.}
\label{fig:maps0316}
\end{figure}

\begin{figure}
\centering
\epsfig{figure=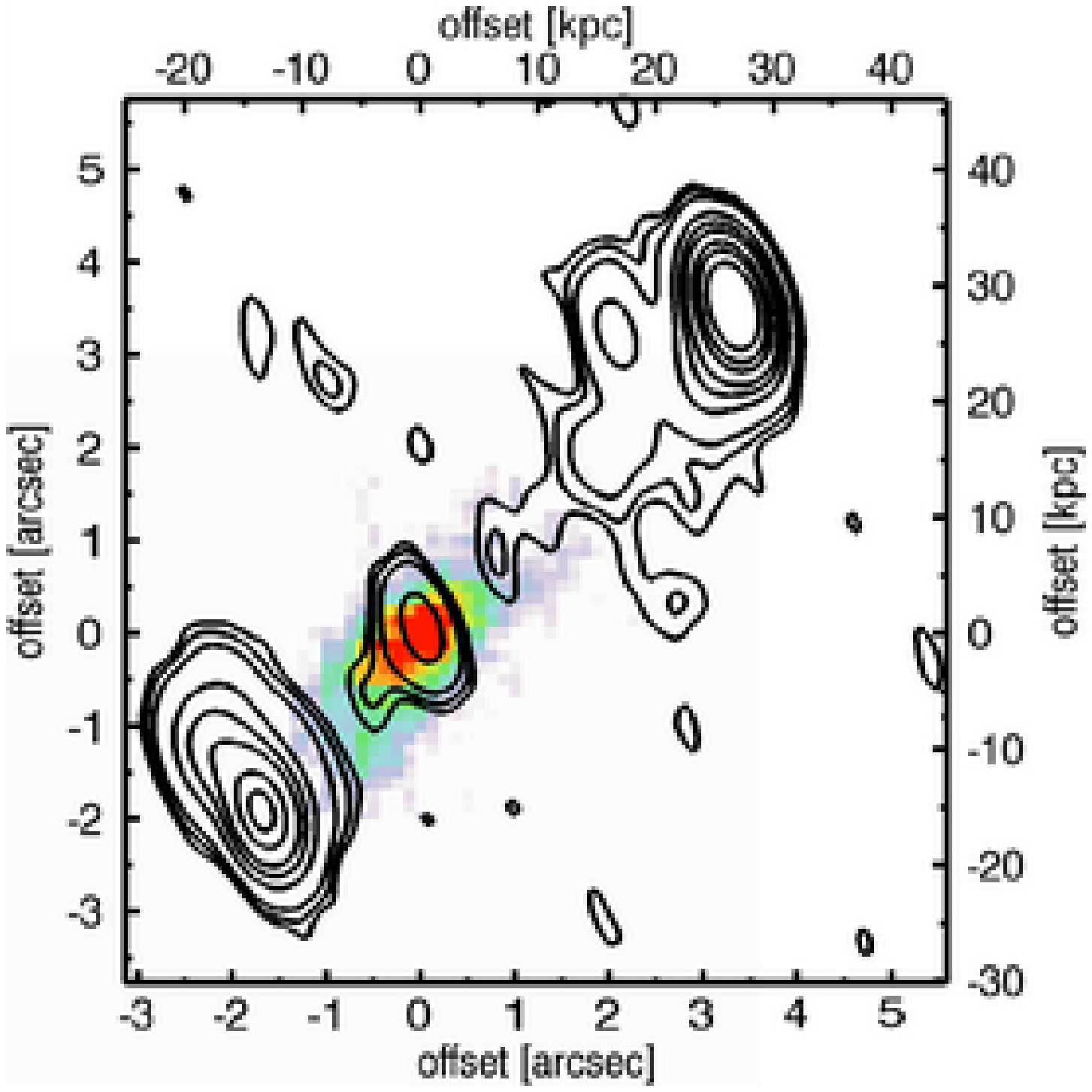,width=0.48\textwidth}
\epsfig{figure=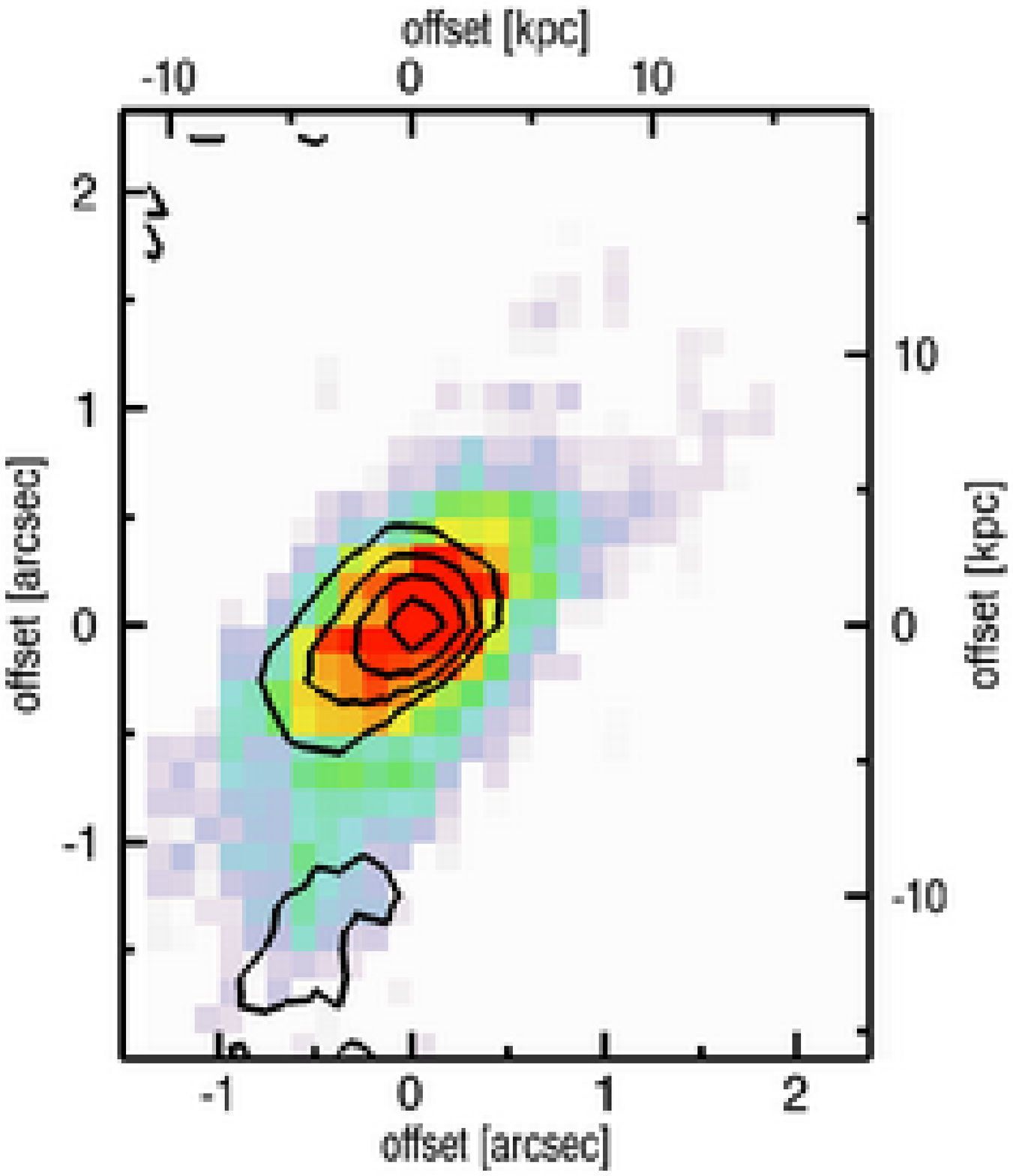,width=0.48\textwidth}\\
\epsfig{figure=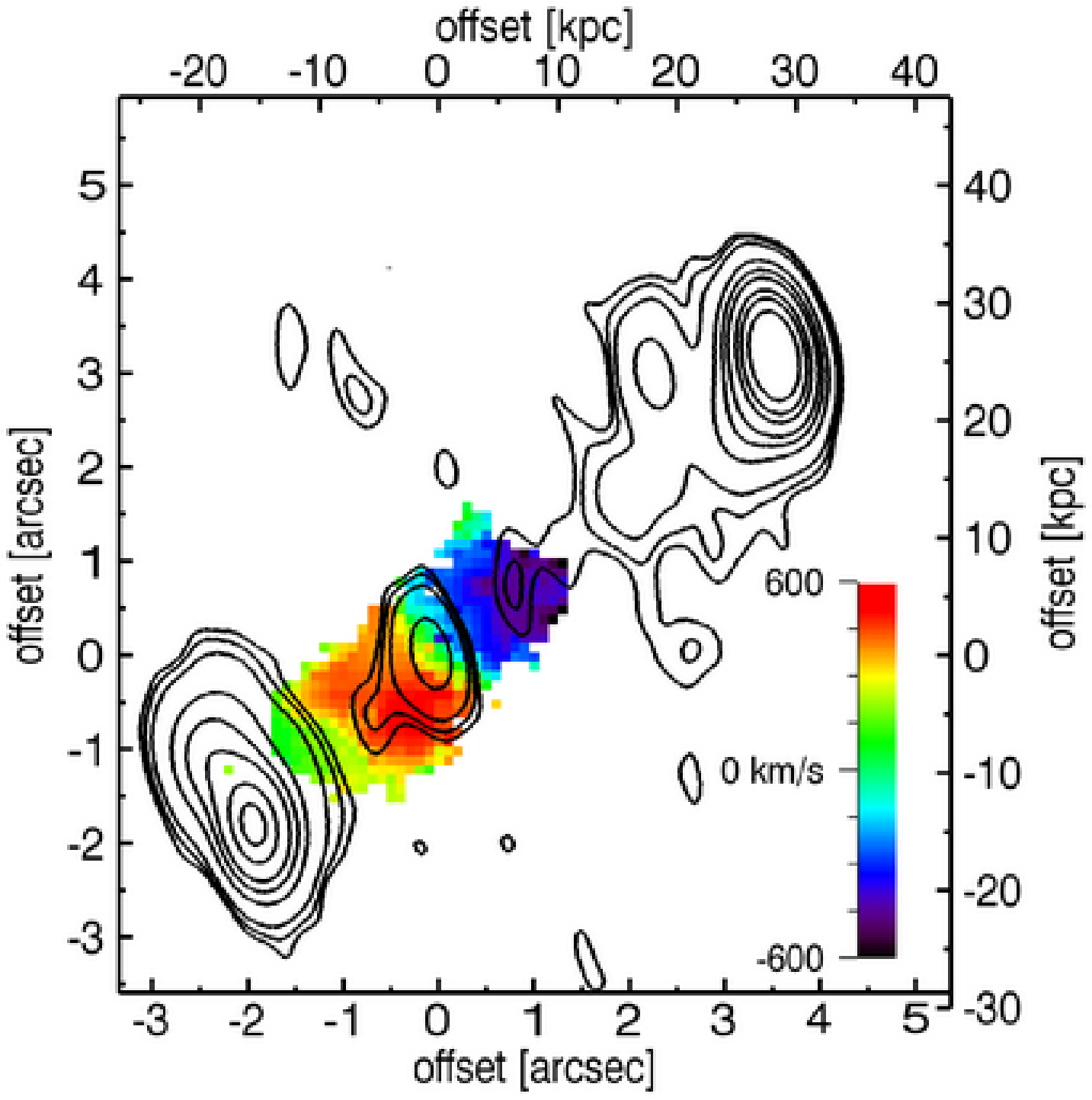,width=0.48\textwidth}
\epsfig{figure=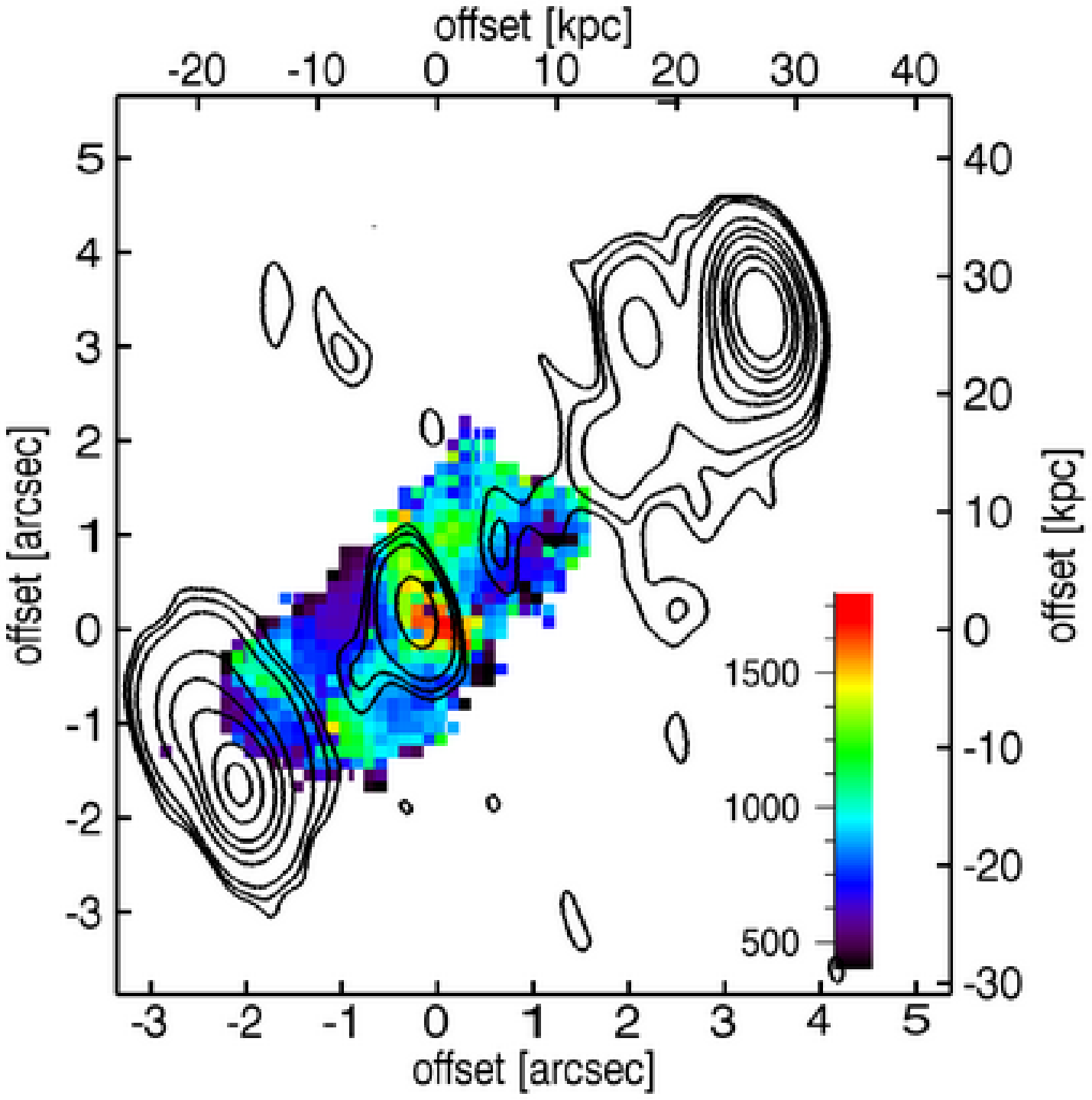,width=0.48\textwidth}
\caption{{\it top left:} [OIII]$\lambda$5007 emission line morphology of
  MRC0406-244 at z$=2.42$ with contours showing the 1.4~GHz morphology. {\it
    top right:} [OIII]$\lambda$5007 emission line morphology with contours
  showing the line-free rest-frame optical continuum. 
  {\it bottom left:} Velocity map. Color bar shows the relative velocities in 
  km s$^{-1}$, contours indicate the 1.4~GHz morphology. {\it bottom right:}
  Maps of the line widths, color bars show the FWHM in km 
  s$^{-1}$, contours indicate the 1.4~GHz morphologies. North is up, east to
  the left in all images.}
\label{fig:maps0406}
\end{figure}

\begin{figure}
\centering
\epsfig{figure=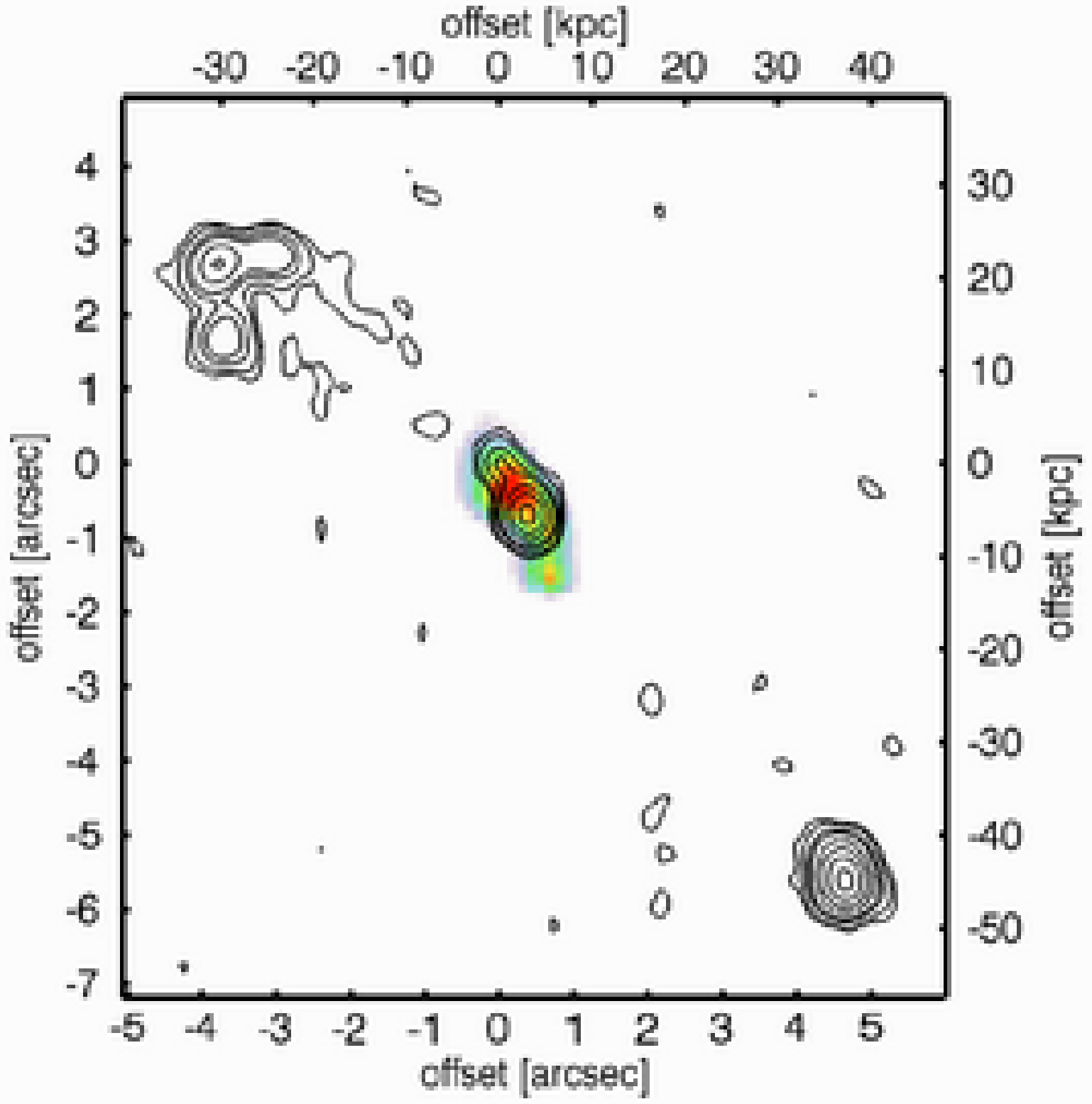,width=0.48\textwidth}
\epsfig{figure=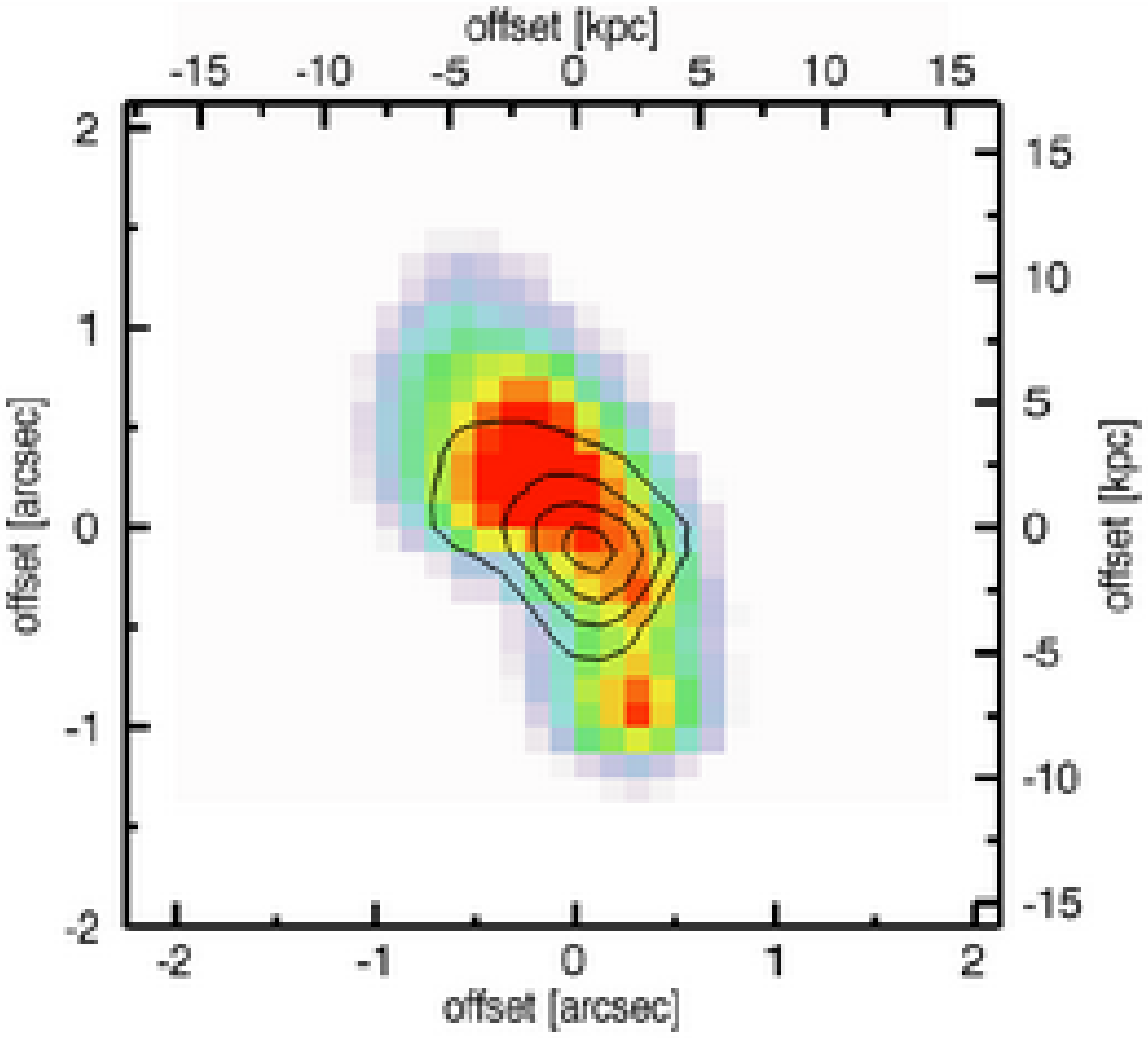,width=0.48\textwidth}\\
\epsfig{figure=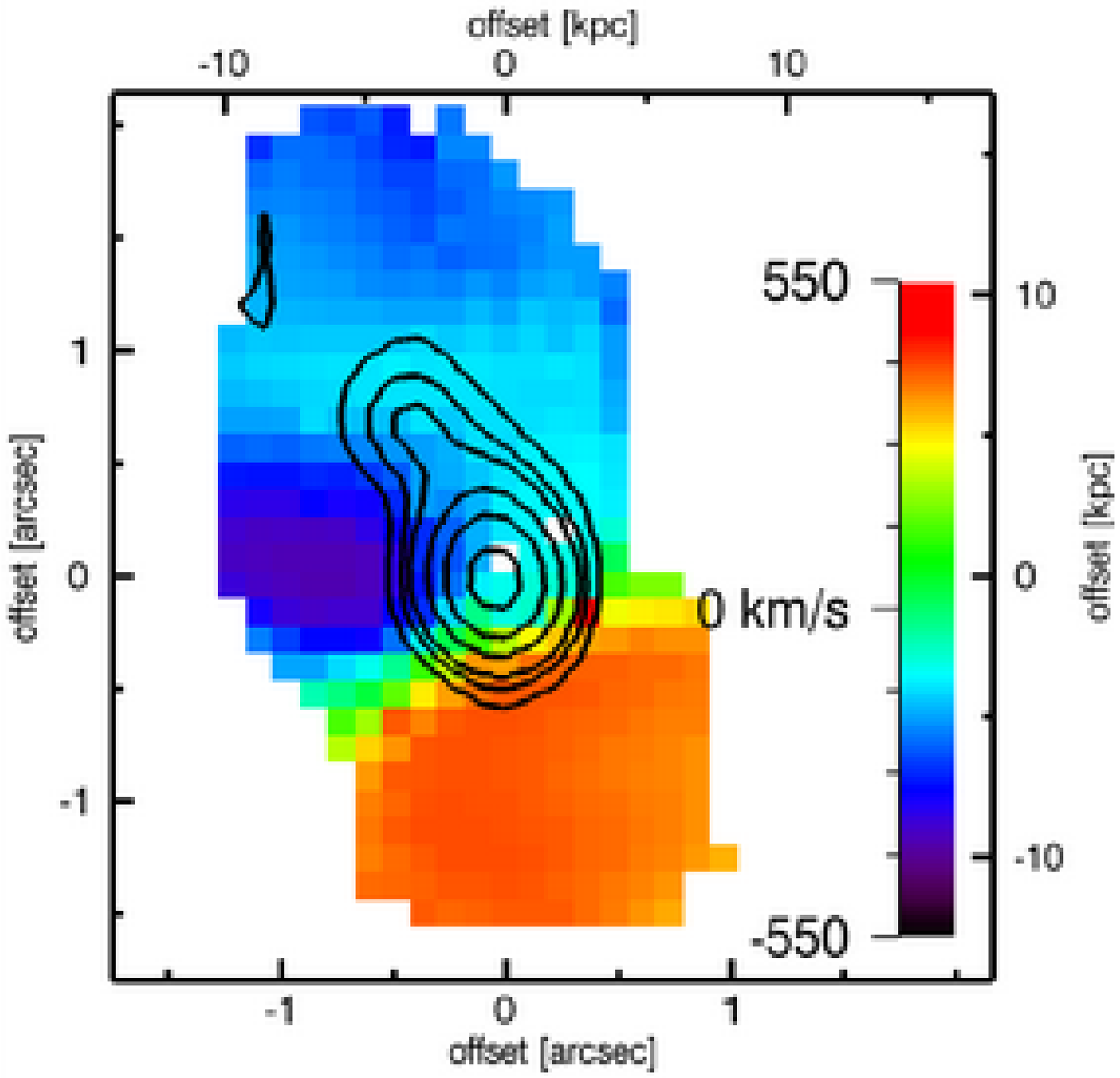,width=0.48\textwidth}
\epsfig{figure=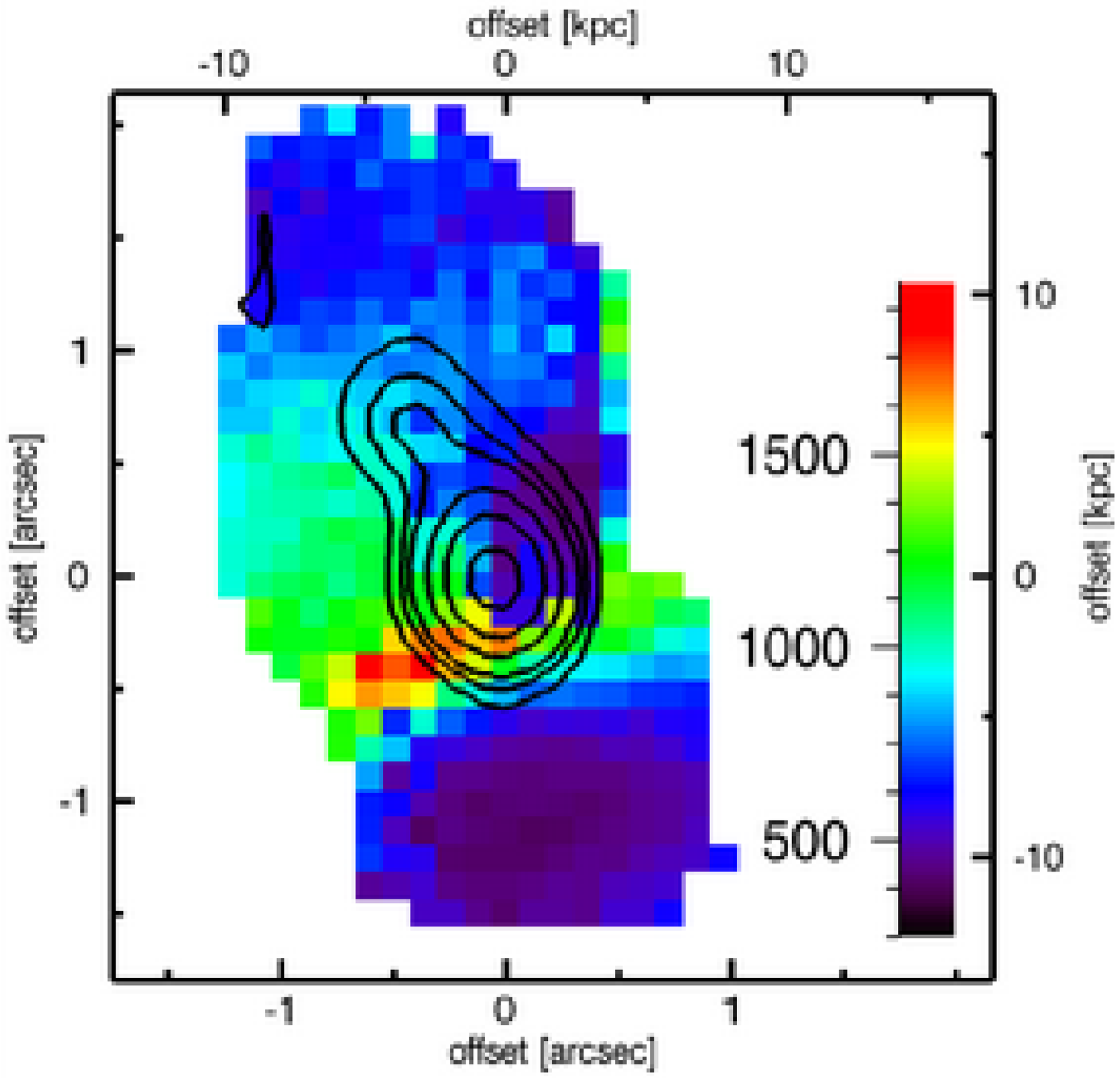,width=0.48\textwidth}
\caption{{\it top left:} [OIII]$\lambda$5007 emission line morphology of
TXS0828+193 at z$=2.57$ with contours showing the 1.4~GHz morphology.  {\it
bottom left:} Velocity map of TXS0828+193. Color bar shows the relative
velocities in km s$^{-1}$, contours indicate the 1.4~GHz morphology. {\it top
right:} [OIII]$\lambda$5007 emission line morphology with contours showing the
line-free, rest-frame optical continuum. {\it bottom right:} Maps of the line
widths, color bars show the FWHM in km s$^{-1}$, contours indicate the 1.4~GHz
morphologies. North is up, east to the left in all images.}
\label{fig:maps0828}
\end{figure}

Velocities and line widths are taken from the kinematic maps
shown in Figures~\ref{fig:maps0316} to~\ref{fig:maps0828} and thus
represent the local values for 
each spatial resolution element. We discarded the very broad lines near the
center of the galaxy, which we believe are largely due to overlaps between
the red and the blue bubble. Within each bubble, line widths are larger than
the velocity changes, so that they give a fair representation of the intrinsic
line widths within the spatial resolution of our data.

Since we neglected the likely presence of other gas phases, as well as
uncertainties related to projection, this can only be a conservative lower
limit. We do not assume pressure equilibrium, but that velocities are
dominated by ram pressure.

\noindent
(2) An energy estimate based on the assumption that the ionized
gas is being entrained and accelerated through adiabatic inflation of a bubble
of hot overpressurized material, which follows, e.g., \citet{dyson80}, and
which yields $\dot{E} = 1.5\times 10^{46} r^{2}_{10} v^3_{1000} n_{0.5}$ erg
s$^{-1}$.  $r_{10}$ indicates the size of the emission line nebulae in units
of 10 kpc, $v_{1000}$ the velocity in units of 1000 km s$^{-1}$, and n$_{0.5}$
the ambient gas density of the warm interstellar medium prior to being
  swept up by the bubble in multiples of 0.5 cm$^{-3}$. This approach is
highly idealized given the complexity of the turbulence and bulk motion
observed and expected from hydrodynamical simulations of radio jets
\citep[e.g.,][]{krause05,saxton05,gaibler07,sutherland07}, but may serve as a
rough analytical surrogate to more detailed estimates.
We used the velocity maps shown in Figures~\ref{fig:maps0316} to
  \ref{fig:maps0828} to estimate
the total velocity offsets between bubbles, and set the expansion velocity in
each bubble to half the total value. Values for individual galaxies are given
in Table~\ref{tab:outflow}.

Although
our estimates of velocities and line widths are precise to 10\%-20\%, each
approach are based on a set of very simple assumptions. While this is
appropriate given the observational difficulties of high-redshift studies, it
does 
introduce astrophysical uncertainties that will make each estimate precise to
factors of a few only.

Applying method (1) to our present data set, we find energy injection rates of
$3\times 10^{43}$ erg s$^{-1}$ and $1.3\times 10^{44}$ erg s$^{-1}$ for
TXS0828+193 and MRC0406-244, respectively, for mass estimates without
extinction correction. This represents a very conservative lower limit on the
energy injection rate. If using extinction-corrected mass estimates instead,
we find energy injection rates of $\sim 2\times 10^{45}$ erg s$^{-1}$ are
required for both galaxies with ionized mass estimates, MRC0406-244 and
TXS0828+193. (Individual values are given in Table~\ref{tab:outflow}.)

For MRC0316-257, we do not have an explicit estimate of the ionized gas mass,
because at the redshift of the source, z$\sim 3.1$, H$\alpha$ cannot be observed
from the ground, and H$\beta$ is very faint. The velocity maps shown in
Figures~\ref{fig:maps0316} to ~\ref{fig:maps0828} , however, suggest
overall very similar 
properties. Under the premise, that the ionized gas mass will not greatly
differ, we would hence expect to also find similar estimates of the kinetic
energy injection rate. For the 10 galaxies in the sample of \cite{baum00} at
z$\ge 1$, we find very similar results. 
Using their measurements of the
ionized gas masses, velocities, and line widths and our method 1, we find
total energy injections of $10^{57-59}$ erg over dynamical timescales
of about $10^7$ yrs.
Note that this is a
strict lower limit. If we correct for extinction as found from our SINFONI
data set, and uncertainties in the alignment between the longslit and the
direction of largest velocity shear, these estimates will easily increase by
about an order of magnitude. This certainly highlights the ubiquity of
powerful outflows of ionized gas in HzRGs.

Method (2) does not depend on the ionized gas mass, and can therefore be
applied to all three galaxies. We used the velocity maps shown in Fig. 8
to estimate the total velocity 
offsets between bubbles, and set the expansion velocity in each bubble to
half the total value. Values for individual galaxies are given in Table
6.

We find that energy injection rates of
$\dot{E} \sim 0.7-4\times 10^{45}$ erg s$^{-1}$ are required to power the
outflows (see Table~\ref{tab:outflow} for individual values). Note that
velocities are not corrected for orientation effects. Assuming that
inclinations will typically be around 45$^{\circ}$ or higher, intrinsic
velocities may be factors $\sim 2-3$ higher, which would result in kinetic
energies of up to several $\times 10^{46}$ erg s$^{-1}$ and total kinetic
energy injections during an AGN lifetime of order $10^{60}$ erg. 

\section{The role of the radio jet}
\subsection{Kinematics}
\label{ssec:whyjet}
Our study provides direct evidence for outflows of ionized gas driven by
powerful AGN during the ``Quasar Era''. \citet{nesvadba06} argued in the case of
MRC1138-262 that the energy injection, mass outflow rates, and dynamical time
scales satisfy the criteria set by observations of massive galaxies at low
redshift. In particular, the outflow time scales are short enough to explain
the observed relative enhancement of $\alpha$ elements over iron as a function
of velocity dispersion in early-type galaxies
\citep[e.g.,][]{pipino04,pipino07}. Direct studies of galaxies undergoing
phases of strong AGN feedback like HzRGs are ideal to investigate by which
mechanism the energy output of the AGN is being converted into thermal and
kinetic energy of the surrounding cold gas. Broadly
speaking, two groups of processes have been proposed in the literature (1)
Outflows driven by the radiative energy output of the AGN
\citep[e.g.,][]{king03,sazonov05,murray05,ciotti07}. (2) Outflows related to
mechanical interactions with the relativistic, synchrotron-emitting plasma of
radio-loud AGN. These were mainly proposed to overcome the cooling-flow
problem \citep[e.g.,][]{binney95,ruszkowski02,kaiser03}, but are also
addressing the impact of the jets in individual radio galaxies
\citep[e.g.,][]{mellema02,fragile04,saxton05,krause05,sutherland07,krause07}.
  
While both classes rely on physically plausible mechanisms, their actual
impact on the surrounding gas depends strongly on the {\it efficiency} by
which the energy output of the AGN is being converted into thermal and kinetic
energy of the surrounding gas. This is best addressed through direct
observations of systems that are in a phase dominated by powerful, AGN driven
outflows, like the galaxies of our sample. Overall our analysis favors a
picture where the radio jet plays a dominant role in driving the gas, based on
arguments related to the observed geometry, timescales, and kinetic energy of
the flows:

\noindent {\bf Orientation relative to the jet axis:} 
In the 3 galaxies of our sample, the extended emission line regions are
elongated along the axis of the radio jets, and are well aligned with the jet
axis. The same is suggested by the overall velocity field of the galaxies --
we clearly distinguish a redshifted and a blueshifted bubble, suggesting a
back-to-back flow, with a sharp transition near the center of the galaxy. The
size of the transition regions appears dominated by the spatial resolution of
our seeing-limited data. 
In the previous section, we also showed that the
blueshifted gas appears to be aligned with the near side of the radio source
in all three cases. This is a prerequisite, if the ionized gas is being
accelerated away from the galaxy by the expanding cocoon of hot gas
surrounding the radio jet.

\noindent {\bf Spatial Extent:} 
The region of high surface brightness emission extends
to radii that are smaller than the size of the radio lobes. Extending this
argument, \citet{nesvadba07b} find that galaxies with {\it compact} radio
sources do not have spatially extended line emission, but kinetic properties
suggesting that the gas kinematics is driven by the same underlying
processes. In agreement with spectral studies of radio jets
\citep[e.g.][]{murgia99,owsianik98}, this is a natural result if compact radio
sources are predominantly young. For similar entrainment rates as found in our
sample, and typical ages of a few$\times 10^{5-6}$ yrs estimated for compact
radio sources, these galaxies should not yet have entrained substantial
fractions of the interstellar medium, in agreement with observations
\citep[see][for details]{nesvadba07b}.

\citet{villar02,villar03,villar06,villar07} find faint rest-frame UV emission
surrounding HzRGs, with line widths and velocity gradients that suggest that
the gas does not participate in the violent processes in the inner halo, that
we are tracing with our rest-frame optical data. These kinematically quiescent
outer halos are always found at radii beyond the radio jets, which is highly
suggestive in a scenario where the outflows are dominated by interactions with
the radio jet.

\noindent
{\bf Time scales:}
In \S\ref{ssec:dynamicaltimes}, we estimated dynamical time scales of a few
$\times 10^7$ yrs from the characteristic velocities and spatial extent of the
flows. In spite of large uncertainties, this is in the typical range of ages
estimated for radio jets \citep[e.g.,][]{kaiser97,kaiser07,blundell99}. 

\noindent
{\bf Energies:}
Anticipating the results of \S\ref{ssec:couplingefficiency}, we estimate that
about 10\% of the kinetic energy flux of the jet is transformed into kinetic
energy of the gas. This is fully consistent with a picture where the radio jet
is the primary driver of the observed gas kinematics. 

These large coupling efficiencies may be somewhat surprising. Early models of
radio jets indicated that powerful jets like the ones in our sample (in the
energy range of FR II radio galaxies), will pass through the ambient medium
rather undisturbed, while low-power jets (in particular FR I galaxies)
may deposit most of their energy in the surrounding medium after being
disrupted \citep[e.g.,][]{young91}. 

More recent models with better resolution, and larger density contrast
between jet plasma, ambient medium, and embedded clouds, suggest a more
subtle picture. Large jet kinetic energies of few $\times 10^{46}$ erg
s$^{-1}$ and highly inhomogeneous ambient media are clearly required to
adequately describe powerful radio galaxies at z$\sim 2$, and several recent
models have now investigated this
\citep[e.g.][]{saxton05,sutherland07,krause05,krause07}. The basic properties
of the simulated cocoons are not unrealistic compared to what is observed,
e.g., internal velocities of order 1000 km s$^{-1}$ or aspect ratios of
$\sim 3$ between the radial and the lateral axis of the cocoon
\citep{saxton05,krause05}.

However, the model parameters are adapted to lower redshift galaxies with
somewhat lower jet eneriges, and lower fractions of cold gas. In particular,
the environments of HzRGs suggest the galaxies are surrounded by halos of cold
gas with complex substructure, rather than well-settled disks of cold gas as
adopted by \citet{sutherland07}.

\citet{saxton05} point out that a clumpy environment may destabilize or even
disrupt the jet by scattering on several clouds, and through Kelvin-Helmholtz
instabilities induced by the overall level of turbulence rather than
individual jet-cloud interactions.  It is plausible that this will enhance the
efficiency with which the kinetic jet energy is being deposited into the cold
gas even for the most powerful jets. Models better adapted to the properties
of powerful radio galaxies at z$\sim 2$ may be an interesting way of
investigating this further, including the kinematic constraints now available
from integral-field spectroscopy.

\subsection{Multiphase gas content of HzRGs}
\label{ssec:massdiscussion}

\subsection{Mass budgets} 
The extended emission line regions of HzRGs do not only have extreme kinematic
properties, but also an unusually large content of ionized gas, few $\times
10^{10}$ M$_{\odot}$ \citep[\S\ref{ssec:ionizedgas}, see also][]{baum00}. This
exceeds the amount of ionized gas in any other high-redshift galaxy population
discussed in the literature by several orders of magnitude.

For example, the same mass estimate applied to the total H$\alpha$
emission of z$\sim 2$ actively star-forming ``BX''-galaxies measured by
\citet{erb06}, would only correspond to a few $\times 10^7$ M$_{\odot}$, and
most of this gas is probably related to star-forming regions and does not
participate in an outflow. \citet{nesvadba07a}
used SINFONI to obtain spatially-resolved data of the starburst-driven wind in
a submillimeter selected, strongly star-forming galaxy (SMG),
SMMJ14011+0252. They find that in spite of high star-formation rates of SFR$\sim
300 M_{\odot}$ yr$^{-1}$, only $\sim 10$\% of the total H$\alpha$ flux is in
a blueshifted component -- the wind -- which would correspond to ionized gas
masses of a few $\times 10^6$ M$_{\odot}$.

Overall, the stellar masses and star-formation rates in z$\sim 2$ HzRGs
closely resemble those of SMGs at similar redshifts. Moreover, several
$10^{10} M_{\odot}$ of cold molecular gas traced by CO millimeter line
emission have been observed in SMGs \citep[e.g.,][]{neri03,greve05} as well as
in HzRGs \citep[e.g.,][]{papadopoulos00,debreuck03,debreuck05,klamer05}. Large
masses of cold gas are a necessary prerequisite to fuel a strong
starburst. However, not all HzRGs have been detected in CO. For TX0828+193
specifically, which appears to have several $10^{10}$ M$_{\odot}$ in ionized
gas(\S\ref{ssec:ionizedgas}), Nesvadba et al. (in prep.) did not detect CO
line emission, but place an upper limit of $\sim 1\times 10^{10} M_{\odot}$ in
cold molecular gas within the galaxy (for a ULIRG-like conversion factor from
CO luminosity to H$_2$ mass, similar to that adopted for other high-redshift
CO-luminous galaxies). Taken at face value, this is less than the amount of
ionized gas we observe with SINFONI.
Thus, it appears that the ratios of ionized to molecular gas in HzRGs, are of
order unity or greater, $R_{\rm{ion,mol}}^{HzRGs} \ge {\cal O}(1)$, compared
to $R_{\rm{ion,mol}}^{SFG} \le {\cal )}(10^{-(3-4)})$ for actively
star-forming galaxies at $z\sim 2$ (SFGs) including SMGs. (Either value is
affected by observational uncertainties of factors of a few, but the large
contrast between star-forming galaxies and radio galaxies makes this
comparison robust nonetheless.) This certainly highlights the importance of
the observed outflows for the evolution of the host galaxy, and it is
interesting in this context that some HzRGs show the signs of strong
star-formation \citep[like large reservoirs of molecular gas, strong PAH and
submillimeter emission,][]{archibald01,reuland04,seymour08}, while others do
not. This is certainly to be expected for a population that marks the
transition from an actively star-forming to a passively evolving evolutionary
stage. A detailed study of the multiphase gas content of HzRGs for a more
representative sample will be crucial to assess this question directly.

\subsection{Ionizing mechanism}
Closely related is the question which mechanism dominates the ionization of
the gas? This is particularly important since our gas mass estimates
\citep[similar to those of ][]{baum00} do assume that most of the gas is
photoionized, and in fact, the bolometric energy output of HzRGs seems
sufficient to postulate that the nebulae may be matter bounded
\citep{nesvadba06}. Photoionization is also favoured by the measured UV line
ratios \citep{villar99}.  However, the gas kinematics suggests that
shocks play a dominant role for the overall energy budget of the outflows, and
hence it would appear natural that shocks would be an important ionizing source
\citep[a well-known puzzle, for more detailed discussions see
e.g.][]{clark98,villar97,allen08}. Using the models of
\citet{dopita95,dopita96} for 
the line emission of fast shocks, we find that by assuming that shock
ionization dominates, gas mass estimates would be about an order of magnitude
larger than what we estimated from case B recombination \citep[see][for
details]{nesvadba06}. In this sense, and in spite of uncertainties of factors
of a few, our gas mass estimates are rather conservative.

Ultimately, photoionization may explain why we observe
larger amounts of ionized gas by several orders of magnitude than what is
found in low-redshift radio galaxies, where fast, AGN-triggered outflows are
predominantly seen in neutral gas through broad, blueshifted ($\sim 1000$ km
s$^{-1}$) HI absorption lines \citep{morganti05}. This dichotomy may be
mainly due to the different radiative properties of the AGN rather than
fundamentally different acceleration mechanisms for the outflows; HzRGs are
harboring luminous, obscured quasars, and are embedded in large quantities of
cold gas, whereas low-redshift radio galaxies seem to be radiatively
inefficient and reside in hot, X-ray emitting halos. If this is the case, then
we may significantly underestimate the frequency of AGN driven outflows,
simply because they are observationally more subtle \citep[see][for a more
detailed discussion]{morganti05}.

\subsection{Entrainment rates}
\label{ssec:entrainment}
The amount of ionized gas and the dynamical time scales provide a rough
estimate of the entrainment rate, as the cocoon plasma interacts with the
ambient medium. The efficiency of the overpressurized plasma in entraining
colder (T$\le 10^4$ K) material is one of the crucial parameters in
characterizing the efficiency by which the radio jet interacts with the ISM
of a galaxy, and is a parameter that is now being addressed by detailed
hydrodynamic simulations \citep[e.g.,][]{krause05,sutherland07}. However, since
the amount of entrained material will strongly depend on the ``macro'' and
``micro''-physics of the jet and ambient material, including several orders of
magnitude in pressure, density, and size scales, it is difficult to derive
this quantity from first principles. 

The ionized gas we see will most likely be on the surface of filaments with
low filling factors \citep{nesvadba06}, that are being swept up, accelerated,
and ionized by the hot medium surrounding the radio jet. We can therefore give
a purely empirical estimate of the entrainment rate, assuming that it roughly
corresponds to the outflow rate of ionized gas,  M$_{ion}$, and the dynamical
timescale, $\tau$. 

Recasting the discussion in \citet{nesvadba07b}, we approximate the rate of
entrained material, $M_{entr,9}$ (in multiples of $10^9 M_{\odot}$), from the
ionized gas masses of $\times 10^{10}$ M$_{\odot}$ and dynamical timescales of
few $\times 10^7$ yrs. This yields a simple scaling, $\rm{M_{entr,9}} = f_c\
f_{\rho_r} f_{geom} \dot{\rm{M}}_{ion}\ \tau_{7}$, where $\tau_{7}$ indicates
the dynamical time scale in units of $10^7$ yrs.  $f_c$, $f_{\rho_r}$, and
$f_{geom}$ parametrize our ignorance of the jet advance speed, radial density
profile, and geometry, respectively. If we assume that all are of order unity
for the relevant radii and time scales, we find an average entrainment rate of
$\dot{M}_{entr}\sim 10^{2-3}$M$_{\odot} \rm{yr}^{-1}$. The cylindrical
shape of the outflows suggests that the outflows do cover large solid angles,
since they are related mainly to the expansion of the cocoon, and contrary to
what may be expected from the pencil-like morphology of the jet.

The large ratio of ionized to molecular gas of order unity will make this
estimate relatively robust. Nonetheless it will be a lower limit, given that
we have no direct constraints on the amount of hot gas in the cocoon, and CO
only traces the cold molecular gas with temperatures of order $\sim 10-100$
K. \citet{krause05} argue that the mass of X-ray emitting gas within the
cocoon may not greatly exceed the mass estimates for the colder gas given
above. \citet{carilli02} estimate that the X-ray emission of MRC1138-262 may
originate from up to $\sim 10^{12}$ M$_{\odot}$ of hot gas, however, more
recent X-ray studies of HzRGs favor the view that much of the emission
originates from Compton scattering of photons from the cosmic microwave
background on the relativistic electrons of the jet
\citep[e.g.][]{overzier05}, and accrodingly less X-ray emitting gas.

\subsection{Coupling Efficiencies}
\label{ssec:couplingefficiency}
If jets are to cause the observed outflows, then they must provide at least
the kinetic energy observed in the gas. The synchrotron emissivity of a radio
source is a strong function of the ambient medium into which a jet expands. It
is therefore difficult to estimate the intrinsic kinetic energy of the
relativistic plasma from the luminosity of the radio source. Several methods
have been proposed in the literature
\citep[e.g.][]{bicknell97,willott99,wan00,birzan04,kaiser07} which suggest
that of order 0.1\%-10\% of the kinetic energy of a jet is emitted in the
radio. Similar to \citet{nesvadba06} we adopt two methods to estimate the jet
kinetic energies of the galaxies in our sample. (1) A simple scaling, assuming
that 10\% of the kinetic energy is radiated at frequencies between 0.1-1 GHz
in the rest-frame. (2) The empirical relationship found by \citet{wan00}
between the observed flux at 178 MHz frequency and the kinetic radio
power. With the two methods, we find that the jets eject several $10^{46}$ erg
s$^{-1}$ in kinetic energy. Given that the adopted values are close to the
lower bound of conversion factors from observed to intrinsic luminosities,
these values serve effectively as lower bounds. For the estimates of
individual galaxies see Table~\ref{tab:radio}.

A critical quantity to gauge the overall impact of jet-driven AGN outflows is
the efficiency with which the jet kinetic energy, $E_{kin,jet}$, is being
translated into kinetic energy of the outflow,$E_{kin,gas}$, the coupling
efficiency $\eta = E_{kin,gas}/E_{kin,jet}$. Obviously, such an estimate will
not be very precise, given the large uncertainties in each
measurement. Nonetheless, we feel that quantifying this value at
order-of-magnitude precision will already be useful. With the values given in
Table~\ref{tab:outflow}, we find overall coupling efficiencies of $\sim
5-80$\%.

\citet{kaiser97} pointed out that for approximately adiabatic expansion, a
radio source will become fainter as it expands. However, in our simple attempt
to estimate jet kinetic energies, we did not take this into account. This may
explain why we find the highest efficiencies in TXS0828+193, which is the
largest source in our sample, by a factor $\sim 2$. We may simply
underestimate the kinetic luminosity of the radio jet by factors of a few. We
therefore favor a coupling efficiency of order 10\%. Similar efficiencies
have been found in radio galaxies with outflows of neutral gas at
significantly lower redshifts \citep{morganti05}, as well as in the X-ray
cavities of massive low-redshift galaxy clusters
\citep[e.g.][]{mcnamara07,birzan04,rafferty06}. This may be a first indication
that the basic mechanism in all these environments may not be fundamentally
different although they act on different scales.

\section{The global role of outflows in HzRGs}

One of the main questions we address in this study is on the frequency of
highly turbulent emission line gas and energetic outflows in HzRGs, similar to
what we observed in our SINFONI study of MRC1138-262. Such outflows may be an
important mechanism in terminating the vigorous starbursts observed in
galaxies near the upper end of the galaxy mass function during the ``Quasar
Era'', if they are powerful enough to efficiently heat and unbind a
significant fraction of the interstellar medium of a massive, gas-rich
galaxy. The presence of such outflows in another three HzRGs with extended
radio sources (among three galaxies observed) suggest that AGN-driven outflows
are a common phenomenon in z$\sim 2-3$ HzRGs.

Overall, the physical conditions and gas kinematics in MRC1138-262 and the
three galaxies of the present sample are very similar. All four galaxies have
typical line widths of FWHM $\sim 800-1000$ km s$^{-1}$ over large areas,
suggesting turbulent energy injection rates of $\sim 10^{44}$ erg
s$^{-1}$. Areas with larger widths may be due to overlap of adjacent
bubbles. H$\alpha$ fluxes indicate fairly similar ionized gas masses in all
four galaxies. The physical conditions within the outflows seem also largely
similar, with electron densities of several hundred cm$^{-3}$ for MRC0406-244
and MRC1138-262. Both TXS0828+193 and MRC0406-244 have [OIII] emission line
ratios consistent with temperatures of $\sim 10^4$ K. Relative velocities
between individual bubbles are $\sim 1000$ km s$^{-1}$ in all cases,
suggesting velocities near the escape velocity of a massive dark matter halo.
Differences are mainly in the emission line morphology where MRC1138-262 seems
particularly irregular. The reason may be different evolutionary stages or
perhaps that MRC1138-262 has a more extreme environment. This is suggested by
the particularly large velocity dispersion of companion galaxies
\citep[e.g.,][]{venemans07,miley06} and an exceptionally large rotation measure
\citep[][]{pentericci00}. However, we cannot rule out that the greater
complexity of MRC1138-262 is simply the
result of different orientations, with the AGN of MRC1138-262 being less
inclined relative to the line of sight. MRC1138-262 hosts an obscured quasar
evidenced by broad nuclear H$\alpha$ line emission \citep{nesvadba06}.
  
The global dynamical properties of the emission line gas however do appear very
similar in all 4 galaxies of our current sample, in spite of this range in the
appearance of individual sources. This suggests that powerful outflows in
HzRGs likely are a common phenomenon. Studying the outflows in a larger sample
of HzRGs with and without broad H$\alpha$ lines may help resolving this issue,
and we have now started observations of a significantly larger sample of $\sim
30$ HzRGs with SINFONI.

\subsection{Subsequent Mass Assembly of HzRGs}
The properties of HzRGs suggest that they are near the end of their main epoch
of active star-formation, as seen, e.g., from the evolution towards very
regular, probably elliptical continuum profiles in HzRGs at z$\sim 2$
\citep[e.g.,][]{vanbreugel98}. This is certainly consistent with postulating
that the AGN winds we observe are truncating the star-formation epoch in the
host galaxy. However, will the fingerprints of this epoch be preserved until
z$=$0? HzRGs reside in overdense environments, and may double their stellar
mass through subsequent satellite accretion \citep[as predicted by galaxy
evolution models,][]{delucia06}.

We can roughly estimate the accretion-dominated growth of HzRGs from the
number of companion galaxies observed around many HzRGs, including MRC0316-257
and MRC1138-262 \citep[e.g.][]{lefevre96,venemans07, venemans05,
kurk04,best03}. Continuum luminosities and H$\alpha$ line widths
\citep[e.g.,][]{kurk04} imply masses of typically few $\times 10^9
M_{\odot}$. With up to $\sim 50$ companions, the central galaxy may gain of
order $10^{11}$ M$_{\odot}$, if all companions are accreted, correspoding
about half its stellar mass. For gas fractions of order 50\%, satellite
accretion may fuel the formation of $\sim 20-30$\% of the total stellar mass
in the descendent of a HzRG at z$\sim 0$.

However, this will not inevitably dilute the fingerprints of previous AGN
feedback, such as the low content of cold and warm gas or truncated
star-formation histories of massive galaxies at low redshift
\citep[e.g.,][]{pipino04}. Dynamical interactions in the dense environment of
HzRGs will lead to premature ageing of the stellar population in the
satellites and efficient gas removal, similar to low-redshift galaxy
clusters. \citet{boselli08} find for dwarf galaxies in the Virgo cluster, that
ram pressure stripping alone may turn strongly star-forming irregular galaxies
into gas-poor ellipticals within 1$-$2 crossing times (few 100 Myrs for a HzRG
halo) and lift the satellite onto the red sequence. \citep[Interestingly, the
red sequence seems to fall in place at z$\sim 2$, as shown][for the
overdensities surrounding MRC1138-262 and MRC0316-257.]{kodama07} The metal
enrichment of the ISM will be particularly efficient, resulting in a
metallicity increase of about 0.5 dex relative to the field
\citep[][]{boselli08}. Comparing with the mass-metallicity relationship of
\citet{tremonti04} we find that this will compensate for the metallicity
difference between a $5\times 10^9$ M$_{\odot}$ galaxy and a $10^{11}$
M$_{\odot}$ galaxy.

AGN outflows may enhance these processes, either through gas stripping when a
satellite crosses the outflow \citep[][]{irwin87,fujita08}, or by increasing
the entropy and pressure in the surrounding hot medium
\citep[][]{nath02,mccarthy08}, where an AGN-driven pre-heating phase at z$\sim
1-3$ appears sufficient to explain the observed temperature profiles of
massive galaxy clusters. In either case, the satellite galaxies will likely
undergo ``dry'' mergers with low dissipation, and may have a surprisingly
small impact on the properties of the low-redshift descendents of HzRGs. Dry
merging may also explain why HzRGs seem to have smaller effective radii than
massive spheroidal galaxies at low redshift, as well as the extended stellar
envelopes of cD galaxies \citep{schombert87}.

\subsection{The cosmological significance of jet-driven outflows} 
\label{ssec:cosmologicalsignificance}
With only 6 HzRGs with rest-frame optical IFU data, our sample is relatively 
small. However, the great similarity of the ionized gas masses and kinetic
properties is intriguing, and raises confidence that similar outflows are 
common among powerful radio galaxies during the era of luminous
quasars. Moreover, large ionized gas masses, line widths, and velocity
gradients have been reported previously, based on observations with longslit
spectrographs of radio galaxies at redshifts z$\sim 1$ and in a few cases below
\citep[e.g.][]{baum00,villar02,villar03,villar99,mccarthy96,tadhunter91,
inskip02,clark98,best97}.

In \S\ref{ssec:couplingefficiency} we argued that of order 10\% of the jet
kinetic energy is transformed into kinetic energy of the gas. We will in the
following assume that this is a typical value for powerful radio sources at
z$\sim 2$. We can then use the observed radio luminosity function of
steep-spectrum radio sources to estimate the global energy release of
jet-driven AGN feedback into the intergalactic medium at high redshift.

We use the radio luminosity function of \citet{willott01}, who find that the
radio-loud population can be split into two classes, a low-luminosity
population associated with galaxies with low-luminosity emission lines
including FRI and FRII radio structures, and a high-luminosity population
consisting of radio galaxies and quasars with strong emission lines, and with
nearly exclusively FRII radio sources. We only include this luminous
population into our estimates, and conservatively use their set of model
parameters which minimizes our estimate. The comoving number density of this
population rises by nearly 3 orders of magnitude between z$=$0 and z$=$2, and
has properties that generally agree very well with the sources we observe.

We computed the luminosity function for redshifts z$=0-4$ in bins of z$=0.1$,
and use the peak of the co-moving number density at a radio-power at 151 MHz
of ${\cal L}_{151} =$ 10$^{27}$ W Hz$^{-1}$ sr$^{-1}$, which is sufficiently
close to the radio power of the sources in our sample.  We then compute the
co-moving number density per redshift bin, $\Phi_{mod}$, converting to our
cosmology. Since the radio-loud phase, $\tau_{jet}$ is significantly shorter
than the cosmic time within these bins, we estimate co-moving number densities
corrected for the duty cycle ,$\Phi_{rl,c}$, i.e.  $\Phi_{rl,c} =
\frac{t_{bin}}{\tau_{jet}}\ \Phi_{mod}$, where $t_{bin}$ indicates the cosmic
time elapsed within each redshift bin, and $\tau_{jet}$ indicates the lifetime
of the radio jet. We find a global energy release at redshifts z$=1-3$
(i.e., near the ``Quasar Era''), $d\rho_{E,kin}/dt$
\begin{equation} d\rho_{E,kin}/ dt = 9 \times 10^{56} \tau_{7}^{-1}\
{dE_{kin}}_{60} / dt_{60}\ {\rm erg\ s^{-1}},
\end{equation}
where $\tau_7$ is given in $10^7$ yrs, and ${dE_{kin}/dt}_{60}$ is given in
$10^{60}$ erg. 
For the redshift evolution of \citet{willott01} this corresponds to about
90\% of the energy release at all redshifts, emitted during $\sim 3.6$
Gyrs.

\subsection{Overall feedback efficiency}
To quantify the overall efficiency of AGN feedback as observed in powerful
radio galaxies, we will now compare the observed kinetic energy injection of
the ionized gas within an activity phase with the possible total energy
release through accretion onto the supermassive black hole. This simply
corresponds to the relationship between the total energy equivalent of the
rest mass of a black hole of a given mass and the kinetic energy of the
outflows, 
\begin{equation}
\eta = 5.6\times 10^{-2}\ E_{kin,60}\ t_{7}\ M_{9}^{-1}
\end{equation}
$E_{kin,60}$ indicates the energy injection in multiples of $10^{60}$ erg,
$t_{7}$ the phase of AGN activity in multiples of $10^7$ yrs, and $M_{BH,9}$
the mass of the supermassive black hole in multiples of $10^9$
M$_{\odot}$. The efficiency $\eta$ is given in percent. 

Black hole masses for HzRGs are not easily derived directly. For an
order-of-mass estimate, we will assume that the low-redshift relationship
between black hole and bulge mass will approximately hold
\citep{magorrian98,ferrarese02,tremaine02,haring04}. The light profiles of
HzRGs at z$\sim 2$ are regular and likely elliptical
\citep{vanbreugel98,pentericci01}, suggesting that these are bulge-dominated
systems. Adopting the stellar mass estimates of \citet{seymour07} for HzRGs,
$\log{M_{bulge}}\sim 11 -11.5$, and the relationship of \citet{haring04},
black hole masses are about $\log{M_{bh}} \sim 8.2-8.8$.  This implies that of
order 0.1\% of the energy equivalent of the black holes in HzRGs are released
in kinetic energy of the outflows. This is a purely empirical estimate, and
does not rely on any assumption regarding the efficiency of accretion or
feedback. 

To investigate this question more globally, we can also relate the global
energy density released by powerful radio sources derived in
\S\ref{ssec:cosmologicalsignificance}, $\rho_{obs}\sim 9\times 10^{56}$ erg
Mpc$^{-3}$, with the local mass density of black holes, estimated by
\citet{yu02}, $\rho_{BH}=2.9\pm 0.5 \times 10^5\ h_{70}^{2}\ M_{\odot}\
Mpc^{-3}$, corresponding to a total energy equivalent of $\rho_{e,kin}\sim
5.2\times 10^{59}$ erg Mpc$^{-3}$.  This will serve as a strict lower limit on
the efficiency with which the rest-frame energy equivalent of the black hole
mass has been translated into outflow energy in the past.  We find an
efficiency of about 0.2\% with which the rest mass of the black hole is being
translated into outflow energy, in agreement with the above estimate based on
the masses of supermassive black holes in HzRGs, and with recent estimates
comparing the optical and radio luminosity function of radio-loud
AGN, but without the direct observations of AGN-driven winds
\citep[][]{shankar08}.

The efficiency of AGN feedback in terms of the black hole accretion is an
important parameter in cosmological models of galaxy evolution and structure
formation, which typically adopt a factor of $\sim 0.5$\%
\citep[e.g.][]{scannapieco04,springel05,dimatteo05,chakrabarti07}. Given the
uncertainties of factors of a few, and the fact that our estimates are based
on the ionized gas only, the agreement is remarkable. This underlines the
potentially significant role of jet-driven outflows for galaxy evolution.

\section{Summary and Conclusions}
We studied the kinematics and physical characteristics of the 
extended emission line regions in 3 powerful radio galaxies at z$\sim 2$,
using integral-field spectroscopy of the rest-frame optical emission
lines. Our results are consistent with what is expected from 
giant, jet-driven AGN outflows, similar to what we found in an earlier study
of the z$\sim 2.2$ galaxy MRC1138-262 \citep{nesvadba06}. Analysis of the
emission and continuum line morphologies, velocity maps, and flow orientation
relative to the orientation of the radio jet does not favor alternative
interpretations, e.g., merging, rotation, or gas infall. Such outflows may be
common  in powerful high-redshift radio galaxies (HzRGs) during the
``Quasar Era''. Overall, we find energetic outflows in all 4 z$\sim 2$ powerful
radio galaxies with extended jets for which we obtained rest-frame optical
integral-field spectroscopy. Comparison with previous longslit spectroscopy
\citep{baum00} suggests similar properties in $> 10$ HzRGs at z$>1$.

The emission line regions extend over $\sim 10$ kpc$\times$30 kpc, and are
elongated along the axis of the radio jets. Most of the complex morphologies
of these galaxies, previously observed with broadband imaging, appears to be
due to line emission. The line-free continuum emission extracted from the data
cube is relatively faint and compact ($\le 10$ kpc). Assuming simple case-B
recombination, we find of order $10^{10}$ M$_{\odot}$ of ionized gas, similar
to the typical cold molecular gas content of HzRGs traced by CO line
emission. From observed line ratios, we can constrain the physical properties 
of the gas. We detect [OIII]$\lambda$4363 in a z$\sim 2$  
galaxy, allowing for a direct estimate of the electron temperature of $\sim
10^4$ K. Electron densities are derived from the
[SII]$\lambda\lambda$6716,6731 line ratio and are $\sim 500$ cm$^{-3}$. From
the measured H$\alpha$/H$\beta$ line ratio we estimate extinction of
A(H$\beta$)$\sim 1-4$ mag. 

The velocity fields are very similar in all 3 sources and are reminiscent of
two bubbles expanding back-to-back from the AGN. Velocity offsets near the
center of the host galaxy are large, 700$-$1000 km s$^{-1}$, but velocities
within each bubble are relatively uniform. Line widths are of order FWHM$\sim
1000$ km s$^{-1}$, indicating strong turbulence. Larger widths in
the central regions are likely due to overlaps between the bubbles. Comparison
with the orientation of the radio jets suggest that this gas is outflowing. 

With kinetic energies of order $10^{60}$ erg required to power the
observed emission line kinematics, these outflows may be sufficient to remove
most of the interstellar medium of the host galaxies in a nearly explosive
event. Observed kinetic energies correspond to $\sim 0.2$\% of the energy
equivalent of the rest mass of the supermassive black holes, in good agreement
with AGN feedback prescriptions used in galaxy evolution models

The following observations lead us to argue that the outflows are dominated by
the radio jets: (1) Geometry: Good alignment with the radio jets and
elongation along the jet axis. The gas is blueshifted on the side of the
approaching radio jet. (2) Size: The sizes of the emission line
regions do not reach beyond the radio lobes. \citet{nesvadba07b} do not find
extended outflows in galaxies with compact radio sources observed with
SINFONI. Emission line gas outside the radio lobes observed in Ly$\alpha$ is
kinematically quiescent \citep{villar03}. (3) Timescales: Dynamical timescales
of the outflows are few $\times 10^7$ yrs, similar to typical ages of radio
jets. (4) Energies: Kinetic energies of the outflows correspond to about 10\%
of the jet kinetic energy. 

Comparing with the z$\sim 2$ radio luminosity function of \citet{willott01},
we find that AGN winds like those observed may release about 9$\times 10^{56}$
erg s$^{-1}$ Mpc$^{-3}$ in energy density from the massive host galaxies of
powerful radio sources at z$\sim 2$. This energy may have a long-term impact
on the subsequent evolution of the AGN host galaxy as well as nearby
galaxies and help maintaining the ``anti-hierarchical'' properties of massive
galaxies in spite of possible subsequent merging with companion
galaxies. Following \citet{nath02,mccarthy08}, much of this energy may
contribute to increasing the entropy and pressure in the environment of
massive early type galaxies and clusters, and enhance the efficiency of
dynamical interactions with galaxies in the environment of HzRGs, such as ram
pressure stripping. Comparing with studies of ram-pressure stripping
\citep{boselli08} we find that the satellite galaxies may already be red,
poor in gas and rich in metals when they are being accreted onto the central
galaxy. In spite of subsequent growth by up to $\sim 50\%$ in stellar mass,
we find that minor ``dry'' mergers with ram-pressure stripped satellites would
not strongly influence the observed evolutionary properties of the central
galaxy.

\acknowledgements 
The authors wish to thank C. Carilli for interesting discussions with and
without good Bavarian beer, and for generously sharing his VLA imaging. NPHN
is grateful to R. Daly, C. Kaiser, A. Pippino, J. Ostriker and
M. Villar-Martin for interesting and inspiring discussions. She also wishes to
thank M. Krause and V. Gaibler for sharing their insight into hydrodynamical
modeling of radio jets and their hospitality at the Landessternwarte in
Heidelberg. Comments by the anonymous referee helped improve the paper. We
thank the ESO OPC for their allocation of observing time and the staff at
Paranal for their help and support in making these observations. NPHN wishes
to acknowledge financial support from the European Commission through a Marie
Curie Postdoctoral Fellowship and MDL wishes to thank the Centre Nationale de
la Recherche Scientifique for its continuing support of his research.  WvB
acknowledges support for radio galaxy studies at UC Merced, including the work
reported here, with the Hubble Space Telescope and the Spitzer Space Telescope
via NASA grants HST \#10127, SST \#1264353, SST \#1265551, SST \#1279182 and
SST \#1281587.
\bibliography{0346}

\begin{thebibliography}{132}
\expandafter\ifx\csname natexlab\endcsname\relax\def\natexlab#1{#1}\fi

\bibitem[{{Alexander} \& {Leahy}(1987)}]{alexander87}
{Alexander}, P. \& {Leahy}, J.~P. 1987, \mnras, 225, 1

\bibitem[{{Alexander} \& {Pooley}(1996)}]{alexander96}
{Alexander}, P. \& {Pooley}, G.~G. 1996, {The large-scale structure, dynamics
  and thermodynamics of Cygnus A} (Cygnus A -- Studay of a Radio Galaxy),
  149--+

\bibitem[{{Allen} {et~al.}(2008){Allen}, {Groves}, {Dopita}, {Sutherland}, \&
  {Kewley}}]{allen08}
{Allen}, M.~G., {Groves}, B.~A., {Dopita}, M.~A., {Sutherland}, R.~S., \&
  {Kewley}, L.~J. 2008, ArXiv e-prints, 805

\bibitem[{{Archibald} {et~al.}(2001){Archibald}, {Dunlop}, {Hughes},
  {Rawlings}, {Eales}, \& {Ivison}}]{archibald01}
{Archibald}, E.~N., {Dunlop}, J.~S., {Hughes}, D.~H., {et~al.} 2001, \mnras,
  323, 417

\bibitem[{{Armus} {et~al.}(1998){Armus}, {Soifer}, {Murphy}, {Neugebauer},
  {Evans}, \& {Matthews}}]{armus98}
{Armus}, L., {Soifer}, B.~T., {Murphy}, Jr., T.~W., {et~al.} 1998, \apj, 495,
  276

\bibitem[{{Baum} \& {McCarthy}(2000)}]{baum00}
{Baum}, S.~A. \& {McCarthy}, P.~J. 2000, \aj, 119, 2634

\bibitem[{{Best} {et~al.}(2006){Best}, {Kaiser}, {Heckman}, \&
  {Kauffmann}}]{best06}
{Best}, P.~N., {Kaiser}, C.~R., {Heckman}, T.~M., \& {Kauffmann}, G. 2006,
  \mnras, 368, L67

\bibitem[{{Best} {et~al.}(2005){Best}, {Kauffmann}, {Heckman}, {Brinchmann},
  {Charlot}, {Ivezi{\'c}}, \& {White}}]{best05}
{Best}, P.~N., {Kauffmann}, G., {Heckman}, T.~M., {et~al.} 2005, \mnras, 362,
  25

\bibitem[{{Best} {et~al.}(2003){Best}, {Lehnert}, {Miley}, \&
  {R{\"o}ttgering}}]{best03}
{Best}, P.~N., {Lehnert}, M.~D., {Miley}, G.~K., \& {R{\"o}ttgering}, H.~J.~A.
  2003, \mnras, 343, 1

\bibitem[{{Best} {et~al.}(1997){Best}, {Longair}, \& {Roettgering}}]{best97}
{Best}, P.~N., {Longair}, M.~S., \& {Roettgering}, J.~H.~A. 1997, \mnras, 292,
  758

\bibitem[{{Bicknell} {et~al.}(1997){Bicknell}, {Dopita}, \&
  {O'Dea}}]{bicknell97}
{Bicknell}, G.~V., {Dopita}, M.~A., \& {O'Dea}, C.~P.~O. 1997, \apj, 485, 112

\bibitem[{{Binney} \& {Tabor}(1995)}]{binney95}
{Binney}, J. \& {Tabor}, G. 1995, \mnras, 276, 663

\bibitem[{{B{\^i}rzan} {et~al.}(2004){B{\^i}rzan}, {Rafferty}, {McNamara},
  {Wise}, \& {Nulsen}}]{birzan04}
{B{\^i}rzan}, L., {Rafferty}, D.~A., {McNamara}, B.~R., {Wise}, M.~W., \&
  {Nulsen}, P.~E.~J. 2004, \apj, 607, 800

\bibitem[{{Blundell} \& {Rawlings}(1999)}]{blundell99}
{Blundell}, K.~M. \& {Rawlings}, S. 1999, \nat, 399, 330

\bibitem[{{Boehringer} {et~al.}(1993){Boehringer}, {Voges}, {Fabian}, {Edge},
  \& {Neumann}}]{boehringer93}
{Boehringer}, H., {Voges}, W., {Fabian}, A.~C., {Edge}, A.~C., \& {Neumann},
  D.~M. 1993, \mnras, 264, L25

\bibitem[{{Bonnet} {et~al.}(2004){Bonnet}, {Abuter}, {Baker}, {Bornemann},
  {Brown}, {Castillo}, {Conzelmann}, {Damster}, {Davies}, {Delabre},
  {Donaldson}, {Dumas}, {Eisenhauer}, {Elswijk}, {Fedrigo}, {Finger},
  {Gemperlein}, {Genzel}, {Gilbert}, {Gillet}, {Goldbrunner}, {Horrobin}, {Ter
  Horst}, {Huber}, {Hubin}, {Iserlohe}, {Kaufer}, {Kissler-Patig}, {Kragt},
  {Kroes}, {Lehnert}, {Lieb}, {Liske}, {Lizon}, {Lutz}, {Modigliani}, {Monnet},
  {Nesvadba}, {Patig}, {Pragt}, {Reunanen}, {R{\"o}hrle}, {Rossi}, {Schmutzer},
  {Schoenmaker}, {Schreiber}, {Stroebele}, {Szeifert}, {Tacconi}, {Tecza},
  {Thatte}, {Tordo}, {van der Werf}, \& {Weisz}}]{bonnet04}
{Bonnet}, H., {Abuter}, R., {Baker}, A., {et~al.} 2004, The Messenger, 117, 17

\bibitem[{{Boselli} {et~al.}(2008){Boselli}, {Boissier}, {Cortese}, \&
  {Gavazzi}}]{boselli08}
{Boselli}, A., {Boissier}, S., {Cortese}, L., \& {Gavazzi}, G. 2008, \apj, 674,
  742

\bibitem[{{Bournaud} {et~al.}(2008){Bournaud}, {Daddi}, {Elmegreen},
  {Elmegreen}, {Nesvadba}, {Vanzella}, {Di Matteo}, {Le Tiran}, {Lehnert}, \&
  {Elbaz}}]{bournaud08}
{Bournaud}, F., {Daddi}, E., {Elmegreen}, B.~G., {et~al.} 2008, ArXiv e-prints,
  803

\bibitem[{{Bower} {et~al.}(2006){Bower}, {Benson}, {Malbon}, {Helly}, {Frenk},
  {Baugh}, {Cole}, \& {Lacey}}]{bower06}
{Bower}, R.~G., {Benson}, A.~J., {Malbon}, R., {et~al.} 2006, \mnras, 370, 645

\bibitem[{{Capetti} {et~al.}(1999){Capetti}, {Axon}, {Macchetto}, {Marconi}, \&
  {Winge}}]{capetti99}
{Capetti}, A., {Axon}, D.~J., {Macchetto}, F.~D., {Marconi}, A., \& {Winge}, C.
  1999, \apj, 516, 187

\bibitem[{{Carilli} {et~al.}(2002){Carilli}, {Harris}, {Pentericci},
  {R{\"o}ttgering}, {Miley}, {Kurk}, \& {van Breugel}}]{carilli02}
{Carilli}, C.~L., {Harris}, D.~E., {Pentericci}, L., {et~al.} 2002, \apj, 567,
  781

\bibitem[{{Carilli} {et~al.}(1997){Carilli}, {Roettgering}, {van Ojik},
  {Miley}, \& {van Breugel}}]{carilli97}
{Carilli}, C.~L., {Roettgering}, H.~J.~A., {van Ojik}, R., {Miley}, G.~K., \&
  {van Breugel}, W.~J.~M. 1997, \apjs, 109, 1

\bibitem[{{Chakrabarti} {et~al.}(2007){Chakrabarti}, {Cox}, {Hernquist},
  {Hopkins}, {Robertson}, \& {Di Matteo}}]{chakrabarti07}
{Chakrabarti}, S., {Cox}, T.~J., {Hernquist}, L., {et~al.} 2007, \apj, 658, 840

\bibitem[{{Ciotti} \& {Ostriker}(2007)}]{ciotti07}
{Ciotti}, L. \& {Ostriker}, J.~P. 2007, \apj, 665, 1038

\bibitem[{{Clark} {et~al.}(1998){Clark}, {Axon}, {Tadhunter}, {Robinson}, \&
  {O'Brien}}]{clark98}
{Clark}, N.~E., {Axon}, D.~J., {Tadhunter}, C.~N., {Robinson}, A., \&
  {O'Brien}, P. 1998, \apj, 494, 546

\bibitem[{{Croton} {et~al.}(2006){Croton}, {Springel}, {White}, {De Lucia},
  {Frenk}, {Gao}, {Jenkins}, {Kauffmann}, {Navarro}, \& {Yoshida}}]{croton06}
{Croton}, D.~J., {Springel}, V., {White}, S.~D.~M., {et~al.} 2006, \mnras, 365,
  11

\bibitem[{{De Breuck} {et~al.}(2005){De Breuck}, {Downes}, {Neri}, {van
  Breugel}, {Reuland}, {Omont}, \& {Ivison}}]{debreuck05}
{De Breuck}, C., {Downes}, D., {Neri}, R., {et~al.} 2005, \aap, 430, L1

\bibitem[{{De Breuck} {et~al.}(2003){De Breuck}, {Neri}, \&
  {Omont}}]{debreuck03}
{De Breuck}, C., {Neri}, R., \& {Omont}, A. 2003, New Astronomy Review, 47, 285

\bibitem[{{De Breuck} {et~al.}(2000){De Breuck}, {R{\"o}ttgering}, {Miley},
  {van Breugel}, \& {Best}}]{debreuck00}
{De Breuck}, C., {R{\"o}ttgering}, H., {Miley}, G., {van Breugel}, W., \&
  {Best}, P. 2000, \aap, 362, 519

\bibitem[{{De Breuck} {et~al.}(2001){De Breuck}, {van Breugel},
  {R{\"o}ttgering}, {Stern}, {Miley}, {de Vries}, {Stanford}, {Kurk}, \&
  {Overzier}}]{debreuck01}
{De Breuck}, C., {van Breugel}, W., {R{\"o}ttgering}, H., {et~al.} 2001, \aj,
  121, 1241

\bibitem[{{De Lucia} {et~al.}(2006){De Lucia}, {Springel}, {White}, {Croton},
  \& {Kauffmann}}]{delucia06}
{De Lucia}, G., {Springel}, V., {White}, S.~D.~M., {Croton}, D., \&
  {Kauffmann}, G. 2006, \mnras, 366, 499

\bibitem[{{De Young}(1991)}]{young91}
{De Young}, D.~S. 1991, \apj, 371, 69

\bibitem[{{Di Matteo} {et~al.}(2005){Di Matteo}, {Springel}, \&
  {Hernquist}}]{dimatteo05}
{Di Matteo}, T., {Springel}, V., \& {Hernquist}, L. 2005, \nat, 433, 604

\bibitem[{{Dopita} \& {Sutherland}(1995)}]{dopita95}
{Dopita}, M.~A. \& {Sutherland}, R.~S. 1995, \apj, 455, 468

\bibitem[{{Dopita} \& {Sutherland}(1996)}]{dopita96}
{Dopita}, M.~A. \& {Sutherland}, R.~S. 1996, \apjs, 102, 161

\bibitem[{{Dyson} \& {Williams}(1980)}]{dyson80}
{Dyson}, J.~E. \& {Williams}, D.~A. 1980, {Physics of the interstellar medium}
  (New York, Halsted Press, 1980.~204 p.)

\bibitem[{{Erb} {et~al.}(2006){Erb}, {Steidel}, {Shapley}, {Pettini}, {Reddy},
  \& {Adelberger}}]{erb06}
{Erb}, D.~K., {Steidel}, C.~C., {Shapley}, A.~E., {et~al.} 2006, \apj, 647, 128

\bibitem[{{Evans}(1998)}]{evans98}
{Evans}, A.~S. 1998, \apj, 498, 553

\bibitem[{{Ferrarese}(2002)}]{ferrarese02}
{Ferrarese}, L. 2002, \apj, 578, 90

\bibitem[{{F{\"o}rster Schreiber} {et~al.}(2006){F{\"o}rster Schreiber},
  {Genzel}, {Lehnert}, {Bouch{\'e}}, {Verma}, {Erb}, {Shapley}, {Steidel},
  {Davies}, {Lutz}, {Nesvadba}, {Tacconi}, {Eisenhauer}, {Abuter}, {Gilbert},
  {Gillessen}, \& {Sternberg}}]{nmfs06}
{F{\"o}rster Schreiber}, N.~M., {Genzel}, R., {Lehnert}, M.~D., {et~al.} 2006,
  \apj, 645, 1062

\bibitem[{{Fragile} {et~al.}(2004){Fragile}, {Murray}, {Anninos}, \& {van
  Breugel}}]{fragile04}
{Fragile}, P.~C., {Murray}, S.~D., {Anninos}, P., \& {van Breugel}, W. 2004,
  \apj, 604, 74

\bibitem[{{Fujita}(2008)}]{fujita08}
{Fujita}, Y. 2008, \mnras, 384, L41

\bibitem[{{Gaibler} {et~al.}(2007){Gaibler}, {Camenzind}, \&
  {Krause}}]{gaibler07}
{Gaibler}, V., {Camenzind}, M., \& {Krause}, M. 2007, ArXiv e-prints, 710

\bibitem[{{Garrington} {et~al.}(1988){Garrington}, {Leahy}, {Conway}, \&
  {Laing}}]{garrington88}
{Garrington}, S.~T., {Leahy}, J.~P., {Conway}, R.~G., \& {Laing}, R.~A. 1988,
  \nat, 331, 147

\bibitem[{{Genzel} {et~al.}(2006){Genzel}, {Tacconi}, {Eisenhauer},
  {F{\"o}rster Schreiber}, {Cimatti}, {Daddi}, {Bouch{\'e}}, {Davies},
  {Lehnert}, {Lutz}, {Nesvadba}, {Verma}, {Abuter}, {Shapiro}, {Sternberg},
  {Renzini}, {Kong}, {Arimoto}, \& {Mignoli}}]{genzel06}
{Genzel}, R., {Tacconi}, L.~J., {Eisenhauer}, F., {et~al.} 2006, \nat, 442, 786

\bibitem[{{Greve} {et~al.}(2005){Greve}, {Bertoldi}, {Smail}, {Neri},
  {Chapman}, {Blain}, {Ivison}, {Genzel}, {Omont}, {Cox}, {Tacconi}, \&
  {Kneib}}]{greve05}
{Greve}, T.~R., {Bertoldi}, F., {Smail}, I., {et~al.} 2005, \mnras, 359, 1165

\bibitem[{{H{\"a}ring} \& {Rix}(2004)}]{haring04}
{H{\"a}ring}, N. \& {Rix}, H.-W. 2004, \apjl, 604, L89

\bibitem[{{Heckman} {et~al.}(1991{\natexlab{a}}){Heckman}, {Lehnert}, {Miley},
  \& {van Breugel}}]{heckman91b}
{Heckman}, T.~M., {Lehnert}, M.~D., {Miley}, G.~K., \& {van Breugel}, W.
  1991{\natexlab{a}}, \apj, 381, 373

\bibitem[{{Heckman} {et~al.}(1991{\natexlab{b}}){Heckman}, {Miley}, {Lehnert},
  \& {van Breugel}}]{heckman91a}
{Heckman}, T.~M., {Miley}, G.~K., {Lehnert}, M.~D., \& {van Breugel}, W.
  1991{\natexlab{b}}, \apj, 370, 78

\bibitem[{{Hopkins} {et~al.}(2006){Hopkins}, {Hernquist}, {Cox}, {Di Matteo},
  {Robertson}, \& {Springel}}]{hopkins06}
{Hopkins}, P.~F., {Hernquist}, L., {Cox}, T.~J., {et~al.} 2006, \apjs, 163, 1

\bibitem[{{Humphrey} {et~al.}(2008){Humphrey}, {Villar-Mart{\'{\i}}n},
  {Vernet}, {Fosbury}, {di Serego Alighieri}, \& {Binette}}]{humphrey08}
{Humphrey}, A., {Villar-Mart{\'{\i}}n}, M., {Vernet}, J., {et~al.} 2008,
  \mnras, 383, 11

\bibitem[{{Inskip} {et~al.}(2002){Inskip}, {Best}, {Rawlings}, {Longair},
  {Cotter}, {R{\"o}ttgering}, \& {Eales}}]{inskip02}
{Inskip}, K.~J., {Best}, P.~N., {Rawlings}, S., {et~al.} 2002, \mnras, 337,
  1381

\bibitem[{{Irwin} {et~al.}(1987){Irwin}, {Seaquist}, {Taylor}, \&
  {Duric}}]{irwin87}
{Irwin}, J.~A., {Seaquist}, E.~R., {Taylor}, A.~R., \& {Duric}, N. 1987, \apjl,
  313, L91

\bibitem[{{Iwamuro} {et~al.}(2003){Iwamuro}, {Motohara}, {Maihara}, {Kimura},
  {Eto}, {Shima}, {Mochida}, {Wada}, {Imai}, \& {Aoki}}]{iwamuro03}
{Iwamuro}, F., {Motohara}, K., {Maihara}, T., {et~al.} 2003, \apj, 598, 178

\bibitem[{{Kaiser} \& {Alexander}(1997)}]{kaiser97}
{Kaiser}, C.~R. \& {Alexander}, P. 1997, \mnras, 286, 215

\bibitem[{{Kaiser} \& {Best}(2007)}]{kaiser07}
{Kaiser}, C.~R. \& {Best}, P.~N. 2007, \mnras, 381, 1548

\bibitem[{{Kaiser} \& {Binney}(2003)}]{kaiser03}
{Kaiser}, C.~R. \& {Binney}, J. 2003, \mnras, 338, 837

\bibitem[{{Kaiser} {et~al.}(1997{\natexlab{a}}){Kaiser}, {Dennett-Thorpe}, \&
  {Alexander}}]{kaiser97a}
{Kaiser}, C.~R., {Dennett-Thorpe}, J., \& {Alexander}, P. 1997{\natexlab{a}},
  \mnras, 292, 723

\bibitem[{{Kaiser} {et~al.}(1997{\natexlab{b}}){Kaiser}, {Dennett-Thorpe}, \&
  {Alexander}}]{kaiser97b}
{Kaiser}, C.~R., {Dennett-Thorpe}, J., \& {Alexander}, P. 1997{\natexlab{b}},
  \mnras, 292, 723

\bibitem[{{King}(2003)}]{king03}
{King}, A. 2003, \apjl, 596, L27

\bibitem[{{Klamer} {et~al.}(2005){Klamer}, {Ekers}, {Sadler}, {Weiss},
  {Hunstead}, \& {De Breuck}}]{klamer05}
{Klamer}, I.~J., {Ekers}, R.~D., {Sadler}, E.~M., {et~al.} 2005, \apjl, 621, L1

\bibitem[{{Kodama} {et~al.}(2007){Kodama}, {Tanaka}, {Kajisawa}, {Kurk},
  {Venemans}, {De Breuck}, {Vernet}, \& {Lidman}}]{kodama07}
{Kodama}, T., {Tanaka}, I., {Kajisawa}, M., {et~al.} 2007, \mnras, 377, 1717

\bibitem[{{Krause}(2005)}]{krause05}
{Krause}, M. 2005, \aap, 431, 45

\bibitem[{{Krause} \& {Alexander}(2007)}]{krause07}
{Krause}, M. \& {Alexander}, P. 2007, \mnras, 376, 465

\bibitem[{{Kronberger} {et~al.}(2007){Kronberger}, {Kapferer}, {Schindler}, \&
  {Ziegler}}]{kronberger07}
{Kronberger}, T., {Kapferer}, W., {Schindler}, S., \& {Ziegler}, B.~L. 2007,
  \aap, 473, 761

\bibitem[{{Kurk} {et~al.}(2004){Kurk}, {Pentericci}, {Overzier},
  {R{\"o}ttgering}, \& {Miley}}]{kurk04}
{Kurk}, J.~D., {Pentericci}, L., {Overzier}, R.~A., {R{\"o}ttgering}, H.~J.~A.,
  \& {Miley}, G.~K. 2004, \aap, 428, 817

\bibitem[{{Laing}(1988)}]{laing88}
{Laing}, R.~A. 1988, \nat, 331, 149

\bibitem[{{Law} {et~al.}(2007){Law}, {Steidel}, {Erb}, {Larkin}, {Pettini},
  {Shapley}, \& {Wright}}]{law07}
{Law}, D.~R., {Steidel}, C.~C., {Erb}, D.~K., {et~al.} 2007, \apj, 669, 929

\bibitem[{{Le Fevre} {et~al.}(1996){Le Fevre}, {Deltorn}, {Crampton}, \&
  {Dickinson}}]{lefevre96}
{Le Fevre}, O., {Deltorn}, J.~M., {Crampton}, D., \& {Dickinson}, M. 1996,
  \apjl, 471, L11+

\bibitem[{{Magorrian} {et~al.}(1998){Magorrian}, {Tremaine}, {Richstone},
  {Bender}, {Bower}, {Dressler}, {Faber}, {Gebhardt}, {Green}, {Grillmair},
  {Kormendy}, \& {Lauer}}]{magorrian98}
{Magorrian}, J., {Tremaine}, S., {Richstone}, D., {et~al.} 1998, \aj, 115, 2285

\bibitem[{{McCarthy} {et~al.}(2008){McCarthy}, {Babul}, {Bower}, \&
  {Balogh}}]{mccarthy08}
{McCarthy}, I.~G., {Babul}, A., {Bower}, R.~G., \& {Balogh}, M.~L. 2008,
  \mnras, 403

\bibitem[{{McCarthy} {et~al.}(1996){McCarthy}, {Baum}, \&
  {Spinrad}}]{mccarthy96}
{McCarthy}, P.~J., {Baum}, S.~A., \& {Spinrad}, H. 1996, \apjs, 106, 281

\bibitem[{{McCarthy} {et~al.}(1992){McCarthy}, {Elston}, \&
  {Eisenhardt}}]{mccarthy92}
{McCarthy}, P.~J., {Elston}, R., \& {Eisenhardt}, P. 1992, \apjl, 387, L29

\bibitem[{{McCarthy} {et~al.}(1991){McCarthy}, {van Breughel}, {Kapahi}, \&
  {Subrahmanya}}]{mccarthy91}
{McCarthy}, P.~J., {van Breughel}, W., {Kapahi}, V.~K., \& {Subrahmanya}, C.~R.
  1991, \aj, 102, 522

\bibitem[{{McNamara} \& {Nulsen}(2007)}]{mcnamara07}
{McNamara}, B.~R. \& {Nulsen}, P.~E.~J. 2007, \araa, 45, 117

\bibitem[{{Mellema} {et~al.}(2002){Mellema}, {Kurk}, \&
  {R{\"o}ttgering}}]{mellema02}
{Mellema}, G., {Kurk}, J.~D., \& {R{\"o}ttgering}, H.~J.~A. 2002, \aap, 395,
  L13

\bibitem[{{Miley} {et~al.}(2006){Miley}, {Overzier}, {Zirm}, {Ford}, {Kurk},
  {Pentericci}, {Blakeslee}, {Franx}, {Illingworth}, {Postman}, {Rosati},
  {R{\"o}ttgering}, {Venemans}, \& {Helder}}]{miley06}
{Miley}, G.~K., {Overzier}, R.~A., {Zirm}, A.~W., {et~al.} 2006, \apjl, 650,
  L29

\bibitem[{{Morganti} {et~al.}(2005){Morganti}, {Tadhunter}, \&
  {Oosterloo}}]{morganti05}
{Morganti}, R., {Tadhunter}, C.~N., \& {Oosterloo}, T.~A. 2005, \aap, 444, L9

\bibitem[{{Murgia} {et~al.}(1999){Murgia}, {Fanti}, {Fanti}, {Gregorini},
  {Klein}, {Mack}, \& {Vigotti}}]{murgia99}
{Murgia}, M., {Fanti}, C., {Fanti}, R., {et~al.} 1999, \aap, 345, 769

\bibitem[{{Murray} {et~al.}(2005){Murray}, {Quataert}, \&
  {Thompson}}]{murray05}
{Murray}, N., {Quataert}, E., \& {Thompson}, T.~A. 2005, \apj, 618, 569

\bibitem[{{Nath} \& {Roychowdhury}(2002)}]{nath02}
{Nath}, B.~B. \& {Roychowdhury}, S. 2002, \mnras, 333, 145

\bibitem[{{Neri} {et~al.}(2003){Neri}, {Genzel}, {Ivison}, {Bertoldi}, {Blain},
  {Chapman}, {Cox}, {Greve}, {Omont}, \& {Frayer}}]{neri03}
{Neri}, R., {Genzel}, R., {Ivison}, R.~J., {et~al.} 2003, \apjl, 597, L113

\bibitem[{{Nesvadba} {et~al.}(2008){Nesvadba}, {Lehnert}, {Davies}, {Verma}, \&
  {Eisenhauer}}]{nesvadba08}
{Nesvadba}, N.~P.~H., {Lehnert}, M.~D., {Davies}, R.~I., {Verma}, A., \&
  {Eisenhauer}, F. 2008, \aap, 479, 67

\bibitem[{{Nesvadba} {et~al.}(2007{\natexlab{a}}){Nesvadba}, {Lehnert}, {De
  Breuck}, {Gilbert}, \& {van Breugel}}]{nesvadba07b}
{Nesvadba}, N.~P.~H., {Lehnert}, M.~D., {De Breuck}, C., {Gilbert}, A., \& {van
  Breugel}, W. 2007{\natexlab{a}}, \aap, 475, 145

\bibitem[{{Nesvadba} {et~al.}(2006{\natexlab{a}}){Nesvadba}, {Lehnert},
  {Eisenhauer}, {Genzel}, {Seitz}, {Davies}, {Saglia}, {Lutz}, {Tacconi},
  {Bender}, \& {Abuter}}]{nesvadba06b}
{Nesvadba}, N.~P.~H., {Lehnert}, M.~D., {Eisenhauer}, F., {et~al.}
  2006{\natexlab{a}}, \apj, 650, 661

\bibitem[{{Nesvadba} {et~al.}(2006{\natexlab{b}}){Nesvadba}, {Lehnert},
  {Eisenhauer}, {Gilbert}, {Tecza}, \& {Abuter}}]{nesvadba06}
{Nesvadba}, N.~P.~H., {Lehnert}, M.~D., {Eisenhauer}, F., {et~al.}
  2006{\natexlab{b}}, \apj, 650, 693

\bibitem[{{Nesvadba} {et~al.}(2007{\natexlab{b}}){Nesvadba}, {Lehnert},
  {Genzel}, {Eisenhauer}, {Baker}, {Seitz}, {Davies}, {Lutz}, {Tacconi},
  {Tecza}, {Bender}, \& {Abuter}}]{nesvadba07a}
{Nesvadba}, N.~P.~H., {Lehnert}, M.~D., {Genzel}, R., {et~al.}
  2007{\natexlab{b}}, \apj, 657, 725

\bibitem[{{Overzier} {et~al.}(2005){Overzier}, {Harris}, {Carilli},
  {Pentericci}, {R{\"o}ttgering}, \& {Miley}}]{overzier05}
{Overzier}, R.~A., {Harris}, D.~E., {Carilli}, C.~L., {et~al.} 2005, \aap, 433,
  87

\bibitem[{{Owsianik} \& {Conway}(1998)}]{owsianik98}
{Owsianik}, I. \& {Conway}, J.~E. 1998, \aap, 337, 69

\bibitem[{{Papadopoulos} {et~al.}(2000){Papadopoulos}, {R{\"o}ttgering}, {van
  der Werf}, {Guilloteau}, {Omont}, {van Breugel}, \&
  {Tilanus}}]{papadopoulos00}
{Papadopoulos}, P.~P., {R{\"o}ttgering}, H.~J.~A., {van der Werf}, P.~P.,
  {et~al.} 2000, \apj, 528, 626

\bibitem[{{Pentericci} {et~al.}(2001){Pentericci}, {McCarthy},
  {R{\"o}ttgering}, {Miley}, {van Breugel}, \& {Fosbury}}]{pentericci01}
{Pentericci}, L., {McCarthy}, P.~J., {R{\"o}ttgering}, H.~J.~A., {et~al.} 2001,
  \apjs, 135, 63

\bibitem[{{Pentericci} {et~al.}(2000){Pentericci}, {Van Reeven}, {Carilli},
  {R{\"o}ttgering}, \& {Miley}}]{pentericci00}
{Pentericci}, L., {Van Reeven}, W., {Carilli}, C.~L., {R{\"o}ttgering},
  H.~J.~A., \& {Miley}, G.~K. 2000, \aaps, 145, 121

\bibitem[{{Pipino} \& {Matteucci}(2004)}]{pipino04}
{Pipino}, A. \& {Matteucci}, F. 2004, \mnras, 347, 968

\bibitem[{{Pipino} {et~al.}(2007){Pipino}, {Matteucci}, \&
  {D'Ercole}}]{pipino07}
{Pipino}, A., {Matteucci}, F., \& {D'Ercole}, A. 2007, ArXiv e-prints, 709

\bibitem[{{Rafferty} {et~al.}(2006){Rafferty}, {McNamara}, {Nulsen}, \&
  {Wise}}]{rafferty06}
{Rafferty}, D.~A., {McNamara}, B.~R., {Nulsen}, P.~E.~J., \& {Wise}, M.~W.
  2006, \apj, 652, 216

\bibitem[{{Reuland} {et~al.}(2004){Reuland}, {R{\"o}ttgering}, {van Breugel},
  \& {De Breuck}}]{reuland04}
{Reuland}, M., {R{\"o}ttgering}, H., {van Breugel}, W., \& {De Breuck}, C.
  2004, \mnras, 353, 377

\bibitem[{{Rocca-Volmerange} {et~al.}(2004){Rocca-Volmerange}, {Le Borgne}, {De
  Breuck}, {Fioc}, \& {Moy}}]{rocca04}
{Rocca-Volmerange}, B., {Le Borgne}, D., {De Breuck}, C., {Fioc}, M., \& {Moy},
  E. 2004, \aap, 415, 931

\bibitem[{{Rudnick} {et~al.}(2003){Rudnick}, {Rix}, {Franx}, {Labb{\'e}},
  {Blanton}, {Daddi}, {F{\"o}rster Schreiber}, {Moorwood}, {R{\"o}ttgering},
  {Trujillo}, {van de Wel}, {van der Werf}, {van Dokkum}, \& {van
  Starkenburg}}]{rudnick03}
{Rudnick}, G., {Rix}, H.-W., {Franx}, M., {et~al.} 2003, \apj, 599, 847

\bibitem[{{Rush} {et~al.}(1997){Rush}, {McCarthy}, {Athreya}, \&
  {Persson}}]{rush97}
{Rush}, B., {McCarthy}, P.~J., {Athreya}, R.~M., \& {Persson}, S.~E. 1997,
  \apj, 484, 163

\bibitem[{{Ruszkowski} \& {Begelman}(2002)}]{ruszkowski02}
{Ruszkowski}, M. \& {Begelman}, M.~C. 2002, \apj, 581, 223

\bibitem[{{Saxton} {et~al.}(2005){Saxton}, {Bicknell}, {Sutherland}, \&
  {Midgley}}]{saxton05}
{Saxton}, C.~J., {Bicknell}, G.~V., {Sutherland}, R.~S., \& {Midgley}, S. 2005,
  \mnras, 359, 781

\bibitem[{{Sazonov} {et~al.}(2005){Sazonov}, {Ostriker}, {Ciotti}, \&
  {Sunyaev}}]{sazonov05}
{Sazonov}, S.~Y., {Ostriker}, J.~P., {Ciotti}, L., \& {Sunyaev}, R.~A. 2005,
  \mnras, 358, 168

\bibitem[{{Scannapieco} \& {Oh}(2004)}]{scannapieco04}
{Scannapieco}, E. \& {Oh}, S.~P. 2004, \apj, 608, 62

\bibitem[{{Schombert}(1987)}]{schombert87}
{Schombert}, J.~M. 1987, \apjs, 64, 643

\bibitem[{{Seymour} {et~al.}(2008){Seymour}, {Ogle}, {De Breuck}, {Fazio},
  {Galametz}, {Haas}, {Lacy}, {Sajina}, {Stern}, {Willner}, \&
  {Vernet}}]{seymour08}
{Seymour}, N., {Ogle}, P., {De Breuck}, C., {et~al.} 2008, \apjl, 681, L1

\bibitem[{{Seymour} {et~al.}(2007){Seymour}, {Stern}, {De Breuck}, {Vernet},
  {Rettura}, {Dickinson}, {Dey}, {Eisenhardt}, {Fosbury}, {Lacy}, {McCarthy},
  {Miley}, {Rocca-Volmerange}, {R{\"o}ttgering}, {Stanford}, {Teplitz}, {van
  Breugel}, \& {Zirm}}]{seymour07}
{Seymour}, N., {Stern}, D., {De Breuck}, C., {et~al.} 2007, \apjs, 171, 353

\bibitem[{{Shankar} {et~al.}(2008){Shankar}, {Cavaliere}, {Cirasuolo}, \&
  {Maraschi}}]{shankar08}
{Shankar}, F., {Cavaliere}, A., {Cirasuolo}, M., \& {Maraschi}, L. 2008, \apj,
  676, 131

\bibitem[{{Silk} \& {Rees}(1998)}]{silk98}
{Silk}, J. \& {Rees}, M.~J. 1998, \aap, 331, L1

\bibitem[{{Smail} {et~al.}(2002){Smail}, {Ivison}, {Blain}, \&
  {Kneib}}]{smail02}
{Smail}, I., {Ivison}, R.~J., {Blain}, A.~W., \& {Kneib}, J.-P. 2002, \mnras,
  331, 495

\bibitem[{{Springel} {et~al.}(2005){Springel}, {Di Matteo}, \&
  {Hernquist}}]{springel05}
{Springel}, V., {Di Matteo}, T., \& {Hernquist}, L. 2005, \mnras, 361, 776

\bibitem[{{Steidel} {et~al.}(1996){Steidel}, {Giavalisco}, {Pettini},
  {Dickinson}, \& {Adelberger}}]{steidel96}
{Steidel}, C.~C., {Giavalisco}, M., {Pettini}, M., {Dickinson}, M., \&
  {Adelberger}, K.~L. 1996, \apjl, 462, L17+

\bibitem[{{Sutherland} \& {Bicknell}(2007)}]{sutherland07}
{Sutherland}, R.~S. \& {Bicknell}, G.~V. 2007, \apjs, 173, 37

\bibitem[{{Swinbank} {et~al.}(2006){Swinbank}, {Chapman}, {Smail}, {Lindner},
  {Borys}, {Blain}, {Ivison}, \& {Lewis}}]{swinbank06}
{Swinbank}, A.~M., {Chapman}, S.~C., {Smail}, I., {et~al.} 2006, \mnras, 371,
  465

\bibitem[{{Tadhunter}(1991)}]{tadhunter91}
{Tadhunter}, C.~N. 1991, \mnras, 251, 46P

\bibitem[{{Tody}(1993)}]{tody93}
{Tody}, D. 1993, in Astronomical Society of the Pacific Conference Series,
  Vol.~52, Astronomical Data Analysis Software and Systems II, ed. R.~J.
  {Hanisch}, R.~J.~V. {Brissenden}, \& J.~{Barnes}, 173--+

\bibitem[{{Tremaine} {et~al.}(2002){Tremaine}, {Gebhardt}, {Bender}, {Bower},
  {Dressler}, {Faber}, {Filippenko}, {Green}, {Grillmair}, {Ho}, {Kormendy},
  {Lauer}, {Magorrian}, {Pinkney}, \& {Richstone}}]{tremaine02}
{Tremaine}, S., {Gebhardt}, K., {Bender}, R., {et~al.} 2002, \apj, 574, 740

\bibitem[{{Tremonti} {et~al.}(2004){Tremonti}, {Heckman}, {Kauffmann},
  {Brinchmann}, {Charlot}, {White}, {Seibert}, {Peng}, {Schlegel}, {Uomoto},
  {Fukugita}, \& {Brinkmann}}]{tremonti04}
{Tremonti}, C.~A., {Heckman}, T.~M., {Kauffmann}, G., {et~al.} 2004, \apj, 613,
  898

\bibitem[{{van Breugel} {et~al.}(1998){van Breugel}, {Stanford}, {Spinrad},
  {Stern}, \& {Graham}}]{vanbreugel98}
{van Breugel}, W.~J.~M., {Stanford}, S.~A., {Spinrad}, H., {Stern}, D., \&
  {Graham}, J.~R. 1998, \apj, 502, 614

\bibitem[{{Venemans} {et~al.}(2005){Venemans}, {R{\"o}ttgering}, {Miley},
  {Kurk}, {De Breuck}, {Overzier}, {van Breugel}, {Carilli}, {Ford}, {Heckman},
  {Pentericci}, \& {McCarthy}}]{venemans05}
{Venemans}, B.~P., {R{\"o}ttgering}, H.~J.~A., {Miley}, G.~K., {et~al.} 2005,
  \aap, 431, 793

\bibitem[{{Venemans} {et~al.}(2007){Venemans}, {R{\"o}ttgering}, {Miley}, {van
  Breugel}, {de Breuck}, {Kurk}, {Pentericci}, {Stanford}, {Overzier}, {Croft},
  \& {Ford}}]{venemans07}
{Venemans}, B.~P., {R{\"o}ttgering}, H.~J.~A., {Miley}, G.~K., {et~al.} 2007,
  \aap, 461, 823

\bibitem[{{Villar-Mart{\'{\i}}n} {et~al.}(2006){Villar-Mart{\'{\i}}n},
  {S{\'a}nchez}, {De Breuck}, {Peletier}, {Vernet}, {Rettura}, {Seymour},
  {Humphrey}, {Stern}, {di Serego Alighieri}, \& {Fosbury}}]{villar06}
{Villar-Mart{\'{\i}}n}, M., {S{\'a}nchez}, S.~F., {De Breuck}, C., {et~al.}
  2006, \mnras, 366, L1

\bibitem[{{Villar-Mart{\'{\i}}n} {et~al.}(2007){Villar-Mart{\'{\i}}n},
  {S{\'a}nchez}, {Humphrey}, {Dijkstra}, {di Serego Alighieri}, {De Breuck}, \&
  {Gonz{\'a}lez Delgado}}]{villar07}
{Villar-Mart{\'{\i}}n}, M., {S{\'a}nchez}, S.~F., {Humphrey}, A., {et~al.}
  2007, \mnras, 378, 416

\bibitem[{{Villar-Martin} {et~al.}(1997){Villar-Martin}, {Tadhunter}, \&
  {Clark}}]{villar97}
{Villar-Martin}, M., {Tadhunter}, C., \& {Clark}, N. 1997, \aap, 323, 21

\bibitem[{{Villar-Mart{\'{\i}}n} {et~al.}(1999){Villar-Mart{\'{\i}}n},
  {Tadhunter}, {Morganti}, {Axon}, \& {Koekemoer}}]{villar99}
{Villar-Mart{\'{\i}}n}, M., {Tadhunter}, C., {Morganti}, R., {Axon}, D., \&
  {Koekemoer}, A. 1999, \mnras, 307, 24

\bibitem[{{Villar-Mart{\'{\i}}n} {et~al.}(2003){Villar-Mart{\'{\i}}n},
  {Vernet}, {di Serego Alighieri}, {Fosbury}, {Humphrey}, \&
  {Pentericci}}]{villar03}
{Villar-Mart{\'{\i}}n}, M., {Vernet}, J., {di Serego Alighieri}, S., {et~al.}
  2003, \mnras, 346, 273

\bibitem[{{Villar-Mart{\'{\i}}n} {et~al.}(2002){Villar-Mart{\'{\i}}n},
  {Vernet}, {di Serego Alighieri}, {Fosbury}, {Pentericci}, {Cohen},
  {Goodrich}, \& {Humphrey}}]{villar02}
{Villar-Mart{\'{\i}}n}, M., {Vernet}, J., {di Serego Alighieri}, S., {et~al.}
  2002, \mnras, 336, 436

\bibitem[{{Wan} {et~al.}(2000){Wan}, {Daly}, \& {Guerra}}]{wan00}
{Wan}, L., {Daly}, R.~A., \& {Guerra}, E.~J. 2000, \apj, 544, 671

\bibitem[{{Willott} {et~al.}(1999){Willott}, {Rawlings}, {Blundell}, \&
  {Lacy}}]{willott99}
{Willott}, C.~J., {Rawlings}, S., {Blundell}, K.~M., \& {Lacy}, M. 1999,
  \mnras, 309, 1017

\bibitem[{{Willott} {et~al.}(2001){Willott}, {Rawlings}, {Blundell}, {Lacy}, \&
  {Eales}}]{willott01}
{Willott}, C.~J., {Rawlings}, S., {Blundell}, K.~M., {Lacy}, M., \& {Eales},
  S.~A. 2001, \mnras, 322, 536

\bibitem[{{Willott} {et~al.}(2003){Willott}, {Rawlings}, {Jarvis}, \&
  {Blundell}}]{willott03}
{Willott}, C.~J., {Rawlings}, S., {Jarvis}, M.~J., \& {Blundell}, K.~M. 2003,
  \mnras, 339, 173

\bibitem[{{Wright} {et~al.}(2007){Wright}, {Larkin}, {Barczys}, {Erb},
  {Iserlohe}, {Krabbe}, {Law}, {McElwain}, {Quirrenbach}, {Steidel}, \&
  {Weiss}}]{wright07}
{Wright}, S.~A., {Larkin}, J.~E., {Barczys}, M., {et~al.} 2007, \apj, 658, 78

\bibitem[{{Yu} \& {Tremaine}(2002)}]{yu02}
{Yu}, Q. \& {Tremaine}, S. 2002, \mnras, 335, 965

\end{thebibliography}

\clearpage
\onecolumn

\clearpage
\onecolumn

\clearpage
\onecolumn

\clearpage
\onecolumn

\begin{table*}
\caption{Sample and observation log}
\label{tab:exposure}
\begin{tabular}{lcccccc}
\hline \hline
Source & $z$ & filter & ToT & seeing  \\
(1)  & (2)         & (3)            & (4)& (5) \\
\hline
0316-257  & 3.13 & K & 100  & 0.7\arcsec$\times$0.4\arcsec \\
\hline
0406-244  & 2.44  & H & 90   & 0.7$\arcsec$$\times$0.6\arcsec \\
0406-244  &      & K & 100  & 0.8\arcsec$\times$0.6\arcsec \\
\hline
0828+193  & 2.57 & H & 135  & 0.7\arcsec $\times$0.5\arcsec \\
0828+193  &      & K & 170  & 0.7\arcsec $\times$0.5\arcsec \\

\hline
\end{tabular}\\
Column (1) -- Source ID. Column (2) -- redshift. Column (3)
-- Total exposure time (on target) in min. Column (4) -- Seeing.  For
TXS0828+193, the seeing was determined from a star within the field of
view. Otherwise, we estimated the seeing from the standard star.
\end{table*}

\begin{table*}
\caption{Emission Lines in MRC0316-257}
\label{tab:emlines0316}
\begin{tabular}{lccccc}
\hline \hline
Line & $\lambda_0$ & $\lambda_{obs}$ & z & FWHM & flux \\
(1)  & (2)         & (3)            & (4)& (5) & (6)  \\
\hline
$[$OIII$]$  & 5007 & 20665.4 $\pm$ 4.2 & 3.1273$\pm$ 0.0008 & 1465$\pm$ 90 & 2.2$\pm$0.1 \\
$[$OIII$]$  & 4959 & 20459.6 $\pm$ 4.3 & 3.1257$\pm$ 0.0009 & 1357$\pm$ 94 & 0.9$\pm$0.1  \\
H$\beta$ & 4861 & 20072.2 $\pm$ 8.1 & 3.1292$\pm$ 0.0017 & 902$\pm$ 266 & 1.2$\pm$0.4  \\ 
\hline
\end{tabular}\\
Column (1) -- line. Column (2) -- Rest-frame
  wavelength in \AA. Column (3) -- Observed wavelength in \AA. Column (4) --
  Redshift. Column (5) -- Rest-frame FWHM, corrected for instrumental
  resolution in km 
  s$^{-1}$. Column (6) -- Line flux in units of 10$^{-15}$ erg s$^{-1}$ cm$^{-2}$. 
\end{table*}

\begin{table*}
\caption{Emission Lines in MRC0406-244}
\label{tab:emlines0406}
\begin{tabular}{lccccc}
\hline \hline
Line & $\lambda_0$ & $\lambda_{obs}$ & z  & FWHM & flux \\
(1)  & (2)         & (3)            & (4)& (5)  & (6) \\
\hline
\hline
H$\beta$    & 4861&1.66532 $\pm$0.00034&2.42587$\pm$0.00071& 1250$\pm$83 & 1.9 $\pm$ 0.1\\
$[$OIII$]$  & 4959&1.69860 $\pm$0.00034&2.42528$\pm$0.00069& 1524$\pm$93 & 2.2 $\pm$ 0.1 \\
$[$OIII$]$  & 5007&1.71495 $\pm$0.00034&2.42510$\pm$0.00068& 1349$\pm$81 & 6.6 $\pm$ 0.4\\
\hline
$[$OI$]$    & 6300& 2.15777$\pm$ 0.00045&2.42503$\pm$0.00072& 826 $\pm$ 65 &0.74 $\pm$ 0.07\\
H$\alpha +[$NII$]$& 6563,6583& 2.24879$\pm$ 0.00045&2.42647$\pm$0.00060& 1756$\pm$ 106&9.75 $\pm$ 0.59\\
$[$SII$]$   & 6716,6731& 2.30292$\pm$ 0.00048&2.42492$\pm$0.00071& 1590$\pm$ 102& 2.40$\pm$ 0.16\\
\hline
\end{tabular}\\
Column (1) -- line. Column (2) -- Rest-frame
  wavelength in \AA. Column (3) -- Observed wavelength in \AA. Column (4) --
  Redshift. Column (5) -- Rest-frame FWHM, corrected for instrumental resolution in km s$^{-1}$. Column
  (6) -- Line flux in units of 10$^{-15}$ erg s$^{-1}$ cm$^{-2}$. For
  H$\alpha$,[NII]$\lambda\lambda$6548,6583 and
  $[$NII$]\lambda\lambda$6716,6731 we give the values for the blended lines, respectively.
\end{table*}

\begin{table*}
\caption{Emission Lines in TXS0828+193}
\label{tab:emlines0828}
\begin{tabular}{lccccc}
\hline \hline
Line & $\lambda_0$ & $\lambda_{obs}$ & z & FWHM & flux \\
(1)  & (2)         & (3)            & (4)& (5) &  (6) \\
\hline
H$\beta$  & 4861& 1.73883$\pm$0.00039& 2.57710$\pm$0.0008& 1448$\pm$112& 0.6$\pm$0.1\\
$[$OIII$]$& 4959& 1.77422$\pm$0.00036& 2.57778$\pm$0.0007& 1239$\pm$76& 2.1$\pm$0.1\\
$[$OIII$]$& 5007& 1.79112$\pm$0.00036& 2.57723$\pm$0.0007& 1185$\pm$71& 5.9$\pm$0.4\\
\hline
$[$OI$]$          & 6300     & 2.2547$\pm$0.0006  & 2.57821$\pm$0.0009&1320$\pm$132&0.5$\pm$0.1\\ 
H$\alpha+[$NII$]$ & 6563,6583& 2.34705$\pm$0.00047& 2.57619$\pm$0.0007&1176$\pm$47&5.1$\pm$0.3\\
\hline
\end{tabular}\\
Column (1) -- line. Column (2) -- Rest-frame
  wavelength in \AA. Column (3) -- Observed wavelength in \AA. Column (4) --
  Redshift. Column (5) -- Rest-frame FWHM, corrected for instrumental resolution in km s$^{-1}$. Column
  (6) -- Line flux in units of 10$^{-15}$ erg s$^{-1}$ cm$^{-2}$. For
  H$\alpha$,[NII]$\lambda\lambda$6548,6583 and
  $[$NII$]\lambda\lambda$6716,6731 we give the values for the blended lines, respectively.
\end{table*}

\begin{table*}
\caption{Radio Properties of HzRGs}
\label{tab:radio}
\begin{tabular}{lcccc}
\hline \hline
Source & D$_{ang}$  & D$_{phys}$ & ${\cal L}_{0.1-1}$ & $\tau_{1.4GHz}$ \\
(1)  & (2)         & (3)            & (4) & (5) \\
\hline
MRC0316-257 & 7.6  & 58  & 45.0 & 0.94\\
MRC0406-244 & 7.3  & 59  & 45.2 & 0.96\\
TXS0828+193 & 12.8 & 103 & 45.4 & 1.7\\
\hline
MRC1138-262 & 11.1 & 92  & 45.2 & 1. \\
\hline
\end{tabular}\\
For comparison we add MRC1138-262 from \citet{nesvadba06}.  Column (1) --
Source name. Column (2) -- Maximal projected angular size of the radio
source. Column (3) -- Maximal projected size in kpc. Column (4) -- Integrated
luminosity between 0.1 and 1.0 GHz in the rest-frame in erg s$^{-1}$
cm$^{-2}$, corresponding to $\sim 0.001 - 0.1 \times$ the kinetic
luminosity. -- Column (5) -- Age of the radio jet for a constant expansion
speed of $0.01 c$ in $10^7$ yrs. Angular size distances are taken from
\citet{debreuck01}, radio luminosities are based on the VLA measurements of 
\citet{carilli97}. 
\end{table*}

\begin{table*}
\caption{Outflow Properties of HzRGs}
\label{tab:outflow}
\begin{tabular}{lcccccc}
\hline \hline
Source & D$_{opt}$ & $\Delta$v & $\tau_{opt}$
& $\dot{E}_{kin,obs}$ & $\dot{E}_{kin,ext}$ & $\dot{E}_{shell}$\\
(1)  & (2)         & (3)            & (4)& (5) & (6) & (7) \\
\hline
MRC0316-257 & 21$\times$9 & 670 & 3.1&      &     & 44.4 \\
MRC0406-244 & 30$\times$9 & 960 & 2.9& 45.3 &45.6 & 45.0 \\
TXS0828+193 & 28$\times$10& 800 & 4.1& 45.3 &45.3 & 45.0 \\
\hline
MRC1138-262 & 53$\times$32& 800 & 1  & 44.2 &     & 45.6 \\
\hline
\end{tabular}\\
For comparison we add MRC1138-262 from \citet{nesvadba06}.  
Column (1) --
Source name. 
Column (2) -- 
Extent of the optical line emission in kpc along
the major and minor axis, respectively. 
Column (3) -- 
Velocity offset between
the blue and redshifted bubble in km s$^{-1}$ in the rest-frame. 
Column (4) --
Age of the outflow for constant $\Delta$v and neglecting projection effects in
$10^7$ yrs. 
Column (5) -- Energy injection necessary to power the outflow in erg
s$^{-1}$('method 1', see \S~\ref{ssec:kinenergy} for details), and mass
estimates based on the measured H$\alpha$ fluxes, uncorrected for
extinction. This serves as a strict lower limit to the kinetic energy
injection. 
Column (6) -- 
The same as Column (5), but using extinction-corrected H$\alpha$ fluxes. 
Column (7) -- 
Energy injection necessary to inflate a hot bubble in erg s$^{-1}$,
('method 2', see \S\ref{ssec:kinenergy} for details).
\end{table*}

\begin{table*}
\caption{H$\alpha$ Luminosities and Ionized Gas Masses}
\label{tab:gasmasses}
\begin{tabular}{lccccc}
\hline \hline
Source & F$_{H\alpha}$ & F$_{H\alpha}^{corr}$ &  ${\cal L}$(H$\alpha$) &
${\cal L}$(H$\alpha$)$^{corr}$ & $\log({\rm M}_{H\alpha})$ \\
(1)  & (2)         & (3)            & (4) & (5) & (6) \\
\hline
0406-244  & -14.1 & -13.4 & 44.5 & 45.2 & 10.6\\
0828+193  & -14.2 & -13.4 & 44.4 & 45.3 & 10.7\\

\hline
\end{tabular}\\
Column (1) -- Source ID. Column (2) -- Measured integrated H$\alpha$
flux in erg s$^{-1}$ cm$^{-2}$. Column (3) -- Extinction corrected H$\alpha$
flux in erg s$^{-1}$ cm$^{-2}$.  Column (4) -- Integrated measured H$\alpha$
luminosity in erg s$^{-1}$. Column (5) -- Integrated extinction corrected
H$\alpha$ luminosity in erg s$^{-1}$.Column (6) -- Ionized gas mass estimate.
\end{table*}

\end{document}